\documentclass{PASAFreshNewFlavour}%

\usepackage{graphicx}
\usepackage{xcolor}
\definecolor{dodgerblue}{RGB}{30, 144, 255}

\usepackage{aas_macros}
\usepackage{hyperref} 
\hypersetup{colorlinks,citecolor=dodgerblue,linkcolor=dodgerblue,urlcolor=dodgerblue}

\usepackage{amsmath}

\usepackage{booktabs}
\usepackage{threeparttable}
\usepackage{etoolbox}
\usepackage{sansmath}
\AtBeginEnvironment{tablenotes}{\sffamily\sansmath}
\usepackage[skip=0.5em]{subcaption}
\usepackage{cleveref}
\usepackage{cprotect}
\usepackage{multirow}

\usepackage{newtxtext,newtxmath}
\usepackage{beramono}
\usepackage{sansmath}

\newcounter{ft}
\def\arcsec{\ensuremath{^{\prime\prime}}}

\def\meanalpha{\ensuremath{-1.4\pm0.2}}

\newcommand{\CORRS}[1]{{#1}}
\newcommand{\CORRSfinal}[1]{{#1}}

\makeatletter\ifdefined\Hy@StartlinkName
\def\addOneNestingLevelStartLink{%
  \gdef\Hy@StartlinkName##1##2{%
    \sbox0{\Hy@StartlinkNameOrig{##1}{##2}}\usebox0
    \global\let\Hy@StartlinkName\Hy@StartlinkNameOrig%
  }%
}
\def\addOneNestingLevelEndLink{%
  \gdef\pdfendlink{%
    \sbox0{\pdfendlinkOrig}\usebox0%
    \global\let\pdfendlink\pdfendlinkOrig%
  }%
}
\let\Hy@StartlinkNameOrig\Hy@StartlinkName
\let\pdfendlinkOrig\pdfendlink
\else
\let\addOneNestingLevelStartLink\relax
\let\addOneNestingLevelEndLink\relax
\fi
\makeatother

\title[Atypical radio-halo--hosting clusters]{MWA and ASKAP observations of atypical radio-halo--hosting galaxy clusters: Abell~141 and Abell~3404}

\author[S.~W.~Duchesne et al.]{S.~W. Duchesne$^1$\thanks{email: \url{stefan.duchesne.astro@gmail.com}},  M. Johnston-Hollitt$^{1,2}$, and A.~G. Wilber$^1$
\affil{$^1$International Centre for Radio Astronomy Research (ICRAR), Curtin University, Bentley, WA 6102, Australia}%
\affil{$^2$Curtin Institute for Computation, Curtin University,
GPO Box U1987, Perth, WA 6845, Australia}%
}%

\jid{PASA}
\doi{10.1017/pas.\the\year.xxx}
\jyear{\the\year}
\documentstatus{\textcolor{green}{accepted} for publication in PASA}

\usepackage{aas_macros}
\begin{document}

\begin{frontmatter}
\maketitle
\rule{\linewidth}{0.75pt}\vspace{11.5pt}
\begin{abstract}
    We report on the detection of a giant radio halo in the cluster Abell~3404 as well as confirmation of the radio halo observed in Abell~141 (with linear extents $\sim 770$~kpc and $\sim 850$~kpc, respectively). We use the Murchison Widefield Array (MWA), the Australian Square Kilometre Array Pathfinder (ASKAP), and the Australia Telescope Compact Array (ATCA) to characterise the emission and intervening radio sources from $\sim100$--1000~MHz; power law models are fit to the spectral energy distributions with spectral indices \CORRS{$\alpha_{88}^{1110} = -1.66 \pm 0.07$} and \CORRS{$\alpha_{88}^{943} = -1.06 \pm 0.09$} for the radio halos in Abell~3404 and Abell~141, respectively. \CORRS{We find strong correlation between radio and X-ray surface brightness for Abell~3404 but little correlation for Abell~141. We note each cluster has an} atypical morphology for a radio-halo--hosting cluster, with Abell~141 having been previously reported to be in a pre-merging state, and Abell~3404 is largely relaxed with only minor evidence for a disturbed morphology. We find that the radio halo powers are consistent with the current radio halo sample and $P_\nu$--$M$ scaling relations, but note that the radio halo in Abell~3404 is an ultra-steep--spectrum radio halo (USSRH) and, as with other USSRHs lies slightly below the best-fit $P_{1.4}$--$M$ relation. We find that an updated scaling relation is consistent with previous results and shifting the frequency to 150~MHz does not significantly alter the best-fit relations with a sample of 86 radio halos. We suggest that the USSRH halo in Abell~3404 represents the faint class of radio halos that will be found in clusters undergoing weak mergers.
\end{abstract}

\rule{\linewidth}{0.75pt}\vspace{11.5pt}
% Current file
% Overview
% 5

\end{frontmatter}

\section{Introduction}
\label{sec:intro}

Galaxy clusters represent ideal laboratories for investigating large-scale structure formation. As the largest virialized systems in the Universe, galaxy clusters are located at the nodes of the Cosmic Web and are assembled through hierarchical structure formation \citep{pee80}; accretion of matter from filaments and mergers between clusters releases energy into the intra-cluster medium (ICM) to be transferred into various non-thermal processes \citep[see e.g.][]{Sarazin2002,Keshet2004,bj14}. Resultant shocks and turbulence in the ICM are thought to energise electrons to emit synchrotron radio emission over large scales with steep power law spectra ($\alpha \lesssim -1$ \footnote{$S_\nu \propto \nu^\alpha$}; see e.g. \citealt{Brunetti2008,bj14,vda+19}) in the micro-Gauss--level magnetic fields permeating the clusters \citep[see e.g.][and \citealt{Donnert2019} for a recent review]{Clarke2001,Bruggen2012}.

The observed radio emission from ICM-based turbulence and shocks can be broken down into four main categories, with somewhat blurred lines between definitions \citep[see e.g.][for taxonomic discussion]{kempner2004,vda+19}. Mega-parsec--scale \textit{radio relics} \footnote{Also called radio \textit{shocks}: \citet{vda+19}.} are found in the low-density environments of cluster outskirts (e.g. in Abell~3667, \citealt{mj-h}; CIZA~J2242.8$+$5301, \citealt{vanWeeren2010}; PSZ1~G096.89$+$24.17 \citealt{dvb+14}; SPT-CL~J2032$-$5627, \citealt{Duchesne2020b})---shocks in the ICM are thought to energise electrons via diffusive-shock acceleration (DSA and related processes, e.g. \citealt{Blandford1987}) and relics have been observed to align with shocks detected via X-ray temperature and surface brightness discontinuities \citep[e.g.][]{Mazzotta2011,Akamatsu2015,Eckert2016b}. Smaller-scale ($\lesssim 400$~kpc) relic sources called \textit{phoenices} are thought to be the revived corpses of ancient, lobed radio galaxies, with the radio plasma re-energised by adiabatic compression via small-scale ICM turbulence and shocks \citep[][furthermore, see \citealt{srm+01,deGasperin2015b} for examples]{eg01} or gentle re-energisation of radio plasma from single \citep{deGasperin2017} or multiple electron populations \citep{Hodgson2021}. Phoenices are typically located closer to the cluster centre, and feature steeper spectra, with steepening beyond $\sim 1$~GHz. \textit{Mini-halos} are smaller ($\lesssim 400$~kpc), centrally-located, steep-spectrum patches of diffuse emission often found surrounding a radio-loud active galactic nucleus (AGN) associated with the brightest cluster galaxy (BCG). These sources are predominantly found in relaxed cool-core clusters, thought to form through the inner sloshing of the ICM gas, with seed electrons fuelled by the embedded AGN \citep[see e.g.][]{Giacintucci2017,Giacintucci2019}. Finally, \textit{giant radio halos} ($\gtrsim 1$~Mpc) are found in the centres of massive, merging clusters and are thought to also form through ICM turbulence generated by the merger process (e.g. in Abell~2255, \citealt{Harris1980}; 1E~0657$-$56, \citealt{Liang2000}; Abell~2163, \citealt{Feretti2001}; Abell~523, \citealt{Giovannini2011}). It has been suggested that mini-halos may transition into giant radio halos during mergers: the emitting cosmic ray electrons of mini-halos being transported throughout the cluster volume and re-accelerated via ICM turbulence \citep{bj14}. Observations suggest radio halos are transient phenomena---\citet{ddbc13} showed that the range of spectral and morphological shapes seen in observed halos can, in part, be attributed to when they occur during a merger.    \par
Historically, radio halo detections have been uncommon in clusters and the number of sources has, until recently, remained low. This was due in part to observational biases and limitations; many historic surveys were performed at 1.4~GHz, missing steep-spectrum emission only visible at lower frequencies. With the current generation of radio telescopes, telescope upgrades, increases in sensitivity, and low-frequency operation are helping to reveal a new population of radio halos \citep[e.g.][]{Duchesne2017,Cassano2019,HyeongHan2020,Wilber2020,DiGennaro2020,Hoeft2020,vanWeeren2020,Knowles2020, Hodgson2021} and to clarify the nature of previously detected systems \citep[e.g.][]{Botteon2020b,Bonafede2020}. \par

In this paper we present new observations of two clusters with candidate radio halo emission, originally detected in MWA data, now followed-up with the Australia Telescope Compact Array \citep[ATCA;][]{fbw92}, the Australian Square Kilometre Array Pathfinder \citep[ASKAP;][]{Hotan2021}, and the recently upgraded Murchison Widefield Array \citep{tgb+13} in its new ``phase 2'' extended baseline configuration \citep[][hereafter ``MWA-2'']{wtt+18}.
In this work we assume a flat $\Lambda$ cold dark matter cosmology with $H_0 = 70$~km\,s$^{-1}$\,Mpc$^{-1}$, $\Omega_\text{M} = 0.3$, and $\Omega_\Lambda = 1-\Omega_\text{M}$.

\subsection{Abell~141}\label{sec:a141}
\begin{figure}
    \centering
    \includegraphics[width=1\linewidth]{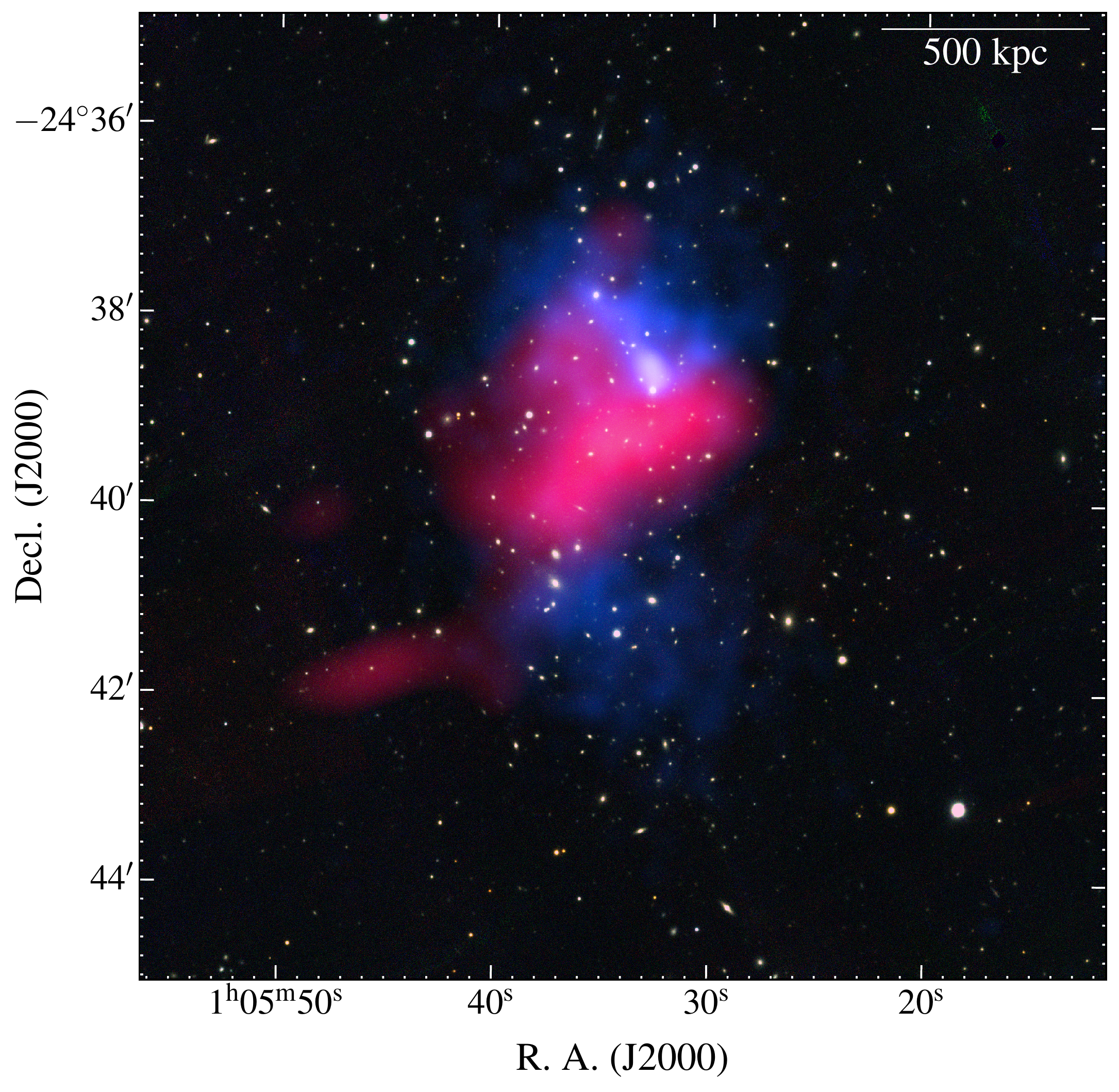}
    \caption{\label{fig:composite:a141} Composite image of Abell~141. Background optical data are from the Pan-STARRS survey, data release 1 \citep[bands \textit{r}, \textit{i}, \text{z};][]{kpc+10,tsl+12} with \textit{Chandra} (blue, Section \ref{sec:data:chandra}) and source-subtracted ASKAP (red, Section \ref{sec:data:askap}) maps overlaid. The linear scale is at the redshift of the cluster.}
\end{figure}

During a search for diffuse, non-thermal emission in a selection of galaxy clusters within a large MWA image at 168~MHz covering the Epoch of Reionization 0-hour field \citep[EoR0;][]{oth+16}, \citet{Duchesne2017} reported the detection of a giant radio halo in the massive, merging galaxy cluster Abell~141 \citep[][]{abell,aco89}. Using the Giant Metrewave Radio Telescope (GMRT), \citet{Venturi2007} reported the non-detection of a radio halo in Abell~141 at 610~MHz. Using the low resolution MWA data and the GMRT limit, \citet{Duchesne2017} reported a spectral index limit of $\alpha_{168~\text{MHz}}^{610~\text{MHz}} < -2.1$, making it one of the steepest-spectrum radio halo detected to date, tied with the radio halo in Abell~521 ($\alpha_{\text{240~MHz}}^{\text{610~MHz}} \approx -2.1$; \citealt{Brunetti2008}). \citet{Caglar2018} investigated the X-ray properties of the cluster, which show both a bi-modal X-ray distribution as well as a bi-modal optical distribution \citep{dki+02}. The cluster is also detected in \textit{Planck} Sunyaev--Zel'dovich (PSZ) surveys as PSZ2~G175.69$-$85.98 \citep{planck16b} and is reported to have an SZ-derived mass of $5.67^{+0.36}_{-0.40}\times 10^{14}$~M$_\odot$. The cluster is reported to have a redshift of $0.23$ \citep{sr99} where 1~arcmin corresponds to 221~kpc in scale. We show an updated composite image of Abell~141 in Fig. \ref{fig:composite:a141} with data described in Section \ref{sec:data:askap} and \ref{sec:data:chandra}.

\subsection{Abell~3404}\label{sec:a3404}

\begin{figure}
    \centering
    \includegraphics[width=1\linewidth]{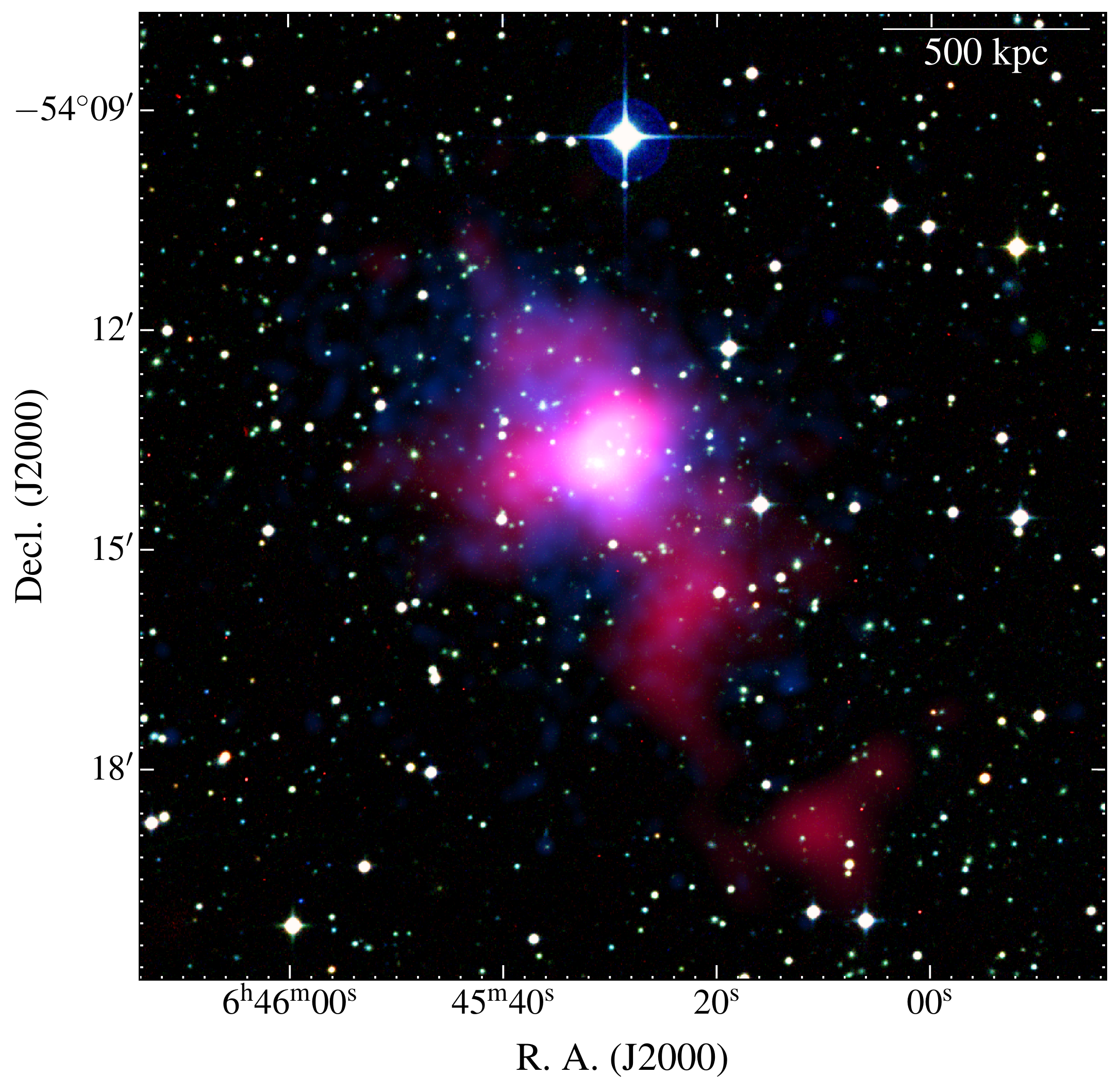}
    \caption{\label{fig:composite:a3404} Composite image of Abell~3404. Background optical data are from the SuperCOSMOS Sky Survey \citep{supercosmos1,supercosmos2,supercosmos3}. Radio and X-ray data overlaid as in Fig.~\ref{fig:composite:a141}. The linear scale is at the redshift of the cluster.}
\end{figure}

Abell~3404 was found to host unclassified extended emission permeating the cluster in  GaLactic and Extragalactic All-sky MWA \addOneNestingLevelStartLink\addOneNestingLevelEndLink\citep[GLEAM;][]{wlb+15,gleamegc} data at 200~MHz from a search for diffuse cluster emission within clusters from the Meta-Catalogue of X-ray detected galaxy Clusters \citep[MCXC;][]{pap+11}. Although the emission is elongated, typical of the morphologies of radio relics, its location at the centre of cluster suggested a halo-type source. The low resolution of the GLEAM data resulted in significant blending of the extended emission with nearby point sources and made it impossible to confirm its nature. \citet{sjp16} investigated the cluster using the Australia Telescope Compact Array \citep[ATCA;][]{fbw92} as part of the ATCA \textsf{REXCESS} \footnote{Representative \textit{XMM-Newton} Cluster Structure Survey \citep{rexcess}.} Diffuse Emission Survey (ARDES), though they found no evidence of a radio halo or other diffuse radio source. In advance of this publication, \citet{Bruggen2020} noted the detection of diffuse emission in Abell~3404 on the edge of a widefield observation of the cluster system Abell 3391-95 but leave the detailed characterisation to this work. The cluster has an SZ-derived mass of $7.96^{+0.23}_{-0.21}\times 10^{14}$~M$_\odot$ and redshift $z=0.1644$ \citep{planck16b}. At the cluster redshift 1~arcmin corresponds to 170~kpc in scale. A composite image of Abell~3404 is shown in Fig. \ref{fig:composite:a3404}. 

\section{Data \& methods}

\begin{table*}
\centering
\begin{threeparttable}
\caption{\label{tab:observing} Details of the radio observations of Abell~141 and Abell~3404.}
\begin{tabular}{cccccc}\toprule
Telescope & $\nu_\text{c}$ \tnote{a} & $\Delta\nu$ \tnote{b} & $\theta_\text{max}$ \tnote{c} & $\tau$ \tnote{d} & Dates  \\
& (MHz) & (MHz) & (arcmin) & (min) & \\\midrule
\multicolumn{6}{c}{Abell~141}\\\midrule
MWA-2 & 88 & 30 & 60 & 64 & 2017 Nov 03,04, Dec 22, 2018 Jan 06 \\
MWA-2 & 118 & 30 & 60 & 68 & 2017 Nov 03,04, Dec 22, 2018 Jan 06\\
MWA-2  & 154 & 30 & 60 & 64 & 2017 Nov 03,04,Dec 22, 2018 Jan 06\\
MWA-2  & 185 & 30 & 60 & 62 & 2017 Nov 03,04, Dec 22, 2018 Jan 06\\
MWA-2  & 216 & 30 & 60 & 60 & 2017 Nov 03,04, Dec 22, 2018 Jan 06\\
ASKAP & 943 & 288 & \CORRS{49} & \CORRS{2145} & \CORRS{2020 Apr 03,04, Jul 03,04, Nov 28}\\
ATCA (CABB) & 2100 & 1500 \tnote{e} & 5.7 & 928 & 2015 May 05,10,15,16 \\\midrule
\multicolumn{6}{c}{Abell~3404}\\\midrule
MWA-2& 88 & 30 & 60 & 58 & 2018 Jan 10, Feb 18,21, Mar 04,13  \\
MWA-2  & 118 & 30 & 60 & 42 & 2018 Jan 10, Feb 18, Mar 04,13  \\
MWA-2  & 154 & 30 & 60 & 58 & 2018 Jan 10, Feb 18,21, Mar 04,13  \\
MWA-2  & 185 & 30 & 60 & 42 & 2018 Jan 10, Feb 18,21, Mar 13 \\
MWA-2  & 216 & 30 & 60 & 38 & 2018 Jan 10, Feb 18,21, Mar 13 \\
ASKAP  & 1013 & 288 & 35 & 689 & 2019 Mar 22 \\
ATCA (pre-CABB)  & 1344 & 128 & 35 & 374 & 2007 Jun 10--16, Jul 26,28--30, Aug 02 \\
ATCA (pre-CABB)  & 1432 & 128 & 35 & 374 & 2007 Jun 10--16, Jul 26,28--30, Aug 02 \\
ATCA (CABB)  & 2100 & 1400 \tnote{e} & 25 & 103 & 2013 Jun 21--23 \\\bottomrule
\end{tabular}
\begin{tablenotes}[flushleft]
\footnotesize \item[a] \CORRS{Central observing frequency.} \item[b] \CORRS{Observation bandwidth.} \item[c] Maximum angular scale observation is sensitive to at the respective central frequency \CORRS{with a $u$--$v$ limit employed during imaging}. \item[d] \CORRS{Total observing time.} \item[e] \CORRS{Originally a 2049~MHz band, significant RFI flagging reduces the usable bandwidth.}
\end{tablenotes}
\end{threeparttable}
\end{table*}
\setcounter{ft}{0}

New data presented in this work are described in the following sections. A summary of observations used in this work and described in this section is presented in Table~\ref{tab:observing}. We also provide representative plots of the $u$--$v$ coverage for all datasets in Appendix \ref{appendix:uv}. 

\begin{figure*}[!t]
    \centering
    \begin{subfigure}{0.5\linewidth}
    \includegraphics[width=1\linewidth]{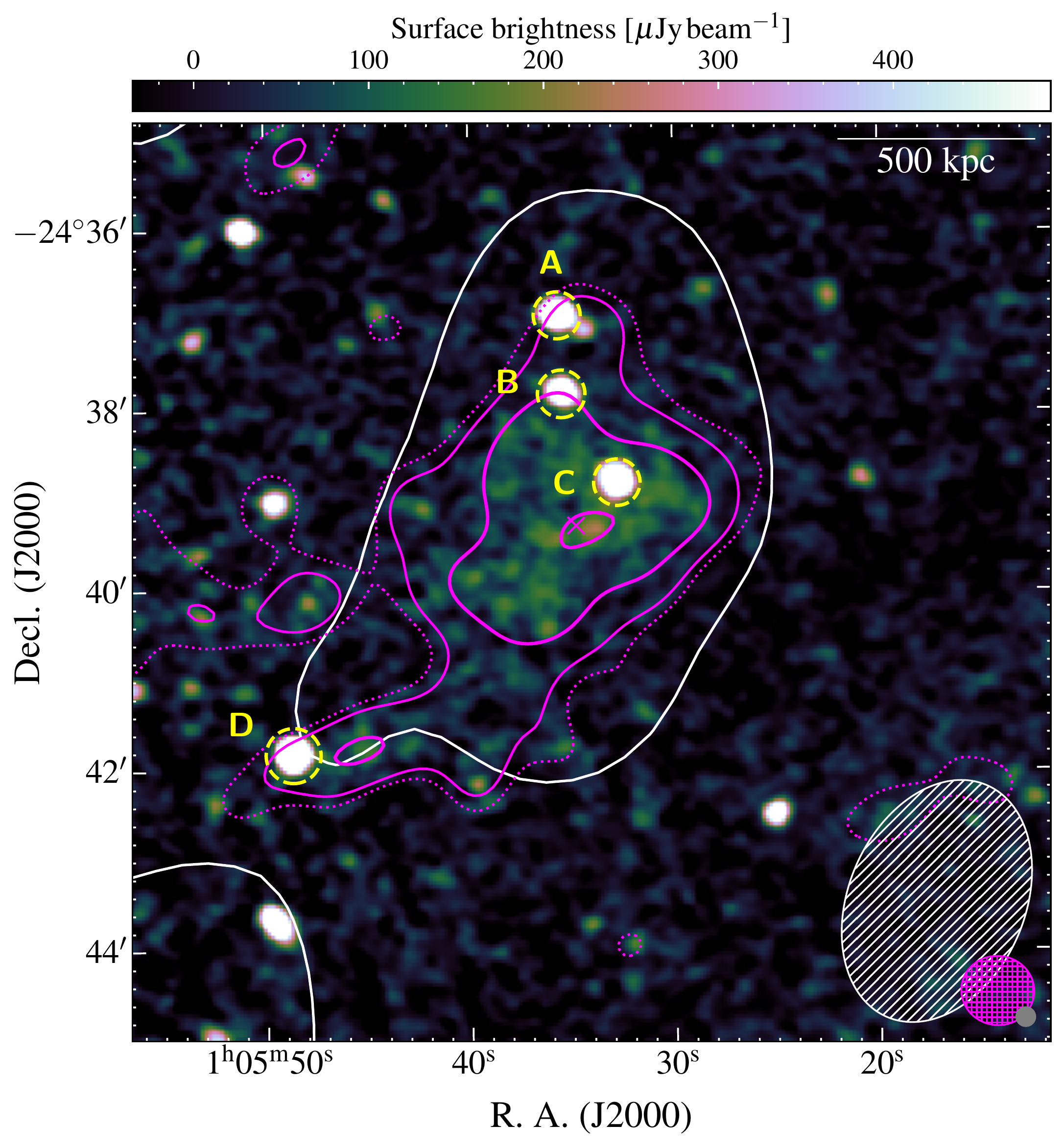}
    \caption{\label{fig:radio:a141:askap}}
    \end{subfigure}\hfill%
    \begin{subfigure}{0.5\linewidth}
    \includegraphics[width=1\linewidth]{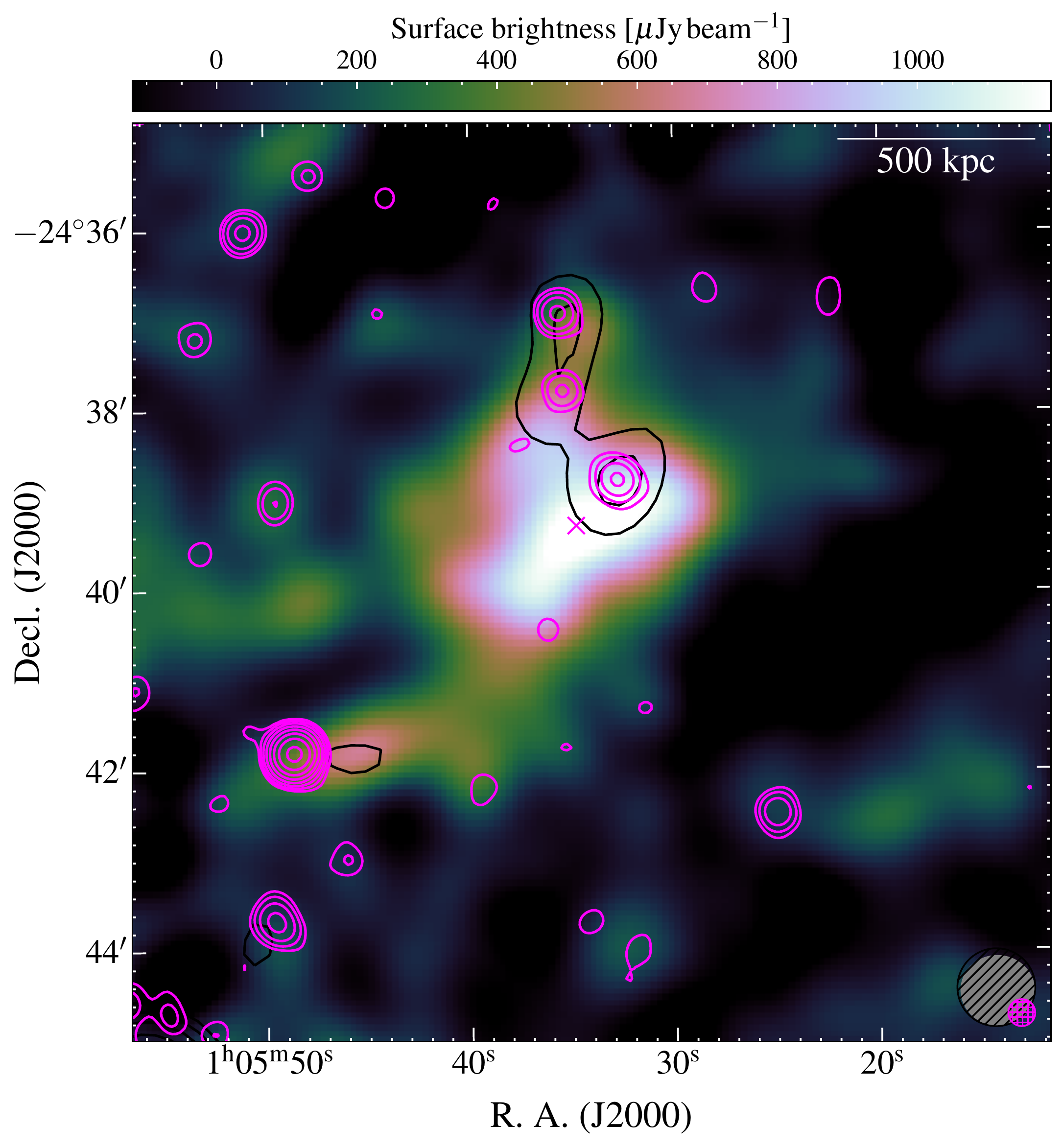}
    \caption{\label{fig:radio:a141:atca}}
    \end{subfigure}
    \caption{\label{fig:radio:a141} Abell~141 radio maps. \subref{fig:radio:a141:askap}: ASKAP, 943-MHz image at \CORRS{robust $=+0.25$}. The overlaid contours are as follows: MWA-2, \CORRS{118-MHz robust $=+2.0$} \CORRS{image}, single white contour at $3\sigma_\text{rms}$ (4.7~mJy\,beam$^{-1}$); ASKAP, \CORRS{943-MHz source-subtracted image}, solid magenta contours starting from $3\sigma_\text{rms}$ ($\sigma_\text{rms} = 0.105$~mJy\,beam$^{-1}$), increasing with increments of $2$ \CORRS{with a} single dotted magenta contour at $2\sigma_\text{rms}$. \CORRS{Sources subtracted after SED modelling are labelled}. \subref{fig:radio:a141:atca}: ASKAP, \CORRS{943-MHz source-subtracted image} with contours as follows: MWA-2, 216-MHz robust \CORRS{$=0.0$}, \CORRS{black contours} starting at $3\sigma_\text{rms}$ (11.4~mJy\,beam$^{-1}$); ATCA full-band \CORRS{image} at robust $=0.0$, magenta contours starting at $3\sigma_\text{rms}$ ($\sigma_\text{rms} = 27$~$\mu$Jy\,beam$^{-1}$\CORRS{)}. For both figures the resolution of each image is shown in the bottom right corner, with the grey ellipse corresponding to the background map. The linear scale in the top right is at the redshift of the cluster.}
\end{figure*}

\begin{figure*}[!t]
    \centering
    \begin{subfigure}{0.5\linewidth}
    \includegraphics[width=1\linewidth]{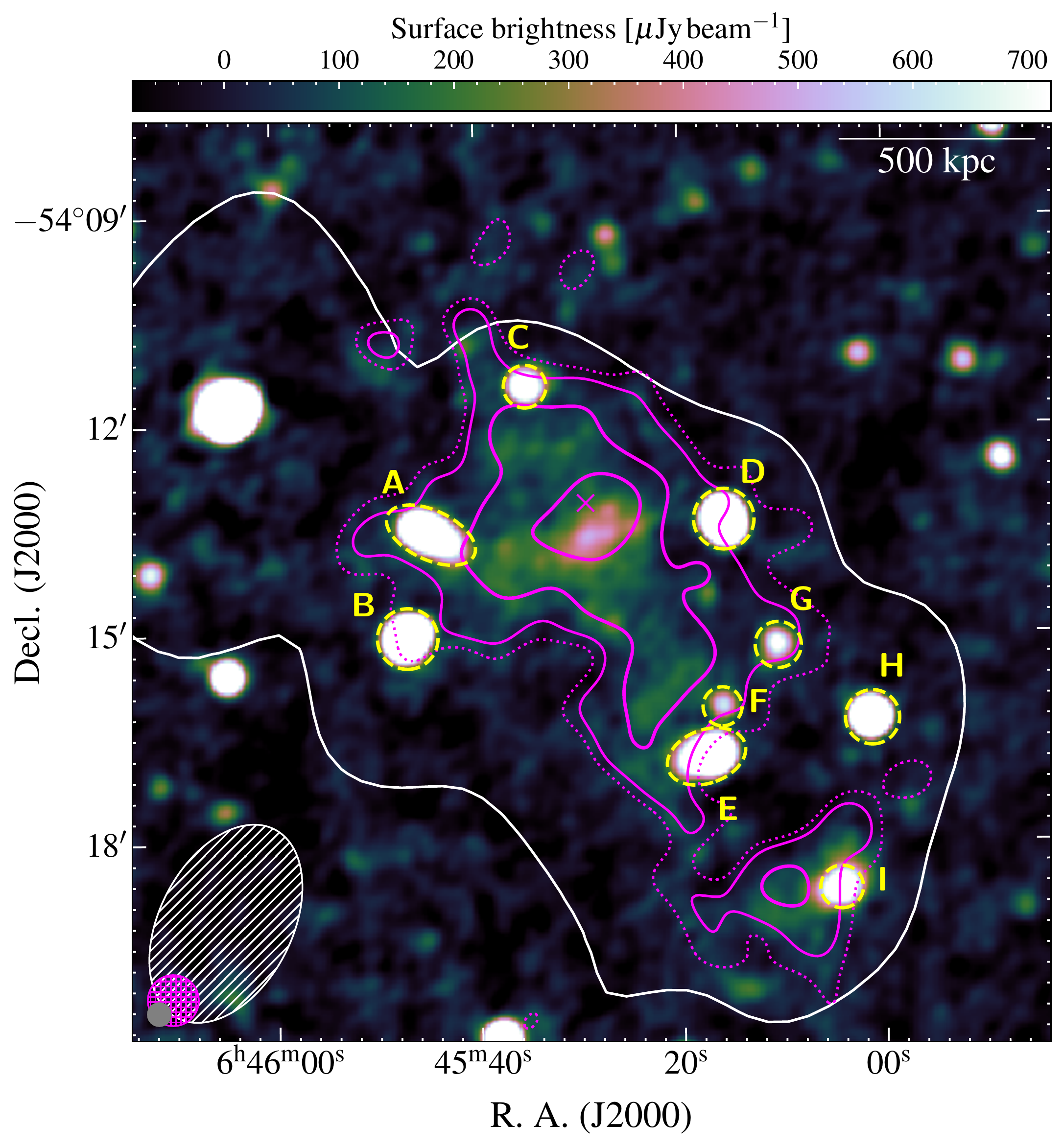}
    \caption{\label{fig:radio:a3404:askap}}
    \end{subfigure}\hfill%
    \begin{subfigure}{0.5\linewidth}
    \includegraphics[width=1\linewidth]{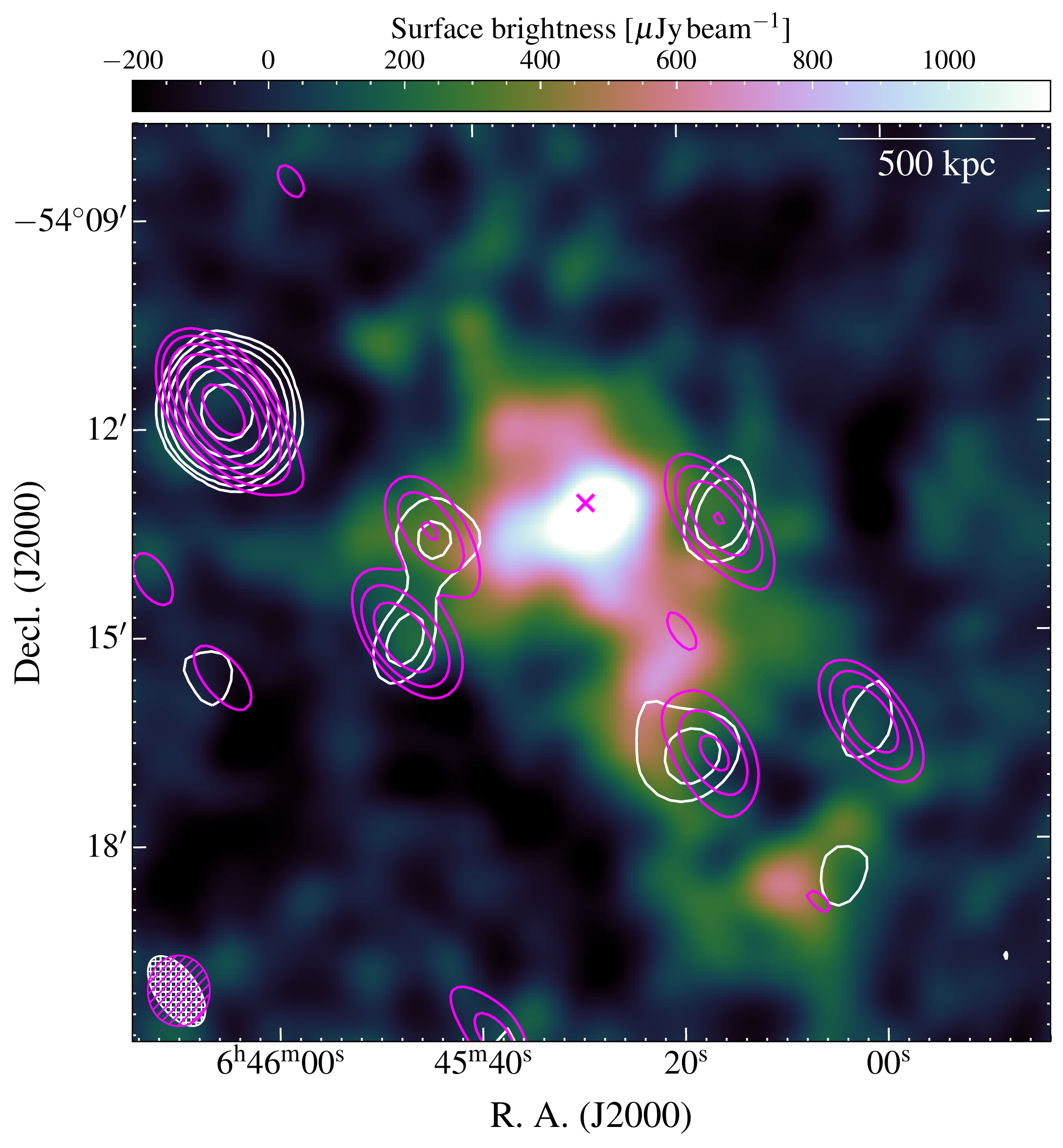}
    \caption{\label{fig:radio:a3404:atca}}
    \end{subfigure}
    \caption{\label{fig:radio:a3404} Abell~3404 radio maps. \subref{fig:radio:a3404:askap}: ASKAP, 1013-MHz image at robust $=+0.5$. The overlaid contours are as follows: MWA-2, 118-MHz robust \CORRS{$=+2.0$}, single white contour at $3\sigma_\text{rms}$ (6.1~mJy\,beam$^{-1}$); ASKAP, 1013-MHz tapered source-subtracted image, solid magenta contours starting from $3\sigma_\text{rms}$ ($\sigma_\text{rms} = 0.84$~mJy\,beam$^{-1}$), increasing with increments of \CORRSfinal{2} with a single dotted magenta contour at $2\sigma_\text{rms}$. \CORRS{Sources subtracted after SED modelling are labelled}. \subref{fig:radio:a3404:atca}: ASKAP source-subtracted, tapered, with contours as follows: MWA-2, 216-MHz robust $=0.0$ image, white contours starting at $3\sigma_\text{rms}$ ($\sigma_\text{rms} = 6.6$~mJy\,beam$^{-1}$); ATCA, 2.4-GHz robust $=0.0$ image, magenta contours starting at $3\sigma_\text{rms}$ ($\sigma_\text{rms} = 0.24$~mJy\,beam$^{-1}$). The ellipses in the the lower left are as in Fig.~\ref{fig:radio:a141} and the linear scale is at the redshift of the cluster.}
\end{figure*}

\subsection{MWA-2}

\subsubsection{Data processing}
Both Abell~141 and Abell~3404 were observed with the MWA-2 as part of a follow-up of candidate radio halos and relics detected with the GLEAM survey. The observations covered five frequency bands (of 30-MHz bandwidth): 88, 118, 154, 185, and 216-MHz. Observation details are presented in Table \ref{tab:observing}. The MWA data are processed following \citet{Duchesne2020}. Briefly, MWA data are observed in a 2-minute snapshot observing mode, with each 2-minute snapshot calibrated and imaged independently and stacked in the image plane at the end.

We use in-field calibration on a global sky model generated from the GLEAM, NVSS, and SUMSS catalogues (where survey coverage is available) using the \texttt{Mitchcal} algorithm \citep[see][]{oth+16}. The calibrated data are then imaged using \href{https://sourceforge.net/p/wsclean/wiki/Home/}{\texttt{WSClean}} \citep{wsclean1,wsclean2} to perform amplitude and phase self-calibration before imaging again to a lower threshold (i.e. CLEANing more deeply). After initial primary beam correction using the most recent Full-Embedded Element model \citep[][]{Sokolowski2017} these CLEANed images are then corrected for ionosphere-related astrometric shifts using \href{https://github.com/nhurleywalker/fits_warp}{\texttt{fits\_warp.py}} \footnote{\url{https://github.com/nhurleywalker/fits_warp}} \citep{hh18} and finally corrected for residual primary beam errors and flux scale errors with \href{https://gitlab.com/Sunmish/flux_warp}{\texttt{flux\_warp}} \footnote{\url{https://gitlab.com/Sunmish/flux_warp}} \citep{Duchesne2020} to ensure a common flux scaling across snapshots. The snapshots are then co-added, weighted by the square of the primary beam response and local noise.

We make a number of image sets for multiple purposes: a) using an image weighting with a `Briggs' robustness parameter of 0.0 \CORRS{for a maximum resolution image}, b) robust \CORRS{$=+2.0$}, and c) robust $=+1.0$ with \CORRS{an additional 120~arcsec Gaussian taper applied} to produce a more common \CORRS{sensitivity} between the five frequencies and to enhance the low surface-brightness emission in individual snapshots further. We find that, because of the slight difference in inner $u$--$v$ sampling between the five frequencies (see Appendix \ref{appendix:uv}, Figs. \ref{fig:uv:a141:mwa:c69}--\subref{fig:uv:a141:mwa:c169} and \ref{fig:uv:a3404:mwa:c69}--\subref{fig:uv:a3404:mwa:c169}) the 88- and 118-MHz images do not require additional tapering, \CORRS{though for Abell~3404 we find we can use the robust $=+2.0$ images for flux density measurement without additional confusion.} \par
Fig. \ref{fig:radio:a141:askap} and \ref{fig:radio:a141:atca} show the $3\sigma_\text{rms}$ contours of the \CORRS{robust $=+2.0$, 118-MHz image and the robust $=0.0$, 216-MHz} image of Abell~141, respectively. Similarly, Fig. \ref{fig:radio:a3404:askap} shows the $3\sigma_\text{rms}$ contour of the robust \CORRS{$=+2.0$}, 118-MHz image of Abell~3404 and Fig. \ref{fig:radio:a3404:atca} shows the $3\sigma_\text{rms}$ contour of the robust $=0.0$, 216-MHz image. 

\subsubsection{Measuring dirty flux density}

\begin{figure}
    \centering
    \begin{subfigure}[b]{1\linewidth}
    \includegraphics[width=1\linewidth]{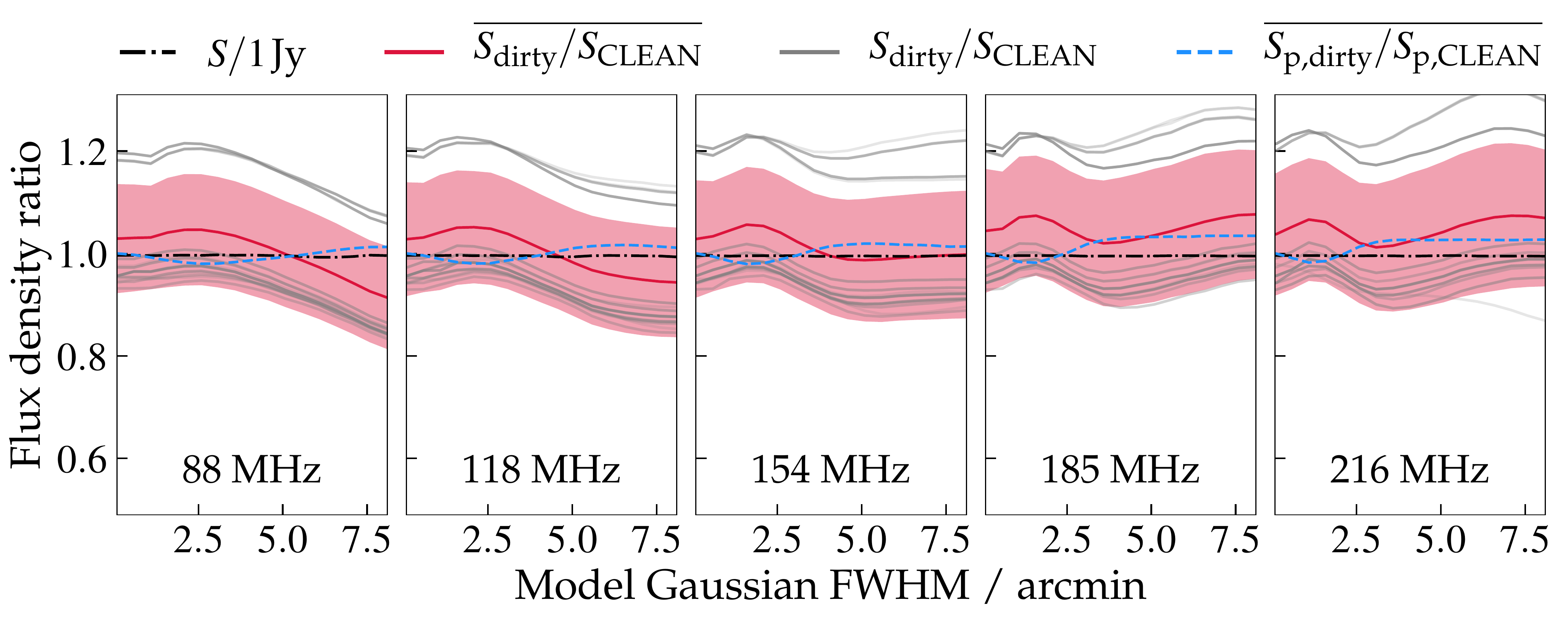}
    \caption{\texttt{FIELD1}, robust $+2.0$.}
    \end{subfigure}\\
    \begin{subfigure}[b]{1\linewidth}
    \includegraphics[width=1\linewidth]{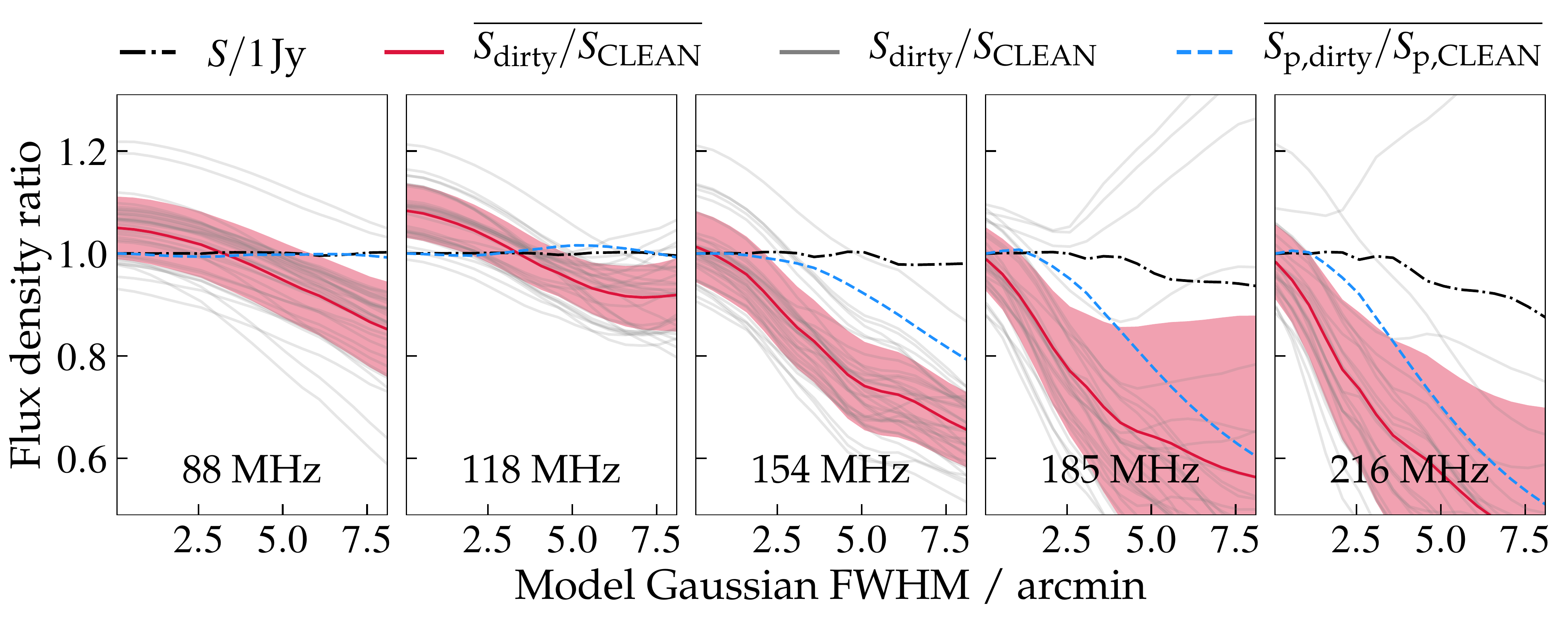}
    \caption{\texttt{FIELD8}, robust $+2.0$.}
    \end{subfigure}
    \caption{\CORRS{Comparison of dirty and CLEANed flux densities in the robust $+2.0$ MWA-2 images as a function of source FWHM for simulated Gaussian sources. Note that individual snapshot $S_\text{dirty}/S_\text{CLEAN}$ ratios are shown as transparent grey lines with the mean value plotted as a solid red line, and a shaded region corresponding to the standard deviation between snapshots.} \label{fig:dirtyclean}}
\end{figure}

Snapshot imaging results in residual dirty flux (i.e. emission that has not been deconvolved by the CLEAN algorithm) that becomes significant in final stacked images as the CLEANing depth is set in the individual snapshot images. Using multiscale CLEAN can mitigate this somewhat as the root-mean-square (rms) noise on larger scales allows CLEANing of large structures below the point-source CLEAN depth. The result is measurement of residual dirty flux, which 1) has a point spread function (PSF) that differs from the restoring beam and 2) the dirty flux may be reduced or increased compared to the equivalent CLEAN flux due to complex PSF sidelobe interactions. We assume that the PSF difference is small---i.e. the fitted restoring beam is an accurate representation of the PSF. To test the difference in measured dirty to CLEAN flux density, we simulate 2-d circular Gaussian sources \CORRS{in all snapshots} with an arbitrary $S = 1$~Jy with varying full-width at half maximum (FWHM) sizes, ranging from 5\arcsec\ to $575$\arcsec\ in 30\arcsec\ intervals and measure the integrated flux density in the dirty and CLEANed maps. We find that the ratios of measured $S_\text{p,dirty}/S_\text{p,CLEAN}$ \CORRSfinal{(peak)} and $S_\text{dirty}/S_\text{CLEAN}$ \CORRSfinal{(integrated)} decrease \CORRS{down to} $\sim$0.5--0.6 for structures up to $10$~arcmin \CORRS{for the Abell~3404 data, but increase by \CORRSfinal{only} a few per cent for Abell~141}. \CORRS{We show the robust $=+2.0$} results in Fig. \ref{fig:dirtyclean}. \par
To correct for this, we create separate CLEAN component model and residual stacked images to match the restored stacked image. When measuring $S$ for real sources, we use the restored image to guide the integration region, but sum the CLEAN model and add the integrated residuals with a correction factor applied to the residual flux density determined by the size of the emission region. The correction factor is determined by the convolved source size and estimated from the nearest simulated ratio of $S_\text{dirty}/S_\text{CLEAN}$.

\subsection{ATCA}\label{sec:atca}

Abell~3404 was observed with both the Compact Array Broadband Backend \citep[CABB;][]{cabb} and the ATCA correlator prior to the CABB installation \citep[hereafter ``pre-CABB'', Project Codes C1683 and C2837;][]{atca:a3404:c1683,atca:a3404:c2837}. The observation details are listed in Table \ref{tab:observing}, though note for both the CABB and pre-CABB observations Abell~3404 was observed in a ``\emph{u--v} cuts'' mode at a range of hour-angles, but not filling in the $u$--$v$ plane significantly. The pre-CABB and CABB data have different phase centres, and the lower end of the CABB data are flagged due to RFI so the two data sets are imaged independently. Abell~141 was observed with the CABB \citep[Project Code C2915;][]{atca:a141}, though two configurations were used: 1.5C and 6A, with maximum/minimum baselines of 4500/77 and 5939/337~m, respectively.

Calibration for all ATCA data follows standard data reduction procedures using \href{https://www.atnf.csiro.au/computing/software/miriad/}{\texttt{miriad}} \citep{stw95}. Bandpass and absolute flux calibration is performed using the standard ATCA centremetre calibrator, PKS~B1934-638, and appropriate secondary calibrators bracket the source observations, used for complex gain and phase calibration. We perform two rounds of self-calibration on each dataset, using \href{https://sourceforge.net/p/wsclean/wiki/Home/}{\texttt{WSClean}} to create initial CLEAN component models and the \href{https://casa.nrao.edu/}{Common Astronomy Software Applications} \citep[\texttt{CASA};][]{casa} task, \texttt{gaincal}, to solve for phase-only gain solutions on successively shorter intervals (i.e. 120~s and 30~s). \CORRSfinal{Additionally, final imaging for the Abell~3404 data removes the longest baselines formed with antenna 6 to achieve a more well-behaved point spread function.} Imaging for all datasets is otherwise similar, utilising a `Briggs' robustness parameter of 0.0, and splitting the full CABB bands into smaller subbands. Note that significant RFI flagging occurs in the 16-cm band for the ATCA data and the final usable band for CABB observations is \CORRS{$\sim 1.5$~GHz and is split into subbands of $\Delta\nu = 300$~MHz for discrete source measurements}. \par

Fig. \ref{fig:radio:a141:atca} shows the robust $=0.0$ full-band \CORRS{2.2-GHz ATCA map \CORRSfinal{for Abell~141} convolved to 18~arcsec.} Fig. \ref{fig:radio:a3404:atca} shows the similar 2.4-GHz full-band map for Abell~3404.

\subsection{ASKAP}\label{sec:data:askap}

ASKAP operates between 700--1800~MHz and features a Phased Array Feed \citep[PAF;][]{askap3,Hotan2014,MCCONNELL2016} allowing the creation of 36 independent primary beams which can be arranged in a number of different footprints to create large $7^\circ \times 7^\circ$ mosaics in ``6 by 6'' primary beam footprints. ASKAP has recently been used to complete some observations for early science and survey projects \citep[e.g. the Evolutionary Map of the Universe, EMU;][]{Norris2011a}. Abell~3404 and Abell~141 feature in early science observations, however, both Abell~141 and Abell~3404 sit towards the edges of primary beams. The ASKAP data are publicly available and are retrieved from the CSIRO \footnote{Commonwealth Scientific and Industrial Research Organisation} ASKAP Science Data Archive \citep[CASDA;][]{casda}. Prior to being made available through CASDA, the \texttt{ASKAPsoft} \footnote{\url{https://www.atnf.csiro.au/computing/software/askapsoft/sdp/docs/current/pipelines/introduction.html}} pipeline uses daily observations of PKS~B1934-638 for bandpass calibration, with each of the 36 beams being calibrated independently. Additionally, the data are averaged to 1~MHz/10~s spectral/temporal resolution. The full bandwidth for each observation is 288~MHz. We summarize additional observation details in Table \ref{tab:observing}.

\subsubsection{ASKAP---Abell~3404}

The ASKAP Scheduling Block (SB)8275 \citep{askap:a3404} has two overlapping beams containing Abell~3404 (beam 17 and 23) with a central observing frequency of $\nu_\text{c} = 1013.5$~MHz. We follow a similar self-calibration process to the ATCA data described in Section \ref{sec:atca}, though this calibration does not reduce the well-known, but not currently understood artefacts that appear around bright sources at a $\sim1$~per~cent level. For the two beams containing Abell~3404, these artefacts are negligible and do not interfere with the cluster. Imaging is performed using \href{https://sourceforge.net/p/wsclean/wiki/Home/}{\texttt{WSClean}}, and we image by splitting the data into 4 subbands of $\Delta\nu = 72$~MHz, jointly CLEANing in the fullband multi-frequency synthesis (MFS) image. Imaging for these data is done by first masking the diffuse emission within the cluster region, ensuring all discrete sources \textit{are} included in the CLEAN process. The initial image weighting is robust $=+0.5$, which we found to be most accurate for modelling the discrete cluster sources. The second image set we produce are the re-imaged residuals convolved with a 25~arcsec beam to highlight diffuse cluster emission. Finally, we re-image the residuals with an additional 35~arcsec taper in 3 subbands of $\Delta\nu = 96$~MHz. 

For each image set, we linearly mosaic beams 17 and 23, applying a correction for primary beam attenuation assuming a 2-dimensional Gaussian model that scales with $1.09\lambda/D$ (A.~Hotan, priv. comms.) with $D=12$~m the diameter of the ASKAP dishes. For quality assurance we compare the spectral energy distribution (SED) of a nearby, bright test source measured by the MWA-2, ATCA, and additional survey data and find that the ASKAP data follow the expected flux density. Additionally, we find an astrometric offset of $\Delta\delta_\text{J2000} \sim -0.0055$, which we correct in the image World Coordinate System (WCS) metadata. Fig. \ref{fig:radio:a3404:askap} shows the ASKAP full-band robust $=+0.5$ image with the source-subtracted 35~arcsec tapered image as contours.

\subsubsection{ASKAP--Abell~141}

\CORRS{Abell~141 is present on the edge of a beam (23) in a number of ASKAP observations \footnote{\CORRS{SB9602, SB9649, SB9910, SB10463, SB12704, SB15191, SB18912, SB18925; \citep{askap:a141}, as part of a gravitational wave follow-up programme, with an average of 8--10-h per observation with similar $u$--$v$ tracks.}}. After an initial round of imaging with all available observations we find most of the observations contain significant radial artefacts crossing the cluster from a nearby bright source which is unable to be removed via direction-independent calibration. We find the SB12704, SB15191, and SB18925 observations do not show significant artefacts and combine those observations for further joint imaging.} We opted to self-calibrate the data, generating a combined, jointly-deconvolved model image of the \CORRSfinal{three} datasets then self-calibrating each dataset individually based on the combined model. Initial imaging was carried out similarly to the Abell~3404 data, masking the cluster diffuse emission. The difference in this imaging process is we use a robust \CORRS{$=+0.25$} weighting scheme, which for these observations provided a better model of the cluster discrete sources. Additionally, we find that since the diffuse emission is much fainter than in Abell~3404, and because the cluster lies further from the primary beam centre, we only consider the fullband re-imaged residual visibilities rather than the subbands produced while CLEANing. As there is only a single primary beam, mosaicking is not required, however, a primary beam correction is applied. Because the source lies to the edge of the beam, we compare the ASKAP flux densities of sources across the image to extrapolated flux densities derived from the ATCA subband images and the catalogue used for MWA-2 flux scaling (independently, see \citealt{Duchesne2020} for details of the calibration catalogue). For comparison with the MWA catalogue, the data are first convolved to a common resolution. \CORRS{We find the flux densities of sources do not deviate beyond $\sim 10$~per cent.} We find no discernible astrometric offset for these data. 

The full-band, robust $=+0.25$ image is shown in Fig.~\ref{fig:radio:a141:askap} with the source-subtracted, 45~arcsec robust $=+0.25$ image as contours in Fig.~\ref{fig:radio:a141:askap} and the background in Fig.~\ref{fig:radio:a141:atca}.

\subsection{\textit{Chandra}}\label{sec:data:chandra}

Both clusters have been observed with the Advanced CDD Imaging Spectrometer (ACIS-I) on the \emph{Chandra} X-ray observatory. We obtain both datasets (Abell~141: Obs.~ID 9410, PI: Smith, 19.91~ks; Abell~3404: Obs.~ID 15301, PI: Murray, 9.96~ks) from the \href{https://cda.harvard.edu/chaser/}{\emph{Chandra} data archive} and use the \href{http://cxc.cfa.harvard.edu/ciao/}{\texttt{CIAO}} software suite \citep[v4.12, with CALDB \footnote{Calibration database.} v4.9.1;][]{Fuscione2006} to process the data following standard \textit{Chandra} data reduction procedures, using the task \verb|chandra_repro| to generate the level-2 event file. From this we generate count and exposure-corrected flux images using the task \texttt{fluximage} applying 1~arcsec binning for the full energy band ([0.5--7]~keV). \CORRS{We use \texttt{wavdetect} to identify point-like sources in the images and remove them, finally creating images in the [0.5--2]~kev band. For Abell~141, as there is an AGN at the centre of the northern sub-cluster that is subtracted, we fill in the removed component using the task \texttt{dmfilth}.} Additional smoothing with a $\sigma = 6$~arcsec 2-dimensional Gaussian kernel is applied to the exposure-corrected image, using the task \texttt{aconvolve}. The \CORRS{[0.5--2]~kev} exposure-corrected, smoothed, \CORRS{and source-subtracted} \textit{Chandra} images are shown in Fig. \ref{fig:xray} for each cluster.

\section{Results \& discussion}\label{sec:results}

\subsection{Morphology}\label{sec:results:morph}

\begin{figure*}[t]
    \centering
    \begin{subfigure}{0.5\linewidth}
    \includegraphics[width=1\linewidth]{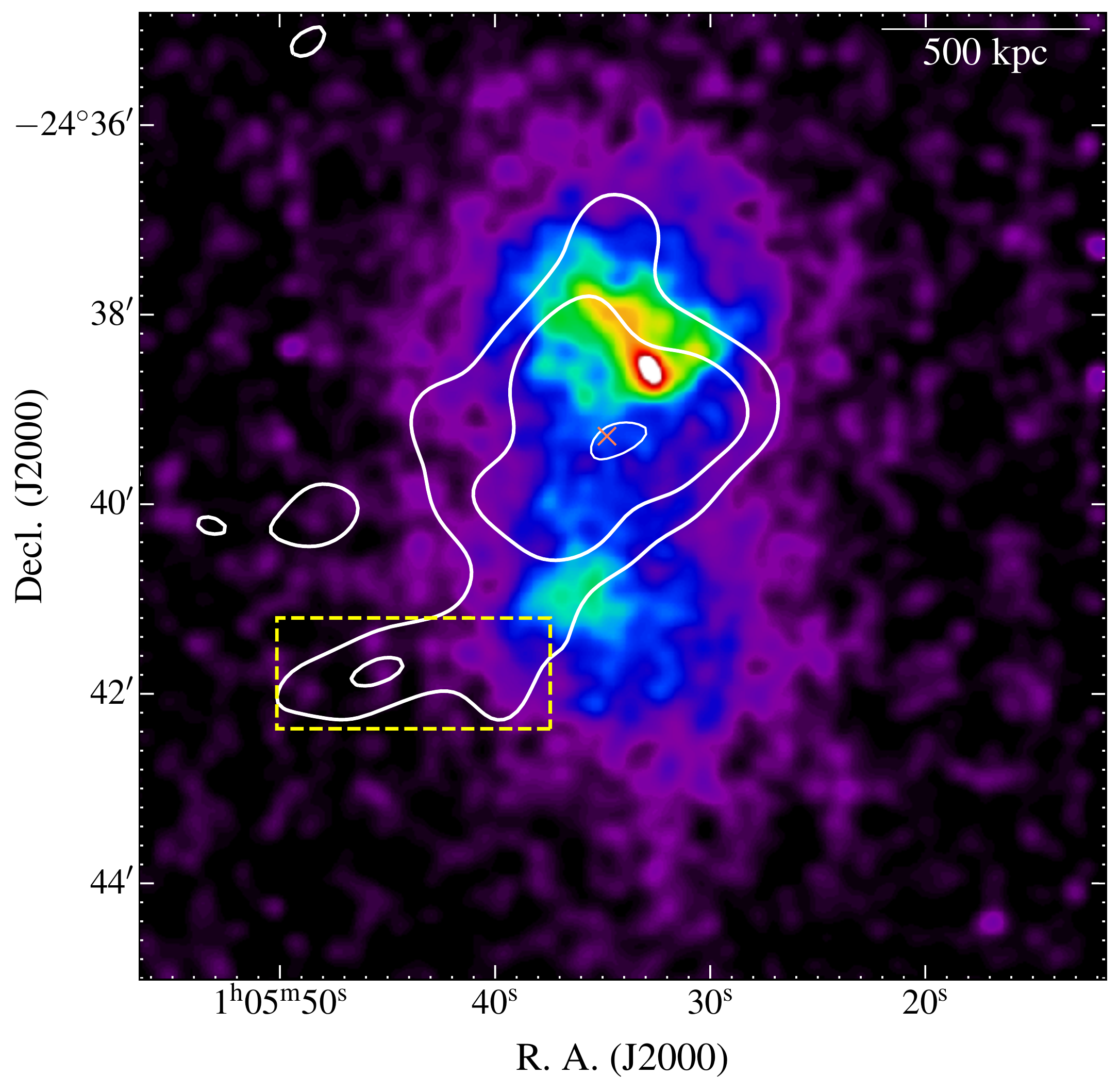}
    \caption{\label{fig:xray:a141} Abell~141.}
    \end{subfigure}%
    \begin{subfigure}{0.5\linewidth}
    \includegraphics[width=1\linewidth]{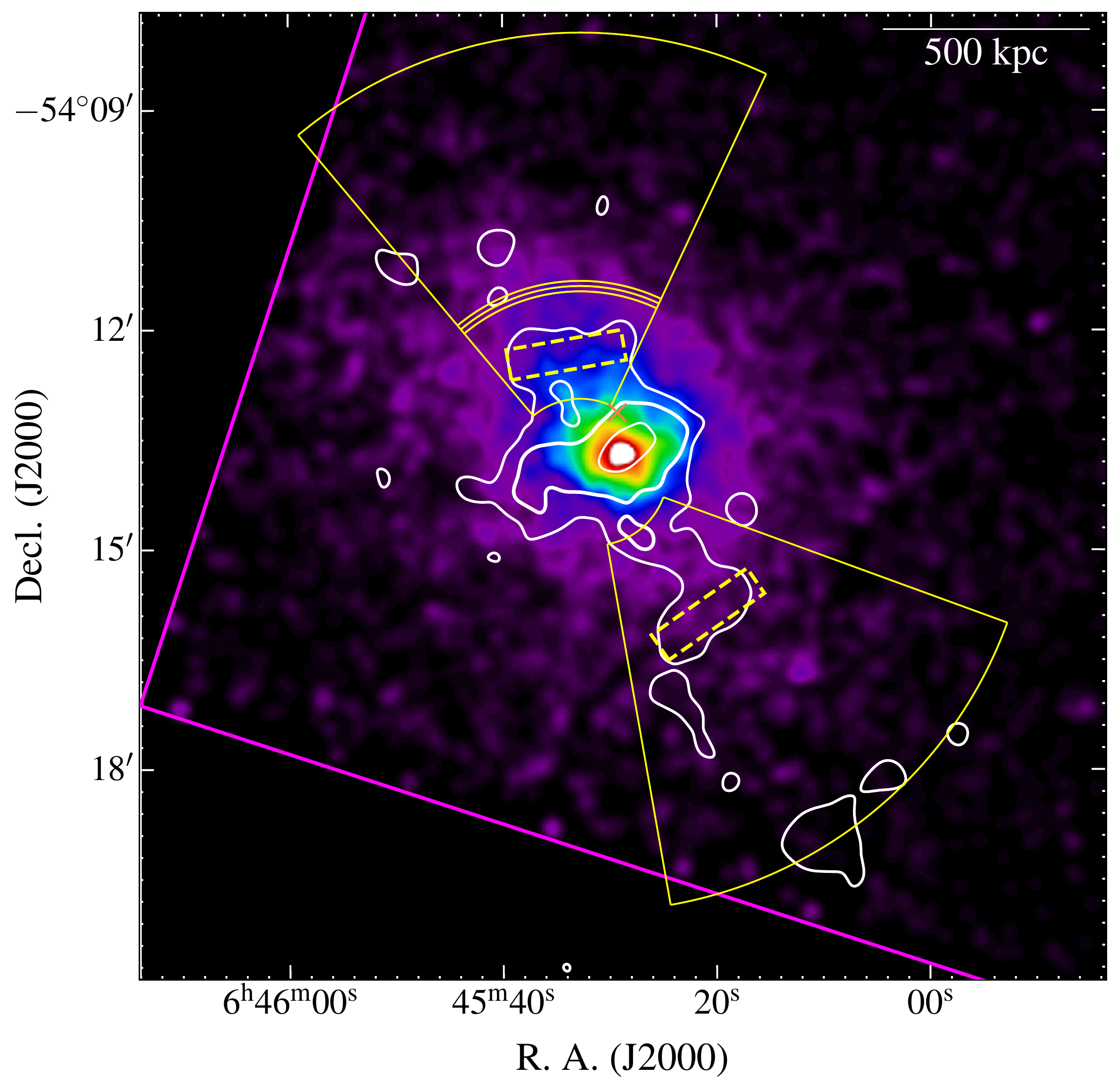}
    \caption{\label{fig:xray:a3404} Abell~3404.}
    \end{subfigure}%
    \cprotect\caption{\label{fig:xray} Exposure-corrected, smoothed, \CORRS{point source-subtracted} \CORRS{[0.5--2]~kev} \emph{Chandra} maps with source-subtracted ASKAP contours overlaid. \subref{fig:xray:a141} Abell~141: ASKAP contours as in Fig. \CORRSfinal{\ref{fig:radio:a141:askap}, but without the $2\sigma_\text{rms}$ contour.} \subref{fig:xray:a3404} Abell~3404: ASKAP contours are the source-subtracted \CORRS{without tapering}, but convolved with a 25~arcsec beam (hence, slightly higher resolution than the \CORRS{tapered}, source-subtracted map). In \subref{fig:xray:a141} the yellow, dashed box indicates the peripheral diffuse source. In \subref{fig:xray:a3404} we show the regions within which we extract radio and X-ray surface brightness profiles (yellow sectors) and indicate extended radio components in those profiles (yellow, dashed rectangles) that may indicate radio shocks discussed in Section \ref{sec:results:xray:a3404:shocks}. \CORRS{The magenta lines indicate the edge of the ACIS-I field-of-view.}}
\end{figure*}

The ASKAP and MWA-2 data show clear evidence of central diffuse emission and additional extended peripheral structures in Abell~141 and Abell~3404 (see Fig. \ref{fig:radio:a141:askap}, \ref{fig:radio:a3404:askap}, \CORRSfinal{\ref{fig:xray:a141}, and \ref{fig:xray:a3404}}), not necessarily associated with any particular radio galaxy, \CORRS{though it is unclear if these components are associated with the central diffuse sources}. No diffuse radio emission is detected in the ATCA data for either cluster (Fig. \ref{fig:radio:a141:atca} and Fig. \ref{fig:radio:a3404:atca}). We show the exposure-corrected, smoothed X-ray maps in Fig.~\ref{fig:xray} to highlight the morphology of the clusters and the co-location of the radio emission.

\subsubsection{Abell~141} \CORRS{The central, diffuse radio emission in Abell~141 has a slightly elongated morphology} as seen in the source-subtracted ASKAP data, with an extension \CORRSfinal{towards the southeast} becoming a distinct peripheral component (yellow, dashed box in Fig. \ref{fig:xray:a141}). The radio emission fills the volume between two X-ray sub-clusters, and extends into the northern sub-cluster. \CORRS{Excluding the peripheral source, w}e measure the size of the central diffuse emission in Abell~141 from N--S and E--W within $2\sigma_\text{rms}$ contours finding deconvolved dimensions of \CORRS{$4.2$~arcmin and $3.7$~arcmin, respectively, corresponding to a linear size of $910$~kpc and $790$~kpc.} We will consider the mean linear extent to be 850~kpc. \CORRS{Additionally, the SE peripheral component has a maximum projected extent of $2.6$~arcmin, corresponding to $550$~kpc.}

\subsubsection{Abell~3404} \CORRS{The central diffuse emission in Abell~3404 is also elongated and we similarly} see peripheral extended components that may not be associated with the central diffuse source---these are most prominent in the 25~arcsec source-subtracted ASKAP image (contours in Fig. \ref{fig:xray:a3404}, indicated by yellow, dashed rectangles). The peak of \CORRSfinal{the} central radio emission is co-located with the X-ray peak, and the radio emission generally fills the X-ray--emitting area. The \CORRS{size of the} central diffuse source is measured in the ASKAP source-subtracted, 35~arcsec \CORRSfinal{tapered} image within $2\sigma_\text{rms}$ contours as above, finding angular sizes in the N--S and E--W directions of $5.3$~arcmin and $3.9$~arcmin, respectively, corresponding to linear sizes of $900$~kpc and $660$~kpc. We note that the the N--S direction is influenced by additional peripheral components, though it is not clear if these are part of the central emission or not (further discussed in Section \ref{sec:results:xray:a3404:shocks}). \CORRSfinal{For the estimate of the size the SE component---which is more distinct---is not included}. We again will consider the mean linear extent of $770$~kpc in the following sections. \CORRS{The NE peripheral component is found to have a maximum projected size of 1.7~arcmin (270~kpc) and the SW component is found to be 1.5~arcmin (230~kpc) in extent.}

\subsection{Radio spectral properties}
\subsubsection{Flux densities}

Fig. \CORRSfinal{\ref{fig:radio:a141:askap} and \ref{fig:radio:a3404:askap}} also show relevant discrete sources that are projected onto the clusters within the MWA-2 emission. For the ASKAP data these were subtracted in the visibilities using CLEAN component models (Section \ref{sec:data:askap}), but in the MWA data their contribution is subtracted from the integrated flux density measurement after extrapolation from \CORRSfinal{their measured} SEDs. The central diffuse emission in Abell~3404 from \CORRS{185--216~MHz} is only barely detected above a $3\sigma_\text{rms}$ significance \CORRS{in the MWA-2 data}, with generally poorer image qualities and lack of detection in individual snapshots. As we cannot guarantee significant enough flux is recovered in these images, we opt not to provide measurements for Abell~3404 \CORRS{in the 185- and 216-MHz MWA-2 images. We instead measure the source using the 200-MHz GLEAM image. For Abell~141, we use all MWA-2 bands and re-measure the source in the 169-MHz EoR-0 image \citep{oth+16} and the 200-MHz GLEAM image. We find the 169-MHz measurement is lower than reported by \citet{Duchesne2017} largely due to additional discrete source-subtraction.} 

Table \ref{tab:sed:discrete} shows the fitted power law properties of the discrete sources in each cluster. We measure the flux density of the peripheral diffuse sources in \CORRS{Abell~141 and }Abell~3404 in the full-band, 25~arcsec ASKAP image, though are unable to provide a spectral index estimate. Individual flux density measurements \CORRS{of the diffuse sources} are provided in Table \ref{tab:sed}, \CORRS{indicating the contributions from discrete sources that} are subtracted from the total integrated flux density in the process. \CORRS{The peripheral components in each cluster contribute to the diffuse source measurements as we cannot subtract them from the MWA data and we only provide measurements of the perierphal sources in the full-band ASKAP images.} \CORRS{Relevant details of images used for flux density measurements are also provided in Table~\ref{tab:sed}.}

\begin{table}[!t]
    \centering
    \begin{threeparttable}
    \caption{\label{tab:sed:discrete} \CORRS{Discrete source SED properties for both clusters.}}
    \begin{tabular}{l c c}\toprule
    Source & $\alpha$ & $\Delta\nu$ \tnote{a} \\
    {}     & {}       & (MHz) \\\midrule
    \multicolumn{3}{c}{Abell~141} \\\midrule
    A & $-1.09\pm0.07$ & 216--2674 \\
    B \tnote{b} & $-1.44\pm0.05$ & 147--2674 \\
    C & $-1.07\pm0.07$ & 147--2674 \\
    D \tnote{c} & $+1.18\pm0.09$ & 943--3000 \\\midrule
    \multicolumn{3}{c}{Abell~3404} \\\midrule
    A & $-0.49\pm0.11$ & 216--1388 \\
    B & $-0.35\pm0.10$ & 216--1388 \\
    C & $-1.02\pm0.14$ & 185--1121 \\
    D \tnote{b} & $-0.67\pm0.18$ & 185--2424 \\
    E & $-0.65\pm0.08$ & 216--2424 \\
    F \tnote{d} & - & - \\
    G \tnote{e} & $-1$ & 905--1121 \\
    \bottomrule
    \end{tabular}
    \begin{tablenotes}[flushleft]
    \footnotesize \item[a] Frequency range over which source is modelled. 
    \item[b] Fit with a curved power law model.
    \item[c] Inverted spectrum.   
    \item[d] Source could not be modelled, but is not distinguishable from \CORRS{``E''} in MWA data. 
    \item[e] No uncertainty is given as the SED is only over the ASKAP subbands and we do not quantify the internal flux scale uncertainty across the band.
    \end{tablenotes}
    \end{threeparttable}
\end{table}
\setcounter{ft}{0}

\begin{table*}[!t]
    \centering
    \begin{threeparttable}
    \caption{\label{tab:sed} \CORRS{Flux density measurements and limits of the diffuse sources.}}
    \begin{tabular}{l l c c c c c}
         \toprule
         \CORRS{Instrument} & \CORRS{Weighting} &  $\nu$ & Resolution & $\langle \sigma_\text{rms} \rangle$ \tnote{a} & \CORRS{$S_{\text{c},\nu}$} \tnote{b} &  $S_\nu$ \\
         & & (MHz) & ($^{\prime\prime} \times ^{\prime\prime}$) & (mJy\,beam$^{-1}$) & (mJy) & (mJy) \\\midrule
         \multicolumn{7}{c}{\CORRS{Abell~141 radio halo ($+$ peripheral source)}} \\\midrule
         MWA-2 & robust 0.0 & 87.7 & $ 130 \times 130$ & 8.6 & $78 \pm 28$ & $182 \pm 46$ \\
         MWA-2 & robust $+2.0$ &118.4 & $168 \times 118$ & 4.7  & $58 \pm 16$ & $135 \pm 28$ \\
         MWA-2 & robust $+1.0$, 120$^{\prime\prime}$ \CORRSfinal{taper}  & 154.2 & $166 \times 147$ & 6.1  & $45 \pm 8$ & $75 \pm 21$ \\
         MWA \tnote{c} & uniform & 169.6 & $109 \times 109$ & 4.9 & $41 \pm 6$ & $80\pm27$ \\
         MWA-2 & robust $+1.0$, 120$^{\prime\prime}$ taper &185.0 & $150 \times 137$ & 5.1  & $ 38 \pm 5$ & $66\pm19$ \\
         MWA & robust $-1.0$ & 200.3 & $133 \times 126$ & 16.3 & $35 \pm 4$ & $79\pm20$ \\
         MWA-2 & robust $+1.0$, 120$^{\prime\prime}$ taper &215.7 & $138 \times 130$ & 5.1 & $32 \pm 3$ & $65 \pm 22$ \\
         GMRT \tnote{d} & natural, 30$^{\prime\prime}$ taper & 617.5 & $71 \times 56$ & 0.45 & - & $< 26$ \tnote{e} \\ 
         ASKAP & robust $+0.25$, 45$^{\prime\prime}$ taper & 943.5 & $49 \times 46$ & 0.14 & - & $13.7 \pm 1.9$ \\
         ATCA & robust 0.0, 45$^{\prime\prime}$ taper & 2224.5 & $44 \times 31$ & 0.052 & - & $< 8.1$ \tnote{e} \\\midrule
         \multicolumn{7}{c}{\CORRS{Abell~141 peripheral source (SE)}}\\\midrule
         ASKAP & robust $+0.25$, 45$^{\prime\prime}$ taper & 943.5 & $49 \times 46$ & 0.15 & - & $1.7 \pm 0.5$ \\\midrule  
         
         \multicolumn{7}{c}{\CORRS{Abell~3404 radio halo ($+$ peripheral sources)}}\\\midrule
         MWA-2 & robust $+2.0$ &87.7 & $242 \times 157$ & 24.7 & $130 \pm 140$ & $990 \pm 190$ \\
         MWA-2 & robust $+2.0$ &118.4 & $183 \times 114$ & 10.4 & $120 \pm 120$ & $620 \pm 140$ \\
         MWA-2 & robust $+2.0$ & 154.2 & $137 \times 86$ & 4.7  & $100 \pm 100$ & $300 \pm110$ \\
         MWA & robust $-1.0$ & 200.3 & $147 \times 134$ & 8.3 & $ 90 \pm 91 $ & $ 204 \pm99$ \\
         ASKAP & robust $+0.5$, 35$^{\prime\prime}$ taper & 917.5 & $36 \times 36$ & 0.15 & - & $18.9 \pm 2.7$ \\
         ASKAP & robust $+0.5$, 35$^{\prime\prime}$ taper & 1013.5 & $36 \times 36$ & 0.12 & - & $16.0 \pm 2.3$ \\
         ASKAP & robust $+0.5$, 35$^{\prime\prime}$ taper & 1109.5 & $36 \times 36$ & 0.16 & - & $14.9 \pm 2.8$ \\
         ATCA & robust 0.0 & 1388.0 & $98 \times 38$ & 0.15 & - &  $< 21$ \tnote{e} \\
         ATCA & robust 0.0 & 2424.0 & $71 \times 37$ & 0.055 & - & $< 5.7$ \tnote{e} \\\midrule
         \multicolumn{7}{c}{Abell~3404 peripheral source (NE)}\\\midrule
         ASKAP & robust $+0.5$ & 1013.5 & $25 \times 25$ & 0.074 & - & $ 1.3 \pm 0.3$ \\\midrule
         \multicolumn{7}{c}{Abell~3404 peripheral source (SW)}\\\midrule
         ASKAP & robust $+0.5$ & 1013.5 & $25 \times 25$ & 0.065 & - & $ 1.4 \pm 0.4$ \\\bottomrule
    \end{tabular}
    \begin{tablenotes}[flushleft]
    \footnotesize \item[a] Average rms noise within the measured source region. \item[b] \CORRS{Confusing source flux that is subtracted from initial measurement based on SEDs reported in Table~\ref{tab:sed:discrete}.} \item[c] \CORRS{EoR-0 field image \citep{oth+16} as used by \citet{Duchesne2017}, re-measured with present integration region and source subtraction, including brightness scaling of 0.69.}  \item[d] \CORRS{Data originally presented by \citet{Venturi2007}.} \item[e] Limit from mock radio halos as described in Section~\ref{sec:results:simulation}.

    \end{tablenotes}
    \end{threeparttable}
\end{table*}
\setcounter{ft}{0}

\subsubsection{Diffuse source spectral indices}\label{sec:results:sed}

\begin{figure}
    \centering
    \includegraphics[width=1\linewidth]{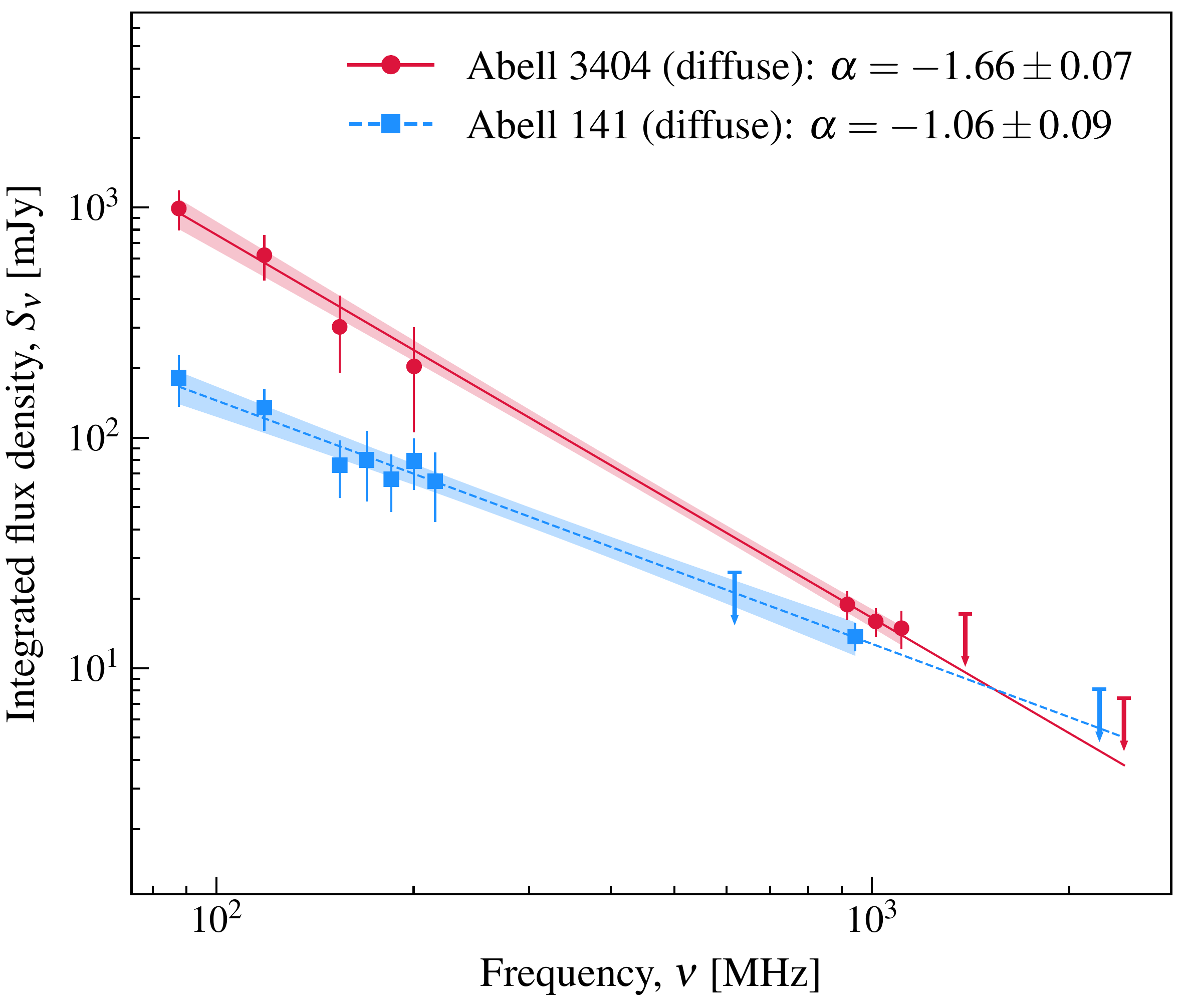}
    \cprotect\caption{\label{fig:seds:diffuse} SEDs of the diffuse emission in Abell~141 and Abell~3404. The lines are power law fits, with 95~per cent confidence intervals represented by the shaded regions. Upper limits are represented by arrows. The fits are extrapolated to ATCA frequencies for ease of comparing to ATCA limits.}
\end{figure}

The central diffuse radio sources \CORRS{can be described by a simple power law between 88 and 943~MHz for Abell~141 and 88 and 1110~MHz for Abell~3404.} Fig. \ref{fig:seds:diffuse} plots the measured data as well as \CORRSfinal{the} best-fit power law models. The spectral index for the central diffuse source in Abell~3404 (\CORRS{$\alpha_{88}^{1013} = -1.66 \pm 0.07$}) pushes it into the ``ultra-steep spectrum radio halo'' category \citep[defined by $\alpha < -1.5$, USSRH;][]{ceb+13}. We find that the central emission in Abell~141 \CORRS{has a flatter spectrum} than reported by \citet{Duchesne2017}, though this is \CORRS{in part} due to the subtraction of three discrete sources with steep spectra. We report all derived source properties in Table~\ref{tab:properties}, including source linear size, spectral index, and extrapolated monochromatic power at 1.4 and 0.15~GHz.

\subsubsection{Limits on high frequency non-detections}\label{sec:results:simulation}
To investigate the non-detections in ATCA data, we obtain upper limits by injecting simulated radio halos into the visibility data. We assume an azimuthally averaged brightness distribution described by \citet[][but see also \citealt{Murgia2009,Bonafede2017}]{Orru2007}, of the form \begin{equation}\label{eq:exp}
    I(r) = I_0 e^{-r/r_e} \quad [\text{Jy}\,\text{arcsec}^{-2}] ,
\end{equation}
with $I_0$ the peak brightness, the $e$-folding radius $r_e = f/R_\text{H}$, and $R_\text{H}$ the radio halo radius. A median value of $f$ is found to be 2.6 by \citet[][based on radio halo samples described by \citealt{Cassano2007,Murgia2009}]{Bonafede2017}. For the purpose of determining limits, we use the calibrated GMRT data presented by \citet{Venturi2007,Venturi2008} of Abell~141, recalculating the limit for consistency with the current method. \par
To determine the initial halo brightness and spectrum, we use the brightness from the ASKAP data and spectral index reported in Section~\ref{sec:results:sed}. Additionally we find values of $f$ (Eq.~\ref{eq:exp}) to be 2.0 and 1.9 \CORRSfinal{that} recover adequate model flux for the Abell~141 and Abell~3404 radio halos, respectively. We use \href{https://sourceforge.net/p/wsclean/wiki/Home/}{\texttt{WSClean}} to inject the simulated halo as a function of frequency into the relevant datasets. During imaging of the mock radio halo, we increase brightness by factors of $\sqrt{2}$ until a detection is made. Imaging is done as a two-part process: first the model data are imaged alone, then imaged with the true calibrated data. The imaging of the model alone allows us to investigate the percentage of flux lost due to the $u$--$v$ sampling. For Abell~141, we image both the ATCA and GMRT data with a 30~arcsec Gaussian taper applied to the visibilities to maximise the likelihood of detection. We find that the ATCA observation of Abell~141 only recovers $\sim 20$~per cent of the model radio halo flux due to the lack of inner spacings limiting sensitivity on larger scales, with the GMRT observation recovering $\sim 60$~per cent. For Abell~3404 the flux recovered is $\sim 70$~per cent and $\sim 90$~per cent for the pre-CABB and CABB data, respectively. We use the same detection criterion as \citet{Bonafede2017}: $D_{2\sigma_\text{rms}}^\text{mock} \geq R_\text{H}$, where $D_{2\sigma_\text{rms}}^\text{mock}$ is the diameter of the mock radio halo within $2\sigma_\text{rms}$ contours. As per \citet{Bonafede2017}, we opt to consider the model radio halo flux density for the limit, as this is the flux density that would be required to make the detection. The resultant limits, along with new limits obtained for the GMRT data \citep{Venturi2007,Venturi2008} are provided in Table \ref{tab:sed}, though are not used in fitting in Section~\ref{sec:results:sed}.\par

We note that the limit found for the GMRT data of Abell~141 is higher than what is reported by \citet{Venturi2007,Venturi2008}: $\sim 7$~mJy \footnote{Note this value is calculated from the reported limit to the luminosity at 610~MHz, requiring an assumption on the spectral index of $-1.3$, though the index does not appreciably change the power calculation here.}. This discrepancy is a result of, in part, the $\sim 60$~per cent of flux lost to $u$--$v$ sampling, and the difference in the modelled brightness profile, where \citeauthor{Venturi2007} use optically thin concentric spheres \citep[see also][]{Brunetti2007}. The remaining difference may be contributed from a different model geometry and spectrum and a bias that occurs when measuring the integrated flux density of low-SNR extended sources \citep[][but see also \citealt{Helfer2003}]{Stroe2016}. 

\citet{Cuciti2018} perform a similar mock halo analysis for GMRT and Karl G.~Jansky Very Large Array (JVLA) data to test flux recovery of incomplete $u$--$v$ sampling. An important note they make is that the recovered flux density fraction decreases as the mock halo brightness decreases. We note that the flux recovery fraction for our mock halos are based on the limits. Appendix \ref{appendix:uv} shows representative $u$--$v$ coverage plots for the observations used in this work (MWA-2, ASKAP, ATCA, and GMRT). These plots highlight that inner $u$--$v$ sampling for the GMRT and ATCA (Abell~141) observations is lacking, whereas the MWA-2, ASKAP, and even the pre-CABB and CABB data for Abell~3404 have much more densely sampled inner $u$--$v$ data. We note, however, that the smaller $\lambda$ values become less populated toward the higher end of the MWA-2 band. We opt not to provide limits on the MWA-2 data between \CORRSfinal{185}--216~MHz for Abell~3404 due to partial detection of the radio halo combined with significant confusion with discrete sources.

\begin{table}
    \centering
    \caption{Derived radio halo properties. \label{tab:properties}}
    \begin{tabular}{c c c}\toprule
         Property & Abell~141 & Abell~3404 \\\midrule
         Linear size (kpc) & 850 & 770 \\
         $\alpha$ & \CORRS{$-1.06 \pm 0.09$} & \CORRS{$-1.66 \pm 0.07$} \\
         $P_{1.4}$ ($10^{23}$\,W\,Hz$^{-1}$) & \CORRS{${}14.4 \pm 2.0 $} & \CORRS{$8.6 \pm 1.7$} \\
         $P_{0.15}$ ($10^{23}$\,W\,Hz$^{-1}$) & \CORRS{${}164 \pm 43 $} & \CORRS{$ 350 \pm 70 $} \\\bottomrule
    \end{tabular}
\end{table}

\subsection{\CORRS{Radio--X-ray correlation}}\label{sec:corr}

\begin{figure*}[t]
    \centering
    \begin{subfigure}[b]{0.5\linewidth}
    \includegraphics[width=1\linewidth]{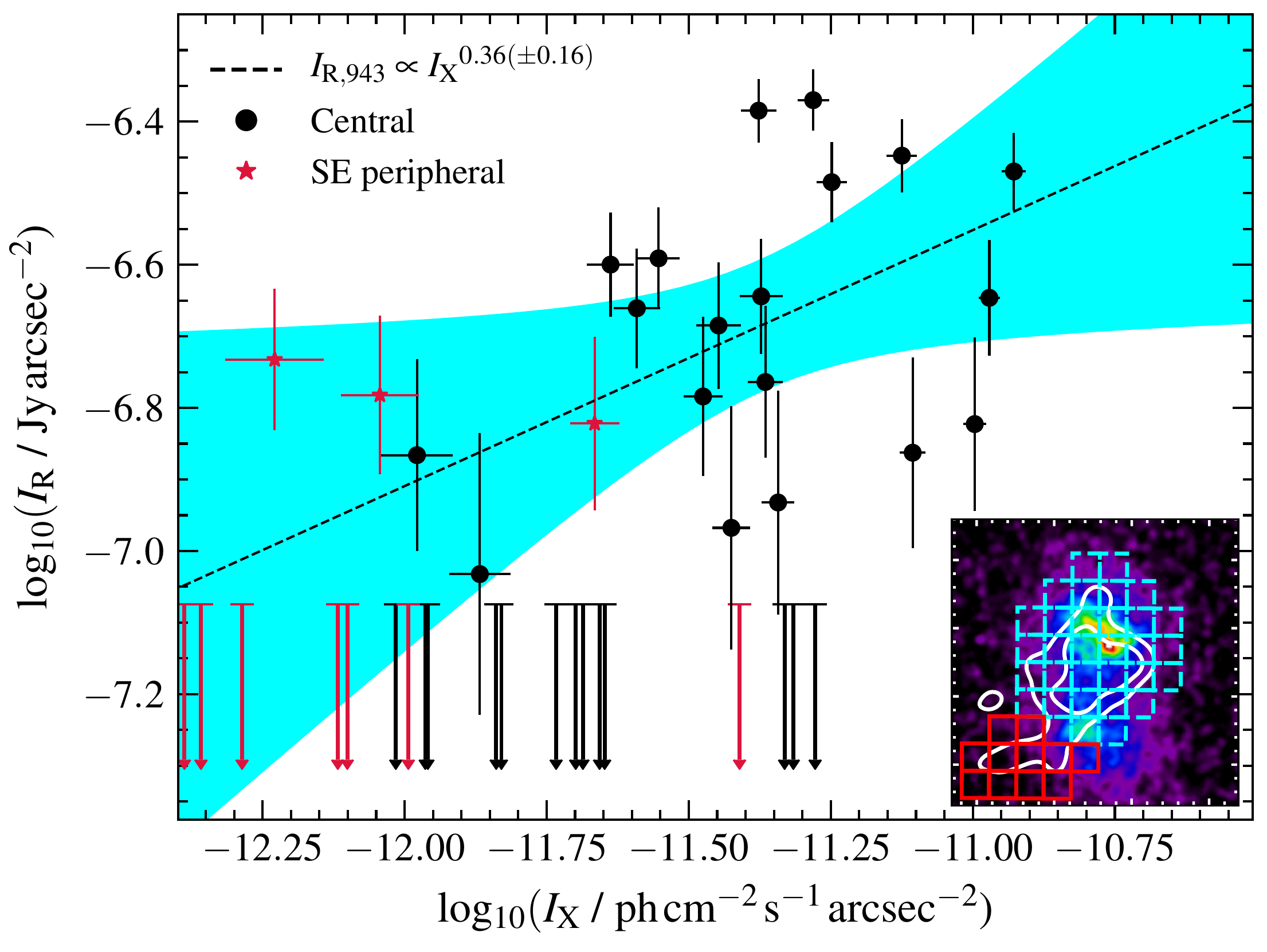}
    \caption{\label{fig:xr:a141} Abell~141.}
    \end{subfigure}\hfill%
    \begin{subfigure}[b]{0.5\linewidth}
    \includegraphics[width=1\linewidth]{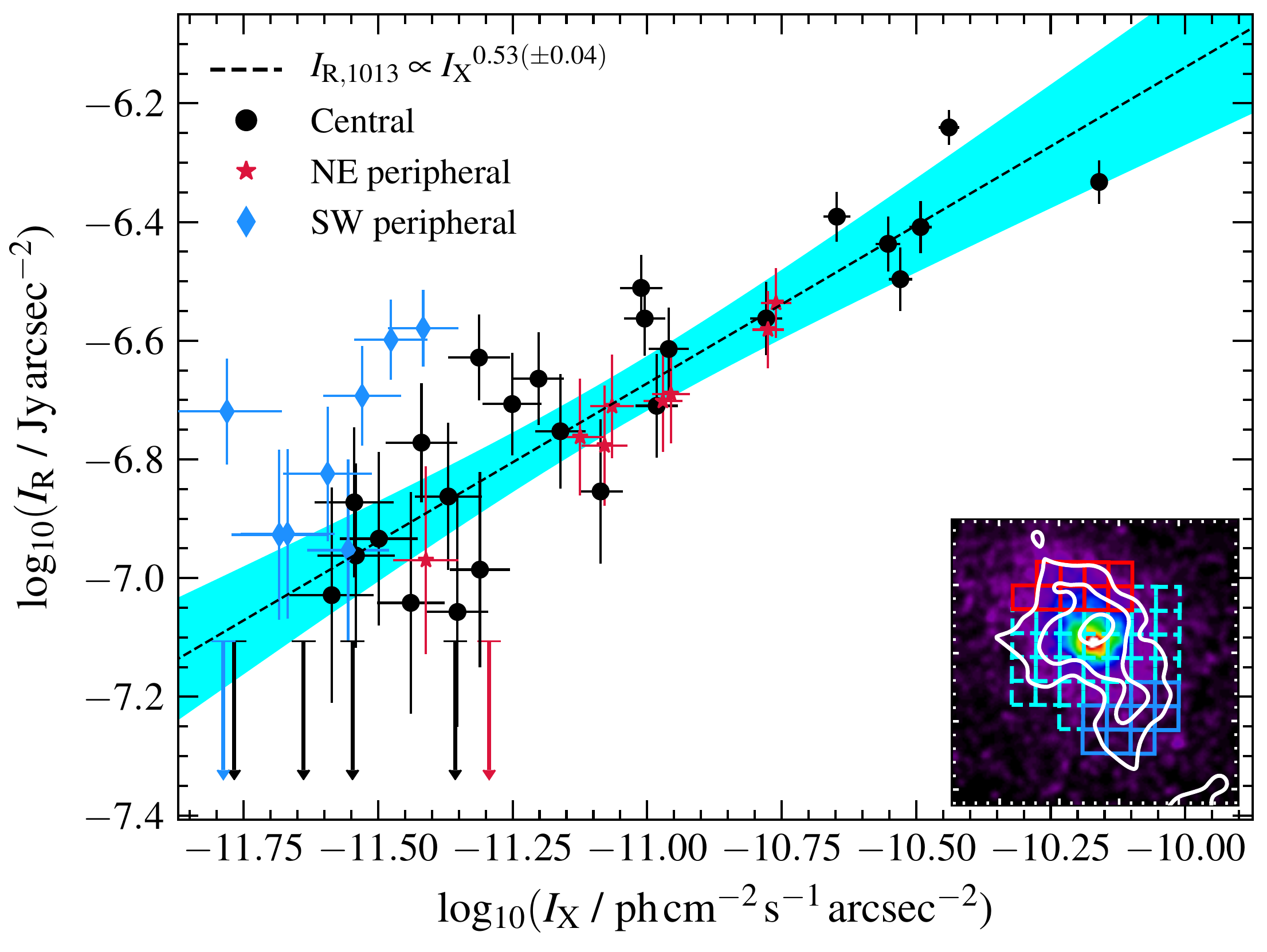}
    \caption{\label{fig:xr:a3404} Abell~3404.}
    \end{subfigure}\\
    \caption{\label{fig:xr} \CORRS{Radio--X-ray point-to-point correlation for \subref{fig:xr:a141} Abell~141 and \subref{fig:xr:a3404} Abell~3404. Upper limits correspond to cells where $I_\text{R} < 2\sigma_\text{rms}$. The black, dashed line is the best-fitting line with a 95~per cent confidence interval shaded in cyan. The insets show the \emph{Chandra} X-ray maps with the source-subtracted ASKAP image overlaid as contours as in Fig.~\ref{fig:xray:a141} and Fig.~\ref{fig:radio:a3404:askap} for Abell~141 and Abell~3404, respectively. The cyan boxes on the insets show the cells within which surface brightesses are calculated, and the red and blue cells indicate the locations of the peripheral components.}}
\end{figure*}

\CORRS{Radio halo brightness ($I_\text{R}$) is often observed to correlate with the X-ray surface brightness ($I_\text{X}$) of the ICM ($I_\text{R} \propto {I_\text{X}}^k$; \citealt{Govoni2001}). Radio halos are typically observed with a sub-linear slope \citep[e.g.][]{Giacintucci2005,Rajpurohit2018,Hoang2019a,Botteon2020b,Rajpurohit2021a,Bruno2021}; conversely mini-halos have been found to have super-linear slopes \citep{Ignesti2020}. We use the exposure-corrected, source-subtracted, smoothed [0.5--2]~kev \textit{Chandra} data along with the low-resolution, source-subtracted ASKAP images to perform a similar analysis for Abell~141 and Abell~3404.}

\CORRS{Following a procedure described by \citet{Ignesti2020}, we construct grids across the diffuse emission with cell sizes corresponding to the ASKAP beam major axis. For Abell~141 this is 49$^{\prime\prime}$, corresponding to 178~kpc and for Abell~3404 this is 44$^{\prime\prime}$, corresponding to 124~kpc. We fit the data as $\text{log}_{10}(I_\text{R}) = k \text{log}_{10}(I_\text{X}) + C$ using the BCES method \footnote{Bivariate Correlated Errors and intrinsic Scatter: \citet{ab96}.} with an orthogonal regression. Fig.~\ref{fig:xr} shows the results for \subref{fig:xr:a141} Abell~141 and \subref{fig:xr:a3404} Abell~3404. For both, we separate out the contribution from the peripheral components, with each grid shown on the inset for each figure. Abell~3404 shows strong correlation (with Spearman rank-order correlation coefficient, $\rho = 0.89$), and \CORRSfinal{we find} a sub-linear trend with $k = 0.53\pm0.04$. While many halos have been found with $k \gtrsim 0.6$ \citep[e.g.][]{Govoni2001,Botteon2020b,Rajpurohit2021a}, we note the steep-spectrum radio halo in MACS~J1149.5$+$2223 is found to have $k \lesssim 0.6$ \citep{Bruno2021}. The SW peripheral component (blue points, Fig.~\subref{fig:xr:a3404}) appears unassociated whereas the NE component follows the correlation tightly. For the fit the NE component is included. For Abell~141, we find no significant correlation ($\rho = 0.31$). This may indicate a mixture of emission components, though \citet{Shimwell2014} notes, in reference to the Bullet Cluster radio halo showing a similar lack of correlation, this may be due to the halo occurring during a specific stage of a complex merger.}

\subsection{Cluster dynamics and source classification}\label{sec:results:xray}

\subsubsection{Abell~141---pre-merger?}\label{sec:results:xray:a141}

\defcitealias{Caglar2018}{C18}
The dynamic nature of Abell~141 has been studied extensively by \citet[][in optical]{dki+02} and \citet[][hereafter \citetalias{Caglar2018}, in X-ray]{Caglar2018}. \citet{dki+02} and \citetalias{Caglar2018} find that the bi-modal distribution (as seen in the X-ray in Fig. \ref{fig:xray:a141}) represents two subclusters (labelled A141N and A141S by \citetalias{Caglar2018}) which have not completed a core-crossing---i.e., the cluster is likely in a pre-merging state. \citetalias{Caglar2018} report that X-ray--emitting gas between the two subclusters features a hotspot, which may imply the presence of a shock or shocks. We note that the central diffuse radio emission coincides more with the A141N subcluster, but extends into the region between the subclusters. Most radio halos have been detected in clusters that are in a merging or dynamic state \citep[see e.g.][]{ceb+13}, with three examples in the literature of likely pre-merger subclusters: Abell~399--401 \citep{Murgia2010}, MACS~J0416.1$-$2403 \citep{Ogrean2015}, and Abell~1758N--S \citep{Botteon2018}. We note in the cases of Abell~399--401 and Abell~1758N--S each subcluster in the corresponding mergers clearly host their own radio halos whereas we do not detect two distinct radio halos in the A141N and A141S subclusters. The emission may represent a bridge between the subclusters rather than a traditional giant radio halo \CORRS{\citep[see][]{Govoni2019,Botteon2020c,Hoeft2020,Bonafede2021}}, though due to the resolution of our data we cannot confirm bridge emission distinct from the radio emission that permeates A141N. 

\emph{A radio relic?} Some radio relics have been found to be co-located with shocks detected via X-ray emission \citep[e.g.][]{Bourdin2013,Akamatsu2015,Eckert2016b,DiGennaro2019}. Assuming the central region temperature jump in the X-ray corresponds to a shock, \citetalias{Caglar2018} derive a Mach number of $\mathcal{M}_\text{X} = 1.69^{+0.41}_{-0.37}$. If we consider that a radio shock traces the same shock structure, and that DSA on a pool of thermal electrons triggers the emission \citep[see e.g.][]{Blandford1987}, a corresponding radio Mach number can be calculated from 
\CORRS{\begin{equation}\label{eq:radiomach}
    \mathcal{M}_\text{R} = \sqrt{\dfrac{2\alpha_\text{inj} - 3}{2\alpha_\text{inj} + 1}} \, ,
\end{equation}}
assuming that \CORRS{$\alpha = \alpha_\text{inj} - 0.5$, with $\alpha_\text{inj}$ the synchrotron injection spectrum index.} We find \CORRS{$\mathcal{M}_\text{R} = 5.9 \pm 0.9$}, inconsistent with the X-ray--derived Mach number. This does not necessarily rule out the possibility of the source being a radio shock, as discussed by \citet{vanWeeren2016,vanWeeren2017} \citep[but see also][]{Hoang2019,Lee2020}, a discrepancy in Mach numbers may indicate that the thermal pool electrons are \textit{not} seed electrons for the emission---i.e., pre-accelerated fossil electrons may be accelerated by the DSA process. In this case the injection spectrum \CORRS{may} resemble the observed emission spectrum \CORRS{\citep{vanWeeren2016}} with $\alpha = \alpha_\text{inj}$ and \CORRS{$\mathcal{M}_\text{R} = 2.1 \pm 0.2$}, in agreement with $\mathcal{M}_\text{X}$. \CORRS{Assuming seed electrons for radio relics originate as fossil electrons rather than thermal pool electrons also alleviates the acceleration efficiency problem for some relics with X-ray detected shocks \citep[see][and references therin]{Botteon2020}.} Additionally, simulations show radio galaxies in cluster environments can supply fossil electrons for re-acceleration \citep{Vazza2021}. \CORRS{A difference in Mach number may also arise from the X-ray and radio emission preferentially tracing different shocks along a line of sight \citep[see e.g.][and references therein]{vanWeeren2016,Rajpurohit2020a} or may arise due to turbulence near the shock front \citep{Dominguez2021}.} \par
\CORRS{The present data (including lack of polarimetry) do not allow us to rule out a radio relic or radio bridge interpretation, or a combination thereof. We consider the source a radio halo because the observed physical characteristics---including its morphology, location, and SED---are consistent with a radio halo classification.}

\subsubsection{Abell~3404---dynamics}\label{sec:results:xray:a34041:dynamics}

Fig. \ref{fig:xray:a3404} shows the exposure-corrected, smoothed, \CORRS{and point-source subtracted [0.5--2]~kev} \emph{Chandra} map for Abell~3404, where the X-ray emission is slightly elongated. We note that using \textit{XMM-Newton} data \citet{pcab09} consider the cluster as non-cool core, but also not morphologically disturbed based on its centroid shift, $w$ \citep[][but see also \citealt{Mohr1993}]{Poole2006}, which gives an indication of the dynamic state of the cluster. We repeat the calculation for the present \textit{Chandra} data, using \href{http://www.isdc.unige.ch/~deckert/newsite/Proffit.html}{\texttt{proffit}} \footnote{\url{http://www.isdc.unige.ch/~deckert/newsite/Proffit.html}} \citep{Eckert2011}, subtracting a fitted background of $I_\text{B} = (1.41 \pm 0.05) \times 10^{-5}$~ph\,cm$^{-2}$\,s$^{-1}$\,arcmin$^{-2}$, finding \CORRS{$w = 0.039$} within $R_{500} = 1280$~kpc \citep{pcab09}, smaller than the \CORRS{expected} $w > 0.075$ for a disturbed system of this radius. We note that the radio halo detected in Abell~S1063 \citep{Xie2020} is similarly detected in a cluster that is considered morphologically relaxed based on this definition.

We also calculate the concentration parameter, $c_{100/500}$ (and $c_{40/400}$; \citealt{Santos2008}), defined via \begin{equation}
    c_{r_1/r_2} = \dfrac{I_\text{X}\left( r < r_1~\text{kpc}\right)}{I_\text{X}\left( r < r_2~\text{kpc}\right)} \, ,
\end{equation}
with $I_\text{X}$ the X-ray surface brightness. We find $c_{100/500} = 0.24$ which, in combination with the centroid shift within 500~kpc (\CORRS{$w_{500~\text{kpc}} = 0.039$}) places the cluster just outside of the halo-hosting quadrant of the $c_{100/500}$--$w_{500~\text{kpc}}$ plane shown by \citet{Cassano2010}---no halos appear in clusters with $c_{100/500} > 0.2$. We also find \CORRS{$c_{40/400} = 0.062$}, consistent with non-cool--core clusters \citep{Santos2008}. {The ambiguity of the X-ray morphological parameters point towards a low-energy/weak merger or a late stage in the merger. For example, a merger between low velocity subclusters or a high mass-ratio between subclusters may result in low-energy mergers. Given the cluster's SZ-derived mass of $7.96_{-0.21}^{+0.23}\times 10^{14}~\text{M}_\odot$ \citep{planck16} we can infer the presence of a massive subcluster prior to the merger. Such scenarios are expected to generate USSRHs \citep{Brunetti2008} as observed here.}

\subsubsection{Abell~3404---radio and X-ray shocks?}\label{sec:results:xray:a3404:shocks}

\begin{figure}
    \centering
    \includegraphics[width=1\linewidth]{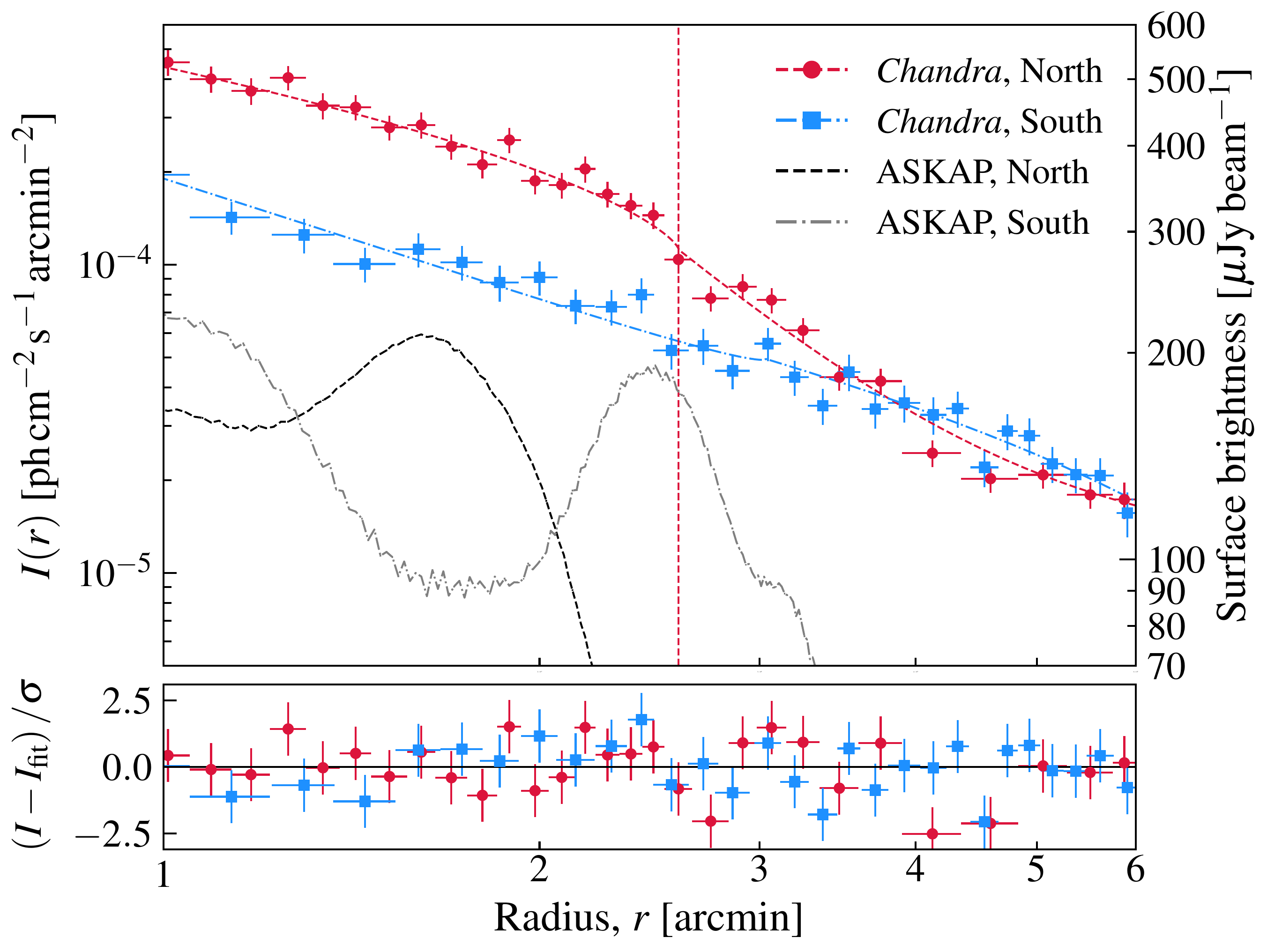}
    \caption{\label{fig:xray:sb} Radio and background-subtracted X-ray surface brightness profile for sectors shown in Fig. \ref{fig:xray:a3404} \CORRSfinal{for Abell~3404}. The radio ordinate clips at $2\sigma_\text{rms}$. The dashed-red vertical line is at location of the discontinuity in the northern profile.}
\end{figure}

Fig. \ref{fig:xray:a3404} shows two extended diffuse emission regions (indicated with dashed, yellow rectangles) either side of the cluster center. We do not have sufficient fidelity in the ASKAP subband data to investigate the spectral properties or to remove the northern component from the radio halo measurement, though for the purpose of this section we consider the possibility that the two extended structures represent radio shocks.

We investigate that possibility by extracting X-ray and radio surface brightness (SB) profiles along the directions of the relic-like sources (extracted regions shown as yellows sectors in Fig. \ref{fig:xray:a3404}). For the X-ray data, we use \href{http://www.isdc.unige.ch/~deckert/newsite/Proffit.html}{\texttt{proffit}}, masking the point sources indicated in Fig. \ref{fig:xray:a3404} with yellow circles. The X-ray SB profiles are binned ensuring each bin is $\geq 10\sigma$ and $\geq 7\sigma$ for the north and south profiles, respectively. The radio SB profile is extracted using in-house \texttt{python} code \href{https://gist.github.com/Sunmish/198ef88e1815d9ba66c0f3ef3b18f74c}{\texttt{fluxtools.py}} \footnote{\url{https://gist.github.com/Sunmish/198ef88e1815d9ba66c0f3ef3b18f74c}}. The extracted profiles in each direction are shown in Fig. \ref{fig:xray:sb}. We find candidate discontinuities in the X-ray SB profile. We fit standard broken power law models, representing an electron density discontinuity either side of a putative shock \citep[see e.g.][]{Owers2009a,Eckert2016b}:
\begin{equation}\label{eq:density}
\rho(r) = 
\begin{cases}
Cr^{-\Gamma_{\text{in}}} , & \text{if } r<r_\text{break} \\
C\dfrac{n_\text{out}}{n_\text{in}} r^{-\Gamma_\text{out}} , & \text{otherwise}
\end{cases} ,
\end{equation}
with $\Gamma$ the power law indices either side of the discontinuity, $j=n_\text{out}/n_\text{in}$ the density jump, and $C$ a normalisation factor, and the SB profile is $\rho(r)$ integrated along the line of sight. For the southern profile we find $j_\text{south} \sim 1$, indicating no evidence for a discontinuity. For the northern profile we obtain \CORRS{$j_\text{north} = 1.11^{+0.09}_{-0.10}$}. The radio source occurs $\sim1$~arcmin ($\sim170$~kpc) from the X-ray discontinuity; in the case of a relic associated with a shock the observed discontinuity in the X-ray profile occurs directly after the relic source (and a hard edge may be seen in the radio map), therefore the northern peripheral radio source is unlikely to be a relic associated with a shock.

\subsection{Radio halo $P_\nu$--$M_{500}$ scaling relations}\label{sec:results:scaling}

\begin{figure*}[!t]
    \centering
    \begin{subfigure}{0.5\linewidth}
    \includegraphics[width=1\linewidth]{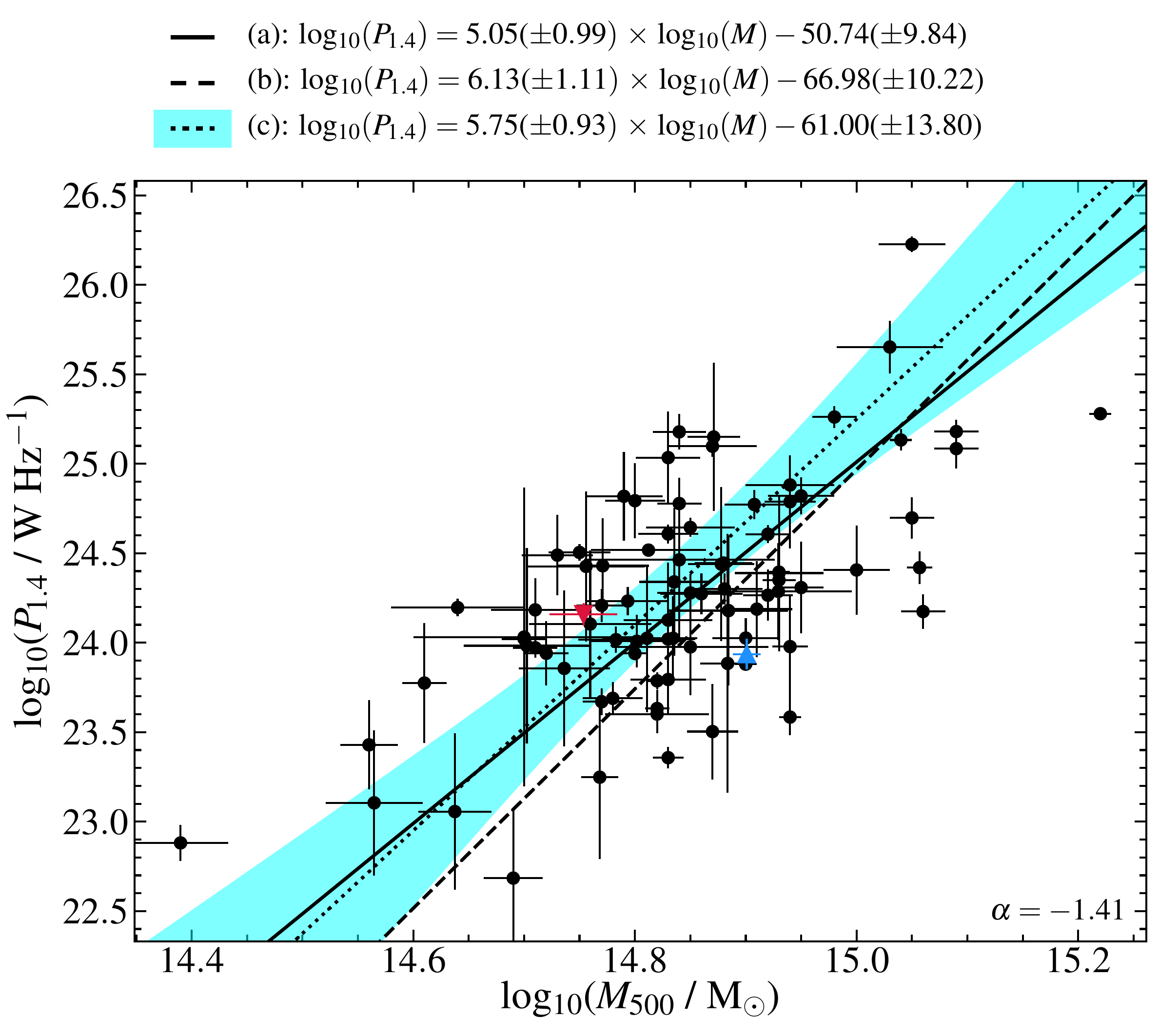}
    \caption{$P_\text{1.4}$--$M_\text{500}$\label{fig:scaling:1400}}
    \end{subfigure}%
    \begin{subfigure}{0.5\linewidth}
    \includegraphics[width=1\linewidth]{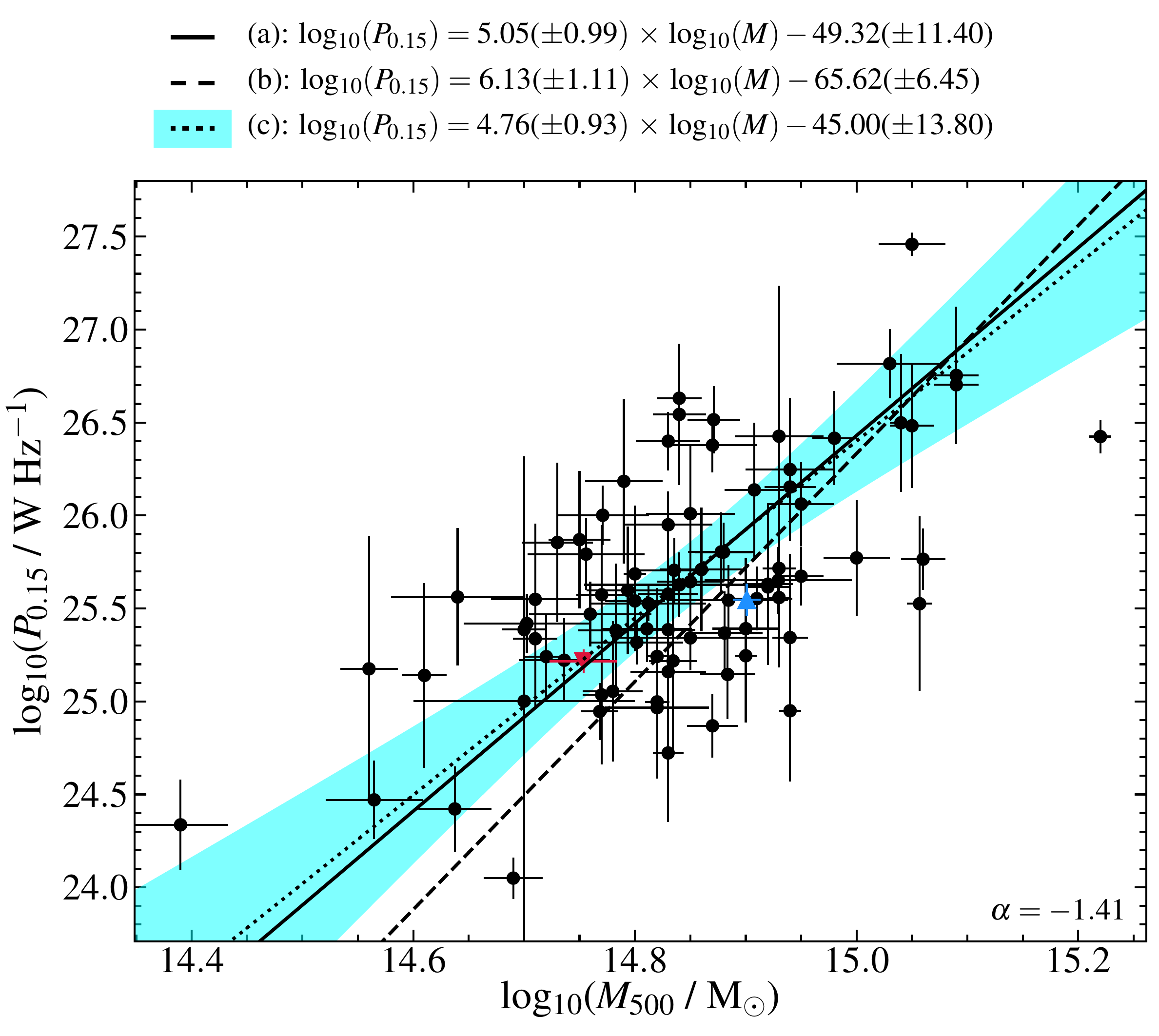}
    \caption{$P_\text{0.15}$--$M_\text{500}$ \label{fig:scaling:150}}
    \end{subfigure}
    \caption{\label{fig:scaling} Radio halo $P_\text{1.4}$--$M_\text{500}$ relation \subref{fig:scaling:1400} and $P_\text{0.15}$--$M_\text{500}$ relation \subref{fig:scaling:150} with best-fitting orthogonal relations from (a) \citet{ceb+13} (solid, black), (b) \citet{vanWeeren2020} (dashed, black), and (c) this work (dotted, black). \CORRSfinal{H}alos are taken from the literature (as discussed in Section \ref{sec:results:scaling}) with the addition of the halo in Abell~3404 (blue, upright triangle) and the updated values for Abell~141 (red, downward triangle), for a total of 86 halos. The shaded regions are 99.7\% confidence intervals for fits from this work.}
\end{figure*}

\defcitealias{ceb+13}{C13}
\defcitealias{Cuciti2021b}{C21}
\defcitealias{vanWeeren2020}{vW20}

\begin{table}
\centering
\caption{Fitted values for the $P_\nu$--$M_{500}$ scaling relations for various methods.\label{tab:scaling:fits}}
\begin{tabular}{lllll} 
\toprule
Method       & $B$    & $\sigma_\text{B}$ & $A$ & $\sigma_\text{A}$  \\ 
\midrule
\multicolumn{5}{c}{$P_{1.4}$--$M_{500}$}\\
\midrule
$P_{1.4} | M_{500}$ & 3.21 & 0.39 & $-23.4$ & 5.8   \\
Orthogonal   & 5.75 & 0.93 & $-61.0$ & 13.8  \\
Bisector     & 4.17 & 0.46 & $-37.7$ & 6.8   \\
\midrule
\multicolumn{5}{c}{$P_{0.15}$--$M_{500}$}\\
\midrule
$P_{0.15} | M_{500}$& 3.15 & 0.41 & $-21.2$ & 6.0   \\
Orthogonal   & 4.76 & 0.93 & $-45.0$ & 13.8  \\
Bisector     & 3.84 & 0.50 & $-31.4$ & 7.4   \\
\bottomrule
\end{tabular}
\end{table}

Despite a somewhat atypical X-ray morphology of each cluster, based on the available data we classify both central diffuse radio sources as giant radio halos (for Abell~141, consistent with the classification from \citealt{Duchesne2017}). Both Abell~141 and Abell~3404 have similar masses, and their constituent radio halos have similar sizes and brightness at $\sim 1$~GHz, however, due to the difference in their SEDs their 1.4- and 0.15-GHz monochromatic luminosities differ by around a factor of two. 

Scaling relationships between radio halo power and various (related) cluster properties have been found \citep[see e.g.][]{Liang2000,Brunetti2007,Basu2012}, somewhat explained physically by turbulent (re-)acceleration models for halo formation \citep[][hereafter \citetalias{ceb+13}]{ceb+13}. A key finding by \citet[][but see also \citealt{Cassano2010,ceb+13}]{Brunetti2009} is the bi-modality to the sample of halo-hosting clusters, where morphologically disturbed (i.e., likely merging) clusters host radio halos, and most relaxed, X-ray--luminous, and massive clusters without halos have upper limits to radio halo power well below the empirical scaling relations.

Recent updates to the power--mass ($P_\nu$--$M_{500}$) relation \citep[][]{ceb+13,mfj+16,Duchesne2017,vanWeeren2020,Cuciti2021b} find results largely consistent within reported uncertainties. We update this relation at $\nu=1.4$~GHz and $\nu=0.15$~GHz following \citet[][hereafter \citetalias{vanWeeren2020}]{vanWeeren2020}. For this work, we incorporate the compiled literature sample of halos reported by \citetalias{ceb+13} and \citet{mfj+16}, halos in Abell~S1121 and Abell~2811 reported by \citet{Duchesne2017}, and new halos reported in the literature from 2017 \citep{Parekh2017,Wilber2018,Cassano2019,Birzan2019,Xie2020,HyeongHan2020,Giovannini2020,Wilber2020,Hoeft2020,Hoang2020,DiGennaro2020,vanWeeren2020,Raja2020,Knowles2020,Raja2021}, with the exception of the halo in ACT-CL~J0528.8$-$3927 reported by \citet{Knowles2020} as its small size and coincidence with a radio-bright BCG suggests a mini-halo. For consistency, cluster masses are obtained from PSZ measurements where available \citep[][with some masses obtained from the South Pole Telescope; \citealt{spt1,spt2}]{Planck2016}, and only clusters with an SZ-derived mass are used. Note all radio halos are scaled to 1.4 and 0.15~GHz with either a measured spectral index or using the full sample mean, $\langle\alpha\rangle = \meanalpha$, following 
\begin{equation}
    P_{\nu_1} = \dfrac{4 \pi D_\text{L}(z)^2}{\left( 1+z\right)^{1+\alpha}} S_{\nu_\text{obs}} \left(\dfrac{\nu_1}{\nu_\text{obs}}\right)^\alpha \quad [\text{W}\,\text{Hz}^{-1}] \, ,
\end{equation}
where $\nu_\text{obs}$ is the closest observed frequency, $\nu_1 \in \{0.15, 1.4\}$, and $D_\text{L}$ is the luminosity distance of the cluster. We do not distinguish between normal giant radio halos and ultra-steep spectrum radio halos defined by $\alpha < -1.5$ (USSRH) (e.g. \citealt{Brunetti2008}; \citetalias{ceb+13}). {We note that integrated flux densities measured by \citetalias{vanWeeren2020} are integrated from a fitted exponential profile (Eq.~\ref{eq:exp}) rather than measured directly from the maps and may be generally larger than those measured via integration directly from the pixel sum, especially for low-SNR halos}. 

We fit the radio halo sample assuming a scaling relation at $\nu=1.4$~GHz and $\nu=0.15$~GHz with a function of the form $\log_\text{10}(P_{\nu}) = B\times \log_\text{10}(M_{500}) + A$ using the BCES method with an orthogonal regression. For comparison with \citetalias{ceb+13} and \citetalias{vanWeeren2020} we shift their fitted relations to the relevant frequencies via 
\begin{equation}\label{eq:scale:rescale}
    \log_{10}(P_{\nu_1}) = B\log_{10}(M_{500}) + A - \langle \alpha \rangle \log_{10}\left(\dfrac{\nu_2}{\nu_1}\right) \, ,
\end{equation}
where $\nu_2$ is the frequency at which the original scaling relation is fit and with $P_{\nu_1} = P_{\nu_2}(\nu_1 / \nu_2)^{\langle \alpha \rangle}$. Fig.~\ref{fig:scaling} shows the $P_{1.4}$--$M_{500}$ (\ref{fig:scaling:1400}) and $P_{0.15}$--$M_{500}$ (\ref{fig:scaling:150}) planes with the relevant fits and aforementioned data. Best-fit values for the scaling relation for various methods are shown in Table~\ref{tab:scaling:fits} for comparison with other works.

The updated position of the radio halo in Abell~141 on both the $P_{1.4}$--$M_{500}$ and $P_{0.15}$--$M_{500}$ planes is in agreement with the fitted relations, and the halo in Abell~3404 lies below the $P_{1.4}$--$M_{500}$ relation consistent with the general population of USSRHs \citep[][]{ceb+13,Cuciti2021b}, but is closer to the fitted $P_{0.15}$--$M_{500}$ relation. We note also that if the halo in Abell~141 was only associated with the A141N subcluster \footnote{With $M_{\text{X},500} = ( 3.79\pm0.3) \times 10^{14}~\text{M}_\odot$ \citepalias{Caglar2018}.} the halo's position on both $P_\nu$--$M_{500}$ relations would be significantly above the best fitting lines. With a sample size of 86 we find (orthogonal) scaling relations consistent with \citetalias{ceb+13} and \citetalias{vanWeeren2020} at both frequencies within respective uncertainties, but note some deviation from the fit reported by \citetalias{vanWeeren2020} at 0.15~GHz. Naturally, scaling radio halo powers to frequencies beyond which they are measured introduces additional uncertainties. We also find that the $P_\nu | M_{500}$ fitting \footnote{$P_{\nu}$ the dependent variable and $M_{500}$ the independent variable.} is the most consistent method (albeit shallower than for orthogonal), with results from \citetalias{vanWeeren2020} ($B = 3.84 \pm 0.69$) and \citet{Cuciti2021b} ($B = 3.26 \pm 0.74$ for their `statistical' sample with USSRHs) with both \citetalias{vanWeeren2020} and \citet{Cuciti2021b} employing smaller, specific samples.

\citetalias{ceb+13} find that inclusion of USSRHs steepens the scaling relation at 1.4~GHz and \citetalias{vanWeeren2020} find generally steeper relations at 0.15~GHz for radio halos detected around 0.15~GHz (and presumably generally steeper in spectrum). In Equation~\ref{eq:scale:rescale} we re-scale the scaling relation with an assumed mean $\langle\alpha\rangle$, where real observations at lower of higher frequencies may bias the sample selection towards steeper or flatter spectrum sources, respectively, finding opposed slopes to the $P_\nu$--$M_{500}$ relation. Uncertainties are still too significant to confirm this as discussed by \citetalias{vanWeeren2020}. 

The concordance between these results and those in the recent literature demonstrates that we are now reaching sufficient sample sizes to have confidence in the scaling relation values. While it has recently been suggested there is little evolution of magnetic field strength as a function of redshift in galaxy clusters and hence a redshift dependence in the power-mass scaling relations are not expected \citep{DiGennaro2020}, an outstanding question remains regarding if there is a difference in the relations for halos and USSRHs which may be expected from differences in their underlying physics. Yet larger samples of halos, expected in the near future, will allow this to be probed. 

\section{Summary}

We have confirmed the detection of centrally-located diffuse, steep-spectrum cluster radio emission in Abell~141, and ultra-steep spectrum emission in Abell~3404. We have presented new observations from the MWA-2, ASKAP, and ATCA of the clusters, with the central diffuse sources detected with the MWA-2 and ASKAP. We conclude that these central \CORRS{diffuse} radio sources can be described as giant radio halos with linear extents of $\sim 850$~kpc and $\sim 770$~kpc, for Abell~141 and Abell~3404, respectively. We find that each source has an SED that can be fitted with a normal power law model, with spectral indices of \CORRS{$\alpha_{88}^{943} = -1.06 \pm 0.09$ and $\alpha_{88}^{1110} = -1.66 \pm 0.07$} for Abell~141 and Abell~3404, respectively, making the radio halo in Abell~3404 the first reported USSRH detected with ASKAP. We find no evidence of these sources in pre-CABB and CABB ATCA data, though find the $u$--$v$ coverage for the observations preclude detection of the radio halos assuming an exponential brightness profile described by the derived spectral index and the measured size and peak brightness in the ASKAP images. \CORRS{Additional peripheral components are detected along with the main central halos, though only for the SW peripheral source in Abell~3404 do we confirm this is not associated with the central halo. No shocks are detected at the locations of the peripheral components in Abell~3404 and \CORRSfinal{they} are unlikely to be relics. These peripheral components may be background/foreground radio galaxies, phoenices, or other extended sources and we do not classify them here.}

We discuss the morphological properties of the two clusters based on their X-ray emission using \textit{Chandra} data, noting that Abell~141 has previously been found to likely be in a pre-merging state, prior to core passage of the sub clusters. Conversely, we find that the dynamic status of Abell~3404 is likely to be in more relaxed state, either in the late stage of a merger or a low-energy merger. \CORRS{The radio--X-ray surface brightness correlation is explored for each cluster, finding strong correlation for Abell~3404 but no significant correlation for Abell~141.} \CORRS{Some of these} properties make them atypical for most radio-halo--hosting clusters, though we find that the radio halos are located in the expected places on the $P$--$M$ scaling relations at both 1.4 and 0.15~GHz, given their \CORRSfinal{respective} spectral properties and cluster masses (considering Abell~141 a single cluster system). We fit the $P_{1.4}$--$M_{500}$ and $P_{0.15}$--$M_{500}$ scaling relations with the current sample of radio halos, finding results consistent with the literature suggesting that we are now converging with a high precision on these relations.

\begin{acknowledgements}

% Personal:
We thank T.~Venturi for providing the calibrated GMRT dataset of Abell~141. \CORRS{We also thank the anonymous referee for providing valuable feedback and their suggestion to explore the radio--X-ray surface brightness correlation.}
% Funding:
SWD acknowledges an Australian Government Research Training Program scholarship administered through Curtin University.
% Data archives/instruments
% ASKAP
The Australian SKA Pathfinder is part of the Australia Telescope National Facility which is managed by CSIRO. Operation of ASKAP is funded by the Australian Government with support from the National Collaborative Research Infrastructure Strategy. ASKAP uses the resources of the Pawsey Supercomputing Centre. Establishment of ASKAP, the Murchison Radio-astronomy Observatory and the Pawsey Supercomputing Centre are initiatives of the Australian Government, with support from the Government of Western Australia and the Science and Industry Endowment Fund.
% MWA
Support for the operation of the MWA is provided by the Australian Government (NCRIS), under a contract to Curtin University administered by Astronomy Australia Limited.
We acknowledge the Pawsey Supercomputing Centre which is supported by the Western Australian and Australian Governments.
This paper includes archived data obtained through the Australia Telescope Online Archive (\url{http://atoa.atnf.csiro.au/}). 
This research has made use of data obtained from the \textit{Chandra} Data Archive and the \textit{Chandra} Source Catalog, and software provided by the \textit{Chandra} X-ray Center (CXC) in the application package \href{https://cxc.cfa.harvard.edu/ciao/}{\texttt{CIAO}}.
This research has made use of the NASA/IPAC Extragalactic Database (NED), which is operated by the Jet Propulsion Laboratory, California Institute of Technology, under contract with the National Aeronautics and Space Administration.
% Python tools not explicitly named in the text
This research made use of a number of \texttt{python} packages not explicitly mentioned in the main text: \href{https://aplpy.github.io/}{\texttt{aplpy}} \citep{Robitaille2012}, \href{https://www.astropy.org/}{\texttt{astropy}} \citep{astropy:2013,astropy:2018}, \href{https://matplotlib.org/}{\texttt{matplotlib}} \citep{Hunter2007}, \href{https://numpy.org/}{\texttt{numpy}} \citep{Numpy2011} and \href{https://scipy.org/}{\texttt{scipy}} \citep{Jones2001}.

\end{acknowledgements}

{\footnotesize
\bibliographystyle{pasa-mnras}
\bibliography{references}

\begin{thebibliography}{}
\makeatletter
\relax
\def\mn@urlcharsother{\let\do\@makeother \do\$\do\&\do\#\do\^\do\_\do\%\do\~}
\definecolor{darkblue}{rgb}{0,0,0.597656}
\def\mndoi{\begingroup\mn@urlcharsother \@ifnextchar [ {\mndoi@} {\mndoi@[]}}
\def\mndoi@[#1]#2{\def\@tempa{#1}\ifx\@tempa\@empty \href
  {http://dx.doi.org/#2} {\textcolor{darkblue}{doi:#2}}\else \href
  {http://dx.doi.org/#2} {\textcolor{darkblue}{#1}}\fi \endgroup}
\def\mn@eprint#1#2{\mn@eprint@#1:#2::\@nil}
\def\mn@eprint@arXiv#1{\href {http://arxiv.org/abs/#1} {{\tt arXiv:#1}}}
\def\mn@eprint@dblp#1{\href {http://dblp.uni-trier.de/rec/bibtex/#1.xml}
  {dblp:#1}}
\def\mn@eprint@#1:#2:#3:#4\@nil{\def\@tempa {#1}\def\@tempb {#2}\def\@tempc
  {#3}\ifx \@tempc \@empty \let \@tempc \@tempb \let \@tempb \@tempa \fi \ifx
  \@tempb \@empty \def\@tempb {arXiv}\fi \@ifundefined
  {mn@eprint@\@tempb}{\@tempb:\@tempc}{\expandafter \expandafter \csname
  mn@eprint@\@tempb\endcsname \expandafter{\@tempc}}}

\bibitem[\protect\citeauthoryear{{Abell}}{{Abell}}{1958}]{abell}
{Abell} G.~O.,  1958, \mndoi [\apjs] {10.1086/190036}, \href
  {http://adsabs.harvard.edu/abs/1958ApJS....3..211A} {3, 211}

\bibitem[\protect\citeauthoryear{{Abell}, {Corwin}  \& {Olowin}}{{Abell}
  et~al.}{1989}]{aco89}
{Abell} G.~O.,  {Corwin} Jr. H.~G.,   {Olowin} R.~P.,  1989, \mndoi [\apjs]
  {10.1086/191333}, \href {http://adsabs.harvard.edu/abs/1989ApJS...70....1A}
  {70, 1}

\bibitem[\protect\citeauthoryear{{Akamatsu} et~al.,}{{Akamatsu}
  et~al.}{2015}]{Akamatsu2015}
{Akamatsu} H.,  et~al., 2015, \mndoi [\aap] {10.1051/0004-6361/201425209},
  \href {https://ui.adsabs.harvard.edu/abs/2015A&A...582A..87A} {582, A87}

\bibitem[\protect\citeauthoryear{{Akritas} \& {Bershady}}{{Akritas} \&
  {Bershady}}{1996}]{ab96}
{Akritas} M.~G.,  {Bershady} M.~A.,  1996, \mndoi [\apj] {10.1086/177901},
  \href {http://adsabs.harvard.edu/abs/1996ApJ...470..706A} {470, 706}

\bibitem[\protect\citeauthoryear{{Astropy Collaboration} et~al.,}{{Astropy
  Collaboration} et~al.}{2013}]{astropy:2013}
{Astropy Collaboration} et~al., 2013, \mndoi [\aap]
  {10.1051/0004-6361/201322068}, \href
  {http://adsabs.harvard.edu/abs/2013A%26A...558A..33A} {558, A33}

\bibitem[\protect\citeauthoryear{{Basu}}{{Basu}}{2012}]{Basu2012}
{Basu} K.,  2012, \mndoi [\mnras] {10.1111/j.1745-3933.2012.01217.x}, \href
  {https://ui.adsabs.harvard.edu/abs/2012MNRAS.421L.112B} {421, L112}

\bibitem[\protect\citeauthoryear{{B{\^\i}rzan} et~al.,}{{B{\^\i}rzan}
  et~al.}{2019}]{Birzan2019}
{B{\^\i}rzan} L.,  et~al., 2019, \mndoi [\mnras] {10.1093/mnras/stz1456}, \href
  {https://ui.adsabs.harvard.edu/abs/2019MNRAS.487.4775B} {487, 4775}

\bibitem[\protect\citeauthoryear{{Blandford} \& {Eichler}}{{Blandford} \&
  {Eichler}}{1987}]{Blandford1987}
{Blandford} R.,  {Eichler} D.,  1987, \mndoi [\physrep]
  {10.1016/0370-1573(87)90134-7}, \href
  {https://ui.adsabs.harvard.edu/abs/1987PhR...154....1B} {154, 1}

\bibitem[\protect\citeauthoryear{{B{\"o}hringer} et~al.,}{{B{\"o}hringer}
  et~al.}{2007}]{rexcess}
{B{\"o}hringer} H.,  et~al., 2007, \mndoi [\aap] {10.1051/0004-6361:20066740},
  \href {http://adsabs.harvard.edu/abs/2007A%26A...469..363B} {469, 363}

\bibitem[\protect\citeauthoryear{{Bonafede} et~al.,}{{Bonafede}
  et~al.}{2017}]{Bonafede2017}
{Bonafede} A.,  et~al., 2017, \mndoi [\mnras] {10.1093/mnras/stx1475}, \href
  {https://ui.adsabs.harvard.edu/abs/2017MNRAS.470.3465B} {470, 3465}

\bibitem[\protect\citeauthoryear{{Bonafede} et~al.,}{{Bonafede}
  et~al.}{2020}]{Bonafede2020}
{Bonafede} A.,  et~al., 2020, arXiv e-prints, \href
  {https://ui.adsabs.harvard.edu/abs/2020arXiv201108856B} {p. arXiv:2011.08856}

\bibitem[\protect\citeauthoryear{{Bonafede} et~al.,}{{Bonafede}
  et~al.}{2021}]{Bonafede2021}
{Bonafede} A.,  et~al., 2021, \mndoi [\apj] {10.3847/1538-4357/abcb8f}, \href
  {https://ui.adsabs.harvard.edu/abs/2021ApJ...907...32B} {907, 32}

\bibitem[\protect\citeauthoryear{{Botteon} et~al.,}{{Botteon}
  et~al.}{2018}]{Botteon2018}
{Botteon} A.,  et~al., 2018, \mndoi [\mnras] {10.1093/mnras/sty1102}, \href
  {https://ui.adsabs.harvard.edu/abs/2018MNRAS.478..885B} {478, 885}

\bibitem[\protect\citeauthoryear{{Botteon} et~al.,}{{Botteon}
  et~al.}{2020a}]{Botteon2020c}
{Botteon} A.,  et~al., 2020a, \mndoi [\mnras] {10.1093/mnrasl/slaa142}, \href
  {https://ui.adsabs.harvard.edu/abs/2020MNRAS.499L..11B} {499, L11}

\bibitem[\protect\citeauthoryear{{Botteon}, {Brunetti}, {Ryu}  \&
  {Roh}}{{Botteon} et~al.}{2020b}]{Botteon2020}
{Botteon} A.,  {Brunetti} G.,  {Ryu} D.,   {Roh} S.,  2020b, \mndoi [\aap]
  {10.1051/0004-6361/201936216}, \href
  {https://ui.adsabs.harvard.edu/abs/2020A&A...634A..64B} {634, A64}

\bibitem[\protect\citeauthoryear{{Botteon} et~al.,}{{Botteon}
  et~al.}{2020c}]{Botteon2020b}
{Botteon} A.,  et~al., 2020c, \mndoi [\apj] {10.3847/1538-4357/ab9a2f}, \href
  {https://ui.adsabs.harvard.edu/abs/2020ApJ...897...93B} {897, 93}

\bibitem[\protect\citeauthoryear{{Bourdin}, {Mazzotta}, {Markevitch},
  {Giacintucci}  \& {Brunetti}}{{Bourdin} et~al.}{2013}]{Bourdin2013}
{Bourdin} H.,  {Mazzotta} P.,  {Markevitch} M.,  {Giacintucci} S.,   {Brunetti}
  G.,  2013, \mndoi [\apj] {10.1088/0004-637X/764/1/82}, \href
  {https://ui.adsabs.harvard.edu/abs/2013ApJ...764...82B} {764, 82}

\bibitem[\protect\citeauthoryear{{Br{\"u}ggen}, {Bykov}, {Ryu}  \&
  {R{\"o}ttgering}}{{Br{\"u}ggen} et~al.}{2012}]{Bruggen2012}
{Br{\"u}ggen} M.,  {Bykov} A.,  {Ryu} D.,   {R{\"o}ttgering} H.,  2012, \mndoi
  [\ssr] {10.1007/s11214-011-9785-9}, \href
  {https://ui.adsabs.harvard.edu/abs/2012SSRv..166..187B} {166, 187}

\bibitem[\protect\citeauthoryear{{Br{\"u}ggen} et~al.,}{{Br{\"u}ggen}
  et~al.}{2021}]{Bruggen2020}
{Br{\"u}ggen} M.,  et~al., 2021, \mndoi [\aap] {10.1051/0004-6361/202039533},
  \href {https://ui.adsabs.harvard.edu/abs/2021A&A...647A...3B} {647, A3}

\bibitem[\protect\citeauthoryear{{Brunetti} \& {Jones}}{{Brunetti} \&
  {Jones}}{2014}]{bj14}
{Brunetti} G.,  {Jones} T.~W.,  2014, \mndoi [International Journal of Modern
  Physics D] {10.1142/S0218271814300079}, \href
  {http://adsabs.harvard.edu/abs/2014IJMPD..2330007B} {23, 1430007}

\bibitem[\protect\citeauthoryear{{Brunetti}, {Venturi}, {Dallacasa}, {Cassano},
  {Dolag}, {Giacintucci}  \& {Setti}}{{Brunetti} et~al.}{2007}]{Brunetti2007}
{Brunetti} G.,  {Venturi} T.,  {Dallacasa} D.,  {Cassano} R.,  {Dolag} K.,
  {Giacintucci} S.,   {Setti} G.,  2007, \mndoi [\apjl] {10.1086/524037}, \href
  {https://ui.adsabs.harvard.edu/abs/2007ApJ...670L...5B} {670, L5}

\bibitem[\protect\citeauthoryear{{Brunetti} et~al.,}{{Brunetti}
  et~al.}{2008}]{Brunetti2008}
{Brunetti} G.,  et~al., 2008, \mndoi [\nat] {10.1038/nature07379}, \href
  {https://ui.adsabs.harvard.edu/abs/2008Natur.455..944B} {455, 944}

\bibitem[\protect\citeauthoryear{{Brunetti}, {Cassano}, {Dolag}  \&
  {Setti}}{{Brunetti} et~al.}{2009}]{Brunetti2009}
{Brunetti} G.,  {Cassano} R.,  {Dolag} K.,   {Setti} G.,  2009, \mndoi [\aap]
  {10.1051/0004-6361/200912751}, \href
  {https://ui.adsabs.harvard.edu/abs/2009A&A...507..661B} {507, 661}

\bibitem[\protect\citeauthoryear{{Bruno} et~al.,}{{Bruno}
  et~al.}{2021}]{Bruno2021}
{Bruno} L.,  et~al., 2021, arXiv e-prints, \href
  {https://ui.adsabs.harvard.edu/abs/2021arXiv210310110B} {p. arXiv:2103.10110}

\bibitem[\protect\citeauthoryear{{Caglar}}{{Caglar}}{2018}]{Caglar2018}
{Caglar} T.,  2018, \mndoi [\mnras] {10.1093/mnras/sty036}, \href
  {https://ui.adsabs.harvard.edu/abs/2018MNRAS.475.2870C} {475, 2870}

\bibitem[\protect\citeauthoryear{{Cassano}, {Brunetti}, {Setti}, {Govoni}  \&
  {Dolag}}{{Cassano} et~al.}{2007}]{Cassano2007}
{Cassano} R.,  {Brunetti} G.,  {Setti} G.,  {Govoni} F.,   {Dolag} K.,  2007,
  \mndoi [\mnras] {10.1111/j.1365-2966.2007.11901.x}, \href
  {https://ui.adsabs.harvard.edu/abs/2007MNRAS.378.1565C} {378, 1565}

\bibitem[\protect\citeauthoryear{{Cassano}, {Ettori}, {Giacintucci},
  {Brunetti}, {Markevitch}, {Venturi}  \& {Gitti}}{{Cassano}
  et~al.}{2010}]{Cassano2010}
{Cassano} R.,  {Ettori} S.,  {Giacintucci} S.,  {Brunetti} G.,  {Markevitch}
  M.,  {Venturi} T.,   {Gitti} M.,  2010, \mndoi [\apjl]
  {10.1088/2041-8205/721/2/L82}, \href
  {https://ui.adsabs.harvard.edu/abs/2010ApJ...721L..82C} {721, L82}

\bibitem[\protect\citeauthoryear{{Cassano} et~al.,}{{Cassano}
  et~al.}{2013}]{ceb+13}
{Cassano} R.,  et~al., 2013, \mndoi [\apj] {10.1088/0004-637X/777/2/141}, \href
  {http://adsabs.harvard.edu/abs/2013ApJ...777..141C} {777, 141}

\bibitem[\protect\citeauthoryear{{Cassano} et~al.,}{{Cassano}
  et~al.}{2019}]{Cassano2019}
{Cassano} R.,  et~al., 2019, \mndoi [\apjl] {10.3847/2041-8213/ab32ed}, \href
  {https://ui.adsabs.harvard.edu/abs/2019ApJ...881L..18C} {881, L18}

\bibitem[\protect\citeauthoryear{{Chapman}, {Dempsey}, {Miller}, {Heywood},
  {Pritchard}, {Sangster}, {Whiting}  \& {Dart}}{{Chapman}
  et~al.}{2017}]{casda}
{Chapman} J.~M.,  {Dempsey} J.,  {Miller} D.,  {Heywood} I.,  {Pritchard} J.,
  {Sangster} E.,  {Whiting} M.,   {Dart} M.,  2017, {CASDA: The CSIRO ASKAP
  Science Data Archive}.
p.~73

\bibitem[\protect\citeauthoryear{{Clarke}, {Kronberg}  \&
  {B{\"o}hringer}}{{Clarke} et~al.}{2001}]{Clarke2001}
{Clarke} T.~E.,  {Kronberg} P.~P.,   {B{\"o}hringer} H.,  2001, \mndoi [\apjl]
  {10.1086/318896}, \href
  {https://ui.adsabs.harvard.edu/abs/2001ApJ...547L.111C} {547, L111}

\bibitem[\protect\citeauthoryear{{Cuciti}, {Brunetti}, {van Weeren},
  {Bonafede}, {Dallacasa}, {Cassano}, {Venturi}  \& {Kale}}{{Cuciti}
  et~al.}{2018}]{Cuciti2018}
{Cuciti} V.,  {Brunetti} G.,  {van Weeren} R.,  {Bonafede} A.,  {Dallacasa} D.,
   {Cassano} R.,  {Venturi} T.,   {Kale} R.,  2018, \mndoi [\aap]
  {10.1051/0004-6361/201731174}, \href
  {https://ui.adsabs.harvard.edu/abs/2018A&A...609A..61C} {609, A61}

\bibitem[\protect\citeauthoryear{{Cuciti} et~al.,}{{Cuciti}
  et~al.}{2021}]{Cuciti2021b}
{Cuciti} V.,  et~al., 2021, \mndoi [\aap] {10.1051/0004-6361/202039208}, \href
  {https://ui.adsabs.harvard.edu/abs/2021A&A...647A..51C} {647, A51}

\bibitem[\protect\citeauthoryear{{Dahle}, {Kaiser}, {Irgens}, {Lilje}  \&
  {Maddox}}{{Dahle} et~al.}{2002}]{dki+02}
{Dahle} H.,  {Kaiser} N.,  {Irgens} R.~J.,  {Lilje} P.~B.,   {Maddox} S.~J.,
  2002, \mndoi [\apjs] {10.1086/338678}, \href
  {http://adsabs.harvard.edu/abs/2002ApJS..139..313D} {139, 313}

\bibitem[\protect\citeauthoryear{{DeBoer} et~al.,}{{DeBoer}
  et~al.}{2009}]{askap3}
{DeBoer} D.~R.,  et~al., 2009, \mndoi [IEEE Proceedings]
  {10.1109/JPROC.2009.2016516}, \href
  {http://adsabs.harvard.edu/abs/2009IEEEP..97.1507D} {97, 1507}

\bibitem[\protect\citeauthoryear{{Di Gennaro} et~al.,}{{Di Gennaro}
  et~al.}{2019}]{DiGennaro2019}
{Di Gennaro} G.,  et~al., 2019, \mndoi [\apj] {10.3847/1538-4357/ab03cd}, \href
  {https://ui.adsabs.harvard.edu/abs/2019ApJ...873...64D} {873, 64}

\bibitem[\protect\citeauthoryear{{Di Gennaro} et~al.,}{{Di Gennaro}
  et~al.}{2020}]{DiGennaro2020}
{Di Gennaro} G.,  et~al., 2020, \mndoi [Nature Astronomy]
  {10.1038/s41550-020-01244-5}, \href
  {https://ui.adsabs.harvard.edu/abs/2020arXiv201101628D} {}

\bibitem[\protect\citeauthoryear{{Dom\'{i}nguez-Fern\'{a}ndez}, {Bruggen},
  {Vazza}, {Banda-Barragan}, {Rajpurohit}, {Mignone}, {Mukherjee}  \&
  {Vaidya}}{{Dom\'{i}nguez-Fern\'{a}ndez} et~al.}{2021}]{Dominguez2021}
{Dom\'{i}nguez-Fern\'{a}ndez} P.,  {Bruggen} M.,  {Vazza} F.,  {Banda-Barragan}
  W.~E.,  {Rajpurohit} K.,  {Mignone} A.,  {Mukherjee} D.,   {Vaidya} B.,
  2021, \mndoi [\mnras] {doi:10.1093/mnras/staa3018}, \href
  {https://ui.adsabs.harvard.edu/abs/2021MNRAS.500..795D} {500, 795}

\bibitem[\protect\citeauthoryear{{Donnert}, {Dolag}, {Brunetti}  \&
  {Cassano}}{{Donnert} et~al.}{2013}]{ddbc13}
{Donnert} J.,  {Dolag} K.,  {Brunetti} G.,   {Cassano} R.,  2013, \mndoi
  [\mnras] {10.1093/mnras/sts628}, \href
  {http://adsabs.harvard.edu/abs/2013MNRAS.429.3564D} {429, 3564}

\bibitem[\protect\citeauthoryear{{Donnert}, {Vazza}, {Br{\"u}ggen}  \&
  {ZuHone}}{{Donnert} et~al.}{2018}]{Donnert2019}
{Donnert} J.,  {Vazza} F.,  {Br{\"u}ggen} M.,   {ZuHone} J.,  2018, \mndoi
  [\ssr] {10.1007/s11214-018-0556-8}, \href
  {https://ui.adsabs.harvard.edu/abs/2018SSRv..214..122D} {214, 122}

\bibitem[\protect\citeauthoryear{{Duchesne}, {Johnston-Hollitt}, {Zhu}, {Wayth}
   \& {Line}}{{Duchesne} et~al.}{2020}]{Duchesne2020}
{Duchesne} S.~W.,  {Johnston-Hollitt} M.,  {Zhu} Z.,  {Wayth} R.~B.,   {Line}
  J.~L.~B.,  2020, \mndoi [\pasa] {10.1017/pasa.2020.29}, \href
  {https://ui.adsabs.harvard.edu/abs/2020PASA...37...37D} {37, e037}

\bibitem[\protect\citeauthoryear{{Duchesne}, {Johnston-Hollitt}, {Bartalucci},
  {Hodgson}  \& {Pratt}}{{Duchesne} et~al.}{2021a}]{Duchesne2020b}
{Duchesne} S.~W.,  {Johnston-Hollitt} M.,  {Bartalucci} I.,  {Hodgson} T.,
  {Pratt} G.~W.,  2021a, \mndoi [\pasa] {10.1017/pasa.2020.51}, \href
  {https://ui.adsabs.harvard.edu/abs/2021PASA...38....5D} {38, e005}

\bibitem[\protect\citeauthoryear{{Duchesne}, {Johnston-Hollitt}, {Offringa},
  {Pratt}, {Zheng}  \& {Dehghan}}{{Duchesne} et~al.}{2021b}]{Duchesne2017}
{Duchesne} S.~W.,  {Johnston-Hollitt} M.,  {Offringa} A.~R.,  {Pratt} G.~W.,
  {Zheng} Q.,   {Dehghan} S.,  2021b, \mndoi [\pasa] {10.1017/pasa.2021.7},
  \href {https://ui.adsabs.harvard.edu/abs/2021PASA...38...10D} {38, e010}

\bibitem[\protect\citeauthoryear{{Eckert}, {Molendi}  \& {Paltani}}{{Eckert}
  et~al.}{2011}]{Eckert2011}
{Eckert} D.,  {Molendi} S.,   {Paltani} S.,  2011, \mndoi [\aap]
  {10.1051/0004-6361/201015856}, \href
  {https://ui.adsabs.harvard.edu/abs/2011A%26A...526A..79E} {526, A79}

\bibitem[\protect\citeauthoryear{{Eckert}, {Jauzac}, {Vazza}, {Owers}, {Kneib},
  {Tchernin}, {Intema}  \& {Knowles}}{{Eckert} et~al.}{2016}]{Eckert2016b}
{Eckert} D.,  {Jauzac} M.,  {Vazza} F.,  {Owers} M.~S.,  {Kneib} J.~P.,
  {Tchernin} C.,  {Intema} H.,   {Knowles} K.,  2016, \mndoi [\mnras]
  {10.1093/mnras/stw1435}, \href
  {https://ui.adsabs.harvard.edu/abs/2016MNRAS.461.1302E} {461, 1302}

\bibitem[\protect\citeauthoryear{{En{\ss}lin} \& {Gopal-Krishna}}{{En{\ss}lin}
  \& {Gopal-Krishna}}{2001}]{eg01}
{En{\ss}lin} T.~A.,  {Gopal-Krishna} 2001, \mndoi [\aap]
  {10.1051/0004-6361:20000198}, \href
  {http://adsabs.harvard.edu/abs/2001A%26A...366...26E} {366, 26}

\bibitem[\protect\citeauthoryear{{Feretti}, {Fusco-Femiano}, {Giovannini}  \&
  {Govoni}}{{Feretti} et~al.}{2001}]{Feretti2001}
{Feretti} L.,  {Fusco-Femiano} R.,  {Giovannini} G.,   {Govoni} F.,  2001,
  \mndoi [\aap] {10.1051/0004-6361:20010581}, \href
  {https://ui.adsabs.harvard.edu/abs/2001A&A...373..106F} {373, 106}

\bibitem[\protect\citeauthoryear{{Frater}, {Brooks}  \& {Whiteoak}}{{Frater}
  et~al.}{1992}]{fbw92}
{Frater} R.~H.,  {Brooks} J.~W.,   {Whiteoak} J.~B.,  1992, Journal of
  Electrical and Electronics Engineering Australia, \href
  {http://adsabs.harvard.edu/abs/1992JEEEA..12..103F} {12, 103}

\bibitem[\protect\citeauthoryear{{Fruscione} et~al.,}{{Fruscione}
  et~al.}{2006}]{Fuscione2006}
{Fruscione} A.,  et~al., 2006, in \procspie. p. 62701V,
  \mndoi{10.1117/12.671760}

\bibitem[\protect\citeauthoryear{{Giacintucci} et~al.,}{{Giacintucci}
  et~al.}{2005}]{Giacintucci2005}
{Giacintucci} S.,  et~al., 2005, \mndoi [\aap] {10.1051/0004-6361:20053016},
  \href {https://ui.adsabs.harvard.edu/abs/2005A&A...440..867G} {440, 867}

\bibitem[\protect\citeauthoryear{{Giacintucci}, {Markevitch}, {Cassano},
  {Venturi}, {Clarke}  \& {Brunetti}}{{Giacintucci}
  et~al.}{2017}]{Giacintucci2017}
{Giacintucci} S.,  {Markevitch} M.,  {Cassano} R.,  {Venturi} T.,  {Clarke}
  T.~E.,   {Brunetti} G.,  2017, \mndoi [\apj] {10.3847/1538-4357/aa7069},
  \href {https://ui.adsabs.harvard.edu/abs/2017ApJ...841...71G} {841, 71}

\bibitem[\protect\citeauthoryear{{Giacintucci}, {Markevitch}, {Cassano},
  {Venturi}, {Clarke}, {Kale}  \& {Cuciti}}{{Giacintucci}
  et~al.}{2019}]{Giacintucci2019}
{Giacintucci} S.,  {Markevitch} M.,  {Cassano} R.,  {Venturi} T.,  {Clarke}
  T.~E.,  {Kale} R.,   {Cuciti} V.,  2019, \mndoi [\apj]
  {10.3847/1538-4357/ab29f1}, \href
  {https://ui.adsabs.harvard.edu/abs/2019ApJ...880...70G} {880, 70}

\bibitem[\protect\citeauthoryear{{Giovannini}, {Feretti}, {Girardi}, {Govoni},
  {Murgia}, {Vacca}  \& {Bagchi}}{{Giovannini} et~al.}{2011}]{Giovannini2011}
{Giovannini} G.,  {Feretti} L.,  {Girardi} M.,  {Govoni} F.,  {Murgia} M.,
  {Vacca} V.,   {Bagchi} J.,  2011, \mndoi [\aap]
  {10.1051/0004-6361/201116930}, \href
  {https://ui.adsabs.harvard.edu/abs/2011A&A...530L...5G} {530, L5}

\bibitem[\protect\citeauthoryear{{Giovannini} et~al.,}{{Giovannini}
  et~al.}{2020}]{Giovannini2020}
{Giovannini} G.,  et~al., 2020, \mndoi [\aap] {10.1051/0004-6361/202038263},
  \href {https://ui.adsabs.harvard.edu/abs/2020A&A...640A.108G} {640, A108}

\bibitem[\protect\citeauthoryear{{Govoni}, {En{\ss}lin}, {Feretti}  \&
  {Giovannini}}{{Govoni} et~al.}{2001}]{Govoni2001}
{Govoni} F.,  {En{\ss}lin} T.~A.,  {Feretti} L.,   {Giovannini} G.,  2001,
  \mndoi [\aap] {10.1051/0004-6361:20010115}, \href
  {https://ui.adsabs.harvard.edu/abs/2001A&A...369..441G} {369, 441}

\bibitem[\protect\citeauthoryear{{Govoni} et~al.,}{{Govoni}
  et~al.}{2019}]{Govoni2019}
{Govoni} F.,  et~al., 2019, \mndoi [Science] {10.1126/science.aat7500}, \href
  {https://ui.adsabs.harvard.edu/abs/2019Sci...364..981G} {364, 981}

\bibitem[\protect\citeauthoryear{{Hambly} et~al.,}{{Hambly}
  et~al.}{2001a}]{supercosmos1}
{Hambly} N.~C.,  et~al., 2001a, \mndoi [\mnras]
  {10.1111/j.1365-2966.2001.04660.x}, \href
  {http://adsabs.harvard.edu/abs/2001MNRAS.326.1279H} {326, 1279}

\bibitem[\protect\citeauthoryear{{Hambly}, {Irwin}  \& {MacGillivray}}{{Hambly}
  et~al.}{2001b}]{supercosmos2}
{Hambly} N.~C.,  {Irwin} M.~J.,   {MacGillivray} H.~T.,  2001b, \mndoi [\mnras]
  {10.1111/j.1365-2966.2001.04661.x}, \href
  {http://adsabs.harvard.edu/abs/2001MNRAS.326.1295H} {326, 1295}

\bibitem[\protect\citeauthoryear{{Hambly}, {Davenhall}, {Irwin}  \&
  {MacGillivray}}{{Hambly} et~al.}{2001c}]{supercosmos3}
{Hambly} N.~C.,  {Davenhall} A.~C.,  {Irwin} M.~J.,   {MacGillivray} H.~T.,
  2001c, \mndoi [\mnras] {10.1111/j.1365-2966.2001.04662.x}, \href
  {http://adsabs.harvard.edu/abs/2001MNRAS.326.1315H} {326, 1315}

\bibitem[\protect\citeauthoryear{{Harris}, {Kapahi}  \& {Ekers}}{{Harris}
  et~al.}{1980}]{Harris1980}
{Harris} D.~E.,  {Kapahi} V.~K.,   {Ekers} R.~D.,  1980, \aaps, \href
  {https://ui.adsabs.harvard.edu/abs/1980A&AS...39..215H} {39, 215}

\bibitem[\protect\citeauthoryear{{Harvey-Smith} et~al.,}{{Harvey-Smith}
  et~al.}{2018}]{askap:a3404}
{Harvey-Smith} L.,  et~al., 2018, ASKAP Data Products for Project AS034 (ASKAP
  Early Science Broadband Survey): images and visibilities. v1. CSIRO. Data
  Collection, \url {httFHp://hdl.handle.net/102.100.100/74037?index=1}

\bibitem[\protect\citeauthoryear{{Helfer}, {Thornley}, {Regan}, {Wong},
  {Sheth}, {Vogel}, {Blitz}  \& {Bock}}{{Helfer} et~al.}{2003}]{Helfer2003}
{Helfer} T.~T.,  {Thornley} M.~D.,  {Regan} M.~W.,  {Wong} T.,  {Sheth} K.,
  {Vogel} S.~N.,  {Blitz} L.,   {Bock} D. C.~J.,  2003, \mndoi [\apjs]
  {10.1086/346076}, \href
  {https://ui.adsabs.harvard.edu/abs/2003ApJS..145..259H} {145, 259}

\bibitem[\protect\citeauthoryear{{Hoang} et~al.,}{{Hoang}
  et~al.}{2019a}]{Hoang2019a}
{Hoang} D.~N.,  et~al., 2019a, \mndoi [\aap] {10.1051/0004-6361/201833900},
  \href {https://ui.adsabs.harvard.edu/abs/2019A&A...622A..20H} {622, A20}

\bibitem[\protect\citeauthoryear{{Hoang} et~al.,}{{Hoang}
  et~al.}{2019b}]{Hoang2019}
{Hoang} D.~N.,  et~al., 2019b, \mndoi [\aap] {10.1051/0004-6361/201834025},
  \href {https://ui.adsabs.harvard.edu/abs/2019A&A...622A..21H} {622, A21}

\bibitem[\protect\citeauthoryear{{Hoang} et~al.,}{{Hoang}
  et~al.}{2021}]{Hoang2020}
{Hoang} D.~N.,  et~al., 2021, \mndoi [\mnras] {10.1093/mnras/staa3581}, \href
  {https://ui.adsabs.harvard.edu/abs/2021MNRAS.501..576H} {501, 576}

\bibitem[\protect\citeauthoryear{{Hodgson}, {Bartalucci}, {Johnston-Hollitt},
  {McKinley}, {Vazza}  \& {Wittor}}{{Hodgson} et~al.}{2021}]{Hodgson2021}
{Hodgson} T.,  {Bartalucci} I.,  {Johnston-Hollitt} M.,  {McKinley} B.,
  {Vazza} F.,   {Wittor} D.,  2021, \mndoi [\apj] {10.3847/1538-4357/abe384},
  \href {https://ui.adsabs.harvard.edu/abs/2021ApJ...909..198H} {909, 198}

\bibitem[\protect\citeauthoryear{{Hoeft} et~al.,}{{Hoeft}
  et~al.}{2020}]{Hoeft2020}
{Hoeft} M.,  et~al., 2020, arXiv e-prints, \href
  {https://ui.adsabs.harvard.edu/abs/2020arXiv201010331H} {p. arXiv:2010.10331}

\bibitem[\protect\citeauthoryear{{Hotan} et~al.,}{{Hotan}
  et~al.}{2014}]{Hotan2014}
{Hotan} A.~W.,  et~al., 2014, \mndoi [\pasa] {10.1017/pasa.2014.36}, \href
  {https://ui.adsabs.harvard.edu/abs/2014PASA...31...41H} {31, e041}

\bibitem[\protect\citeauthoryear{{Hotan} et~al.,}{{Hotan}
  et~al.}{2021}]{Hotan2021}
{Hotan} A.~W.,  et~al., 2021, \mndoi [\pasa] {10.1017/pasa.2021.1}, \href
  {https://ui.adsabs.harvard.edu/abs/2021PASA...38....9H} {38, e009}

\bibitem[\protect\citeauthoryear{{Hunter}}{{Hunter}}{2007}]{Hunter2007}
{Hunter} J.~D.,  2007, \mndoi [Computing in Science and Engineering]
  {10.1109/MCSE.2007.55}, \href
  {http://adsabs.harvard.edu/abs/2007CSE.....9...90H} {9, 90}

\bibitem[\protect\citeauthoryear{{Hurley-Walker} \& {Hancock}}{{Hurley-Walker}
  \& {Hancock}}{2018}]{hh18}
{Hurley-Walker} N.,  {Hancock} P.~J.,  2018, \mndoi [Astronomy and Computing]
  {10.1016/j.ascom.2018.08.006}, \href
  {http://adsabs.harvard.edu/abs/2018A%26C....25...94H} {25, 94}

\bibitem[\protect\citeauthoryear{{Hurley-Walker} et~al.,}{{Hurley-Walker}
  et~al.}{2017}]{gleamegc}
{Hurley-Walker} N.,  et~al., 2017, \mndoi [\mnras] {10.1093/mnras/stw2337},
  \href {http://adsabs.harvard.edu/abs/2017MNRAS.464.1146H} {464, 1146}

\bibitem[\protect\citeauthoryear{{HyeongHan} et~al.,}{{HyeongHan}
  et~al.}{2020}]{HyeongHan2020}
{HyeongHan} K.,  et~al., 2020, \mndoi [\apj] {10.3847/1538-4357/aba742}, \href
  {https://ui.adsabs.harvard.edu/abs/2020ApJ...900..127H} {900, 127}

\bibitem[\protect\citeauthoryear{{Ignesti}, {Brunetti}, {Gitti}  \&
  {Giacintucci}}{{Ignesti} et~al.}{2020}]{Ignesti2020}
{Ignesti} A.,  {Brunetti} G.,  {Gitti} M.,   {Giacintucci} S.,  2020, \mndoi
  [\aap] {10.1051/0004-6361/201937207}, \href
  {https://ui.adsabs.harvard.edu/abs/2020A&A...640A..37I} {640, A37}

\bibitem[\protect\citeauthoryear{{Johnston-Hollitt}}{{Johnston-Hollitt}}{2003}]{mj-h}
{Johnston-Hollitt} M.,  2003, PhD thesis, University of Adelaide

\bibitem[\protect\citeauthoryear{{Johnston-Hollitt}, {Finoguenov},
  {B{\"o}hringer}, {Pratt}  \& {Croston}}{{Johnston-Hollitt}
  et~al.}{2007}]{atca:a3404:c1683}
{Johnston-Hollitt} M.,  {Finoguenov} A.,  {B{\"o}hringer} H.,  {Pratt} G.,
  {Croston} J.,  2007, Radio Imaging of an X-ray Luminosity Selected Galaxy
  Cluster Sample (C1683), \url {https://atoa.atnf.csiro.au/}

\bibitem[\protect\citeauthoryear{{Johnston-Hollitt}, {Basu}, {Nord}, {Hindson}
  \& {Shakouri}}{{Johnston-Hollitt} et~al.}{2013}]{atca:a3404:c2837}
{Johnston-Hollitt} M.,  {Basu} K.,  {Nord} M.,  {Hindson} L.,   {Shakouri} S.,
  2013, A census of radio emission in a complete SZ-derived cluster sample
  (C2837), \url {https://atoa.atnf.csiro.au/}

\bibitem[\protect\citeauthoryear{Jones, Oliphant, Peterson  et~al.}{Jones
  et~al.}{2001}]{Jones2001}
Jones E.,  Oliphant T.,  Peterson P.,   et~al., 2001, {SciPy}: Open source
  scientific tools for {Python}, \url {http://www.scipy.org/}

\bibitem[\protect\citeauthoryear{{Kaiser} et~al.,}{{Kaiser}
  et~al.}{2010}]{kpc+10}
{Kaiser} N.,  et~al., 2010, in Ground-based and Airborne Telescopes III. p.
  77330E, \mndoi{10.1117/12.859188}

\bibitem[\protect\citeauthoryear{{Kempner}, {Blanton}, {Clarke}, {En{\ss}lin},
  {Johnston-Hollitt}  \& {Rudnick}}{{Kempner} et~al.}{2004}]{kempner2004}
{Kempner} J.~C.,  {Blanton} E.~L.,  {Clarke} T.~E.,  {En{\ss}lin} T.~A.,
  {Johnston-Hollitt} M.,   {Rudnick} L.,  2004, in {Reiprich} T.,  {Kempner}
  J.,   {Soker} N.,  eds, The Riddle of Cooling Flows in Galaxies and Clusters
  of galaxies. p.~335 (\mn@eprint {arXiv} {astro-ph/0310263})

\bibitem[\protect\citeauthoryear{{Keshet}, {Waxman}  \& {Loeb}}{{Keshet}
  et~al.}{2004}]{Keshet2004}
{Keshet} U.,  {Waxman} E.,   {Loeb} A.,  2004, \mndoi [\apj] {10.1086/424837},
  \href {https://ui.adsabs.harvard.edu/abs/2004ApJ...617..281K} {617, 281}

\bibitem[\protect\citeauthoryear{{Knowles} et~al.,}{{Knowles}
  et~al.}{2021}]{Knowles2020}
{Knowles} K.,  et~al., 2021, \mndoi [\mnras] {10.1093/mnras/stab939}, \href
  {https://ui.adsabs.harvard.edu/abs/2021MNRAS.504.1749K} {504, 1749}

\bibitem[\protect\citeauthoryear{{Lee}, {Jee}, {Kang}, {Ryu}, {Kimm}  \&
  {Br{\"u}ggen}}{{Lee} et~al.}{2020}]{Lee2020}
{Lee} W.,  {Jee} M.~J.,  {Kang} H.,  {Ryu} D.,  {Kimm} T.,   {Br{\"u}ggen} M.,
  2020, \mndoi [\apj] {10.3847/1538-4357/ab855f}, \href
  {https://ui.adsabs.harvard.edu/abs/2020ApJ...894...60L} {894, 60}

\bibitem[\protect\citeauthoryear{{Liang}, {Hunstead}, {Birkinshaw}  \&
  {Andreani}}{{Liang} et~al.}{2000}]{Liang2000}
{Liang} H.,  {Hunstead} R.~W.,  {Birkinshaw} M.,   {Andreani} P.,  2000, \mndoi
  [\apj] {10.1086/317223}, \href
  {https://ui.adsabs.harvard.edu/abs/2000ApJ...544..686L} {544, 686}

\bibitem[\protect\citeauthoryear{{Martinez Aviles} et~al.,}{{Martinez Aviles}
  et~al.}{2016}]{mfj+16}
{Martinez Aviles} G.,  et~al., 2016, \mndoi [\aap]
  {10.1051/0004-6361/201628788}, \href
  {http://adsabs.harvard.edu/abs/2016A%26A...595A.116M} {595, A116}

\bibitem[\protect\citeauthoryear{{Mazzotta}, {Bourdin}, {Giacintucci},
  {Markevitch}  \& {Venturi}}{{Mazzotta} et~al.}{2011}]{Mazzotta2011}
{Mazzotta} P.,  {Bourdin} H.,  {Giacintucci} S.,  {Markevitch} M.,   {Venturi}
  T.,  2011, \memsai, \href
  {https://ui.adsabs.harvard.edu/abs/2011MmSAI..82..495M} {82, 495}

\bibitem[\protect\citeauthoryear{{McConnell} et~al.,}{{McConnell}
  et~al.}{2016}]{MCCONNELL2016}
{McConnell} D.,  et~al., 2016, \mndoi [\pasa] {10.1017/pasa.2016.37}, \href
  {https://ui.adsabs.harvard.edu/abs/2016PASA...33...42M} {33, e042}

\bibitem[\protect\citeauthoryear{{McMullin}, {Waters}, {Schiebel}, {Young}  \&
  {Golap}}{{McMullin} et~al.}{2007}]{casa}
{McMullin} J.~P.,  {Waters} B.,  {Schiebel} D.,  {Young} W.,   {Golap} K.,
  2007, in {Shaw} R.~A.,  {Hill} F.,   {Bell} D.~J.,  eds,  Astronomical
  Society of the Pacific Conference Series Vol. 376, Astronomical Data Analysis
  Software and Systems XVI. p.~127

\bibitem[\protect\citeauthoryear{{Mohr}, {Fabricant}  \& {Geller}}{{Mohr}
  et~al.}{1993}]{Mohr1993}
{Mohr} J.~J.,  {Fabricant} D.~G.,   {Geller} M.~J.,  1993, \mndoi [\apj]
  {10.1086/173019}, \href
  {https://ui.adsabs.harvard.edu/abs/1993ApJ...413..492M} {413, 492}

\bibitem[\protect\citeauthoryear{{Murgia}, {Govoni}, {Markevitch}, {Feretti},
  {Giovannini}, {Taylor}  \& {Carretti}}{{Murgia} et~al.}{2009}]{Murgia2009}
{Murgia} M.,  {Govoni} F.,  {Markevitch} M.,  {Feretti} L.,  {Giovannini} G.,
  {Taylor} G.~B.,   {Carretti} E.,  2009, \mndoi [\aap]
  {10.1051/0004-6361/200911659}, \href
  {https://ui.adsabs.harvard.edu/abs/2009A&A...499..679M} {499, 679}

\bibitem[\protect\citeauthoryear{{Murgia}, {Govoni}, {Feretti}  \&
  {Giovannini}}{{Murgia} et~al.}{2010}]{Murgia2010}
{Murgia} M.,  {Govoni} F.,  {Feretti} L.,   {Giovannini} G.,  2010, \mndoi
  [\aap] {10.1051/0004-6361/200913414}, \href
  {https://ui.adsabs.harvard.edu/abs/2010A&A...509A..86M} {509, A86}

\bibitem[\protect\citeauthoryear{{Murphy}, {Lenc}, {Whiting}, {Huynh}  \&
  {Hotan}}{{Murphy} et~al.}{2019}]{askap:a141}
{Murphy} T.,  {Lenc} E.,  {Whiting} M.,  {Huynh} M.,   {Hotan} A.,  2019, ASKAP
  Data Products for Project AS111 (ASKAP Pilot Survey for Gravitational Wave
  Counterparts): images and visibilities. v1. CSIRO. Data Collection, \url
  {http://hdl.handle.net/102.100.100/175570?index=1}

\bibitem[\protect\citeauthoryear{{Norris} et~al.,}{{Norris}
  et~al.}{2011}]{Norris2011a}
{Norris} R.~P.,  et~al., 2011, \mndoi [\pasa] {10.1071/AS11021}, \href
  {https://ui.adsabs.harvard.edu/abs/2011PASA...28..215N} {28, 215}

\bibitem[\protect\citeauthoryear{{Offringa} \& {Smirnov}}{{Offringa} \&
  {Smirnov}}{2017}]{wsclean2}
{Offringa} A.~R.,  {Smirnov} O.,  2017, \mndoi [\mnras]
  {10.1093/mnras/stx1547}, \href
  {https://ui.adsabs.harvard.edu/abs/2017MNRAS.471..301O} {471, 301}

\bibitem[\protect\citeauthoryear{{Offringa} et~al.,}{{Offringa}
  et~al.}{2014}]{wsclean1}
{Offringa} A.~R.,  et~al., 2014, \mndoi [\mnras] {10.1093/mnras/stu1368}, \href
  {https://ui.adsabs.harvard.edu/abs/2014MNRAS.444..606O} {444, 606}

\bibitem[\protect\citeauthoryear{{Offringa} et~al.,}{{Offringa}
  et~al.}{2016}]{oth+16}
{Offringa} A.~R.,  et~al., 2016, \mndoi [\mnras] {10.1093/mnras/stw310}, \href
  {http://adsabs.harvard.edu/abs/2016MNRAS.458.1057O} {458, 1057}

\bibitem[\protect\citeauthoryear{{Ogrean} et~al.,}{{Ogrean}
  et~al.}{2015}]{Ogrean2015}
{Ogrean} G.~A.,  et~al., 2015, \mndoi [\apj] {10.1088/0004-637X/812/2/153},
  \href {https://ui.adsabs.harvard.edu/abs/2015ApJ...812..153O} {812, 153}

\bibitem[\protect\citeauthoryear{{Orr{\'u}}, {Murgia}, {Feretti}, {Govoni},
  {Brunetti}, {Giovannini}, {Girardi}  \& {Setti}}{{Orr{\'u}}
  et~al.}{2007}]{Orru2007}
{Orr{\'u}} E.,  {Murgia} M.,  {Feretti} L.,  {Govoni} F.,  {Brunetti} G.,
  {Giovannini} G.,  {Girardi} M.,   {Setti} G.,  2007, \mndoi [\aap]
  {10.1051/0004-6361:20066118}, \href
  {https://ui.adsabs.harvard.edu/abs/2007A&A...467..943O} {467, 943}

\bibitem[\protect\citeauthoryear{{Owers}, {Nulsen}, {Couch}, {Markevitch}  \&
  {Poole}}{{Owers} et~al.}{2009}]{Owers2009a}
{Owers} M.~S.,  {Nulsen} P. E.~J.,  {Couch} W.~J.,  {Markevitch} M.,   {Poole}
  G.~B.,  2009, \mndoi [\apj] {10.1088/0004-637X/692/1/702}, \href
  {https://ui.adsabs.harvard.edu/abs/2009ApJ...692..702O} {692, 702}

\bibitem[\protect\citeauthoryear{{Parekh}, {Dwarakanath}, {Kale}  \&
  {Intema}}{{Parekh} et~al.}{2017}]{Parekh2017}
{Parekh} V.,  {Dwarakanath} K.~S.,  {Kale} R.,   {Intema} H.,  2017, \mndoi
  [\mnras] {10.1093/mnras/stw2521}, \href
  {https://ui.adsabs.harvard.edu/abs/2017MNRAS.464.2752P} {464, 2752}

\bibitem[\protect\citeauthoryear{{Peebles}}{{Peebles}}{1980}]{pee80}
{Peebles} P.~J.~E.,  1980, {The large-scale structure of the universe}.
Princeton Univ. Press, Princeton, N.~J.

\bibitem[\protect\citeauthoryear{{Piffaretti}, {Arnaud}, {Pratt},
  {Pointecouteau}  \& {Melin}}{{Piffaretti} et~al.}{2011}]{pap+11}
{Piffaretti} R.,  {Arnaud} M.,  {Pratt} G.~W.,  {Pointecouteau} E.,   {Melin}
  J.-B.,  2011, \mndoi [\aap] {10.1051/0004-6361/201015377}, \href
  {http://adsabs.harvard.edu/abs/2011A%26A...534A.109P} {534, A109}

\bibitem[\protect\citeauthoryear{{Planck Collaboration} et~al.,}{{Planck
  Collaboration} et~al.}{2016a}]{planck16b}
{Planck Collaboration} et~al., 2016a, \mndoi [\aap]
  {10.1051/0004-6361/201525823}, \href
  {https://ui.adsabs.harvard.edu/abs/2016A&A...594A..27P} {594, A27}

\bibitem[\protect\citeauthoryear{{Planck Collaboration} et~al.,}{{Planck
  Collaboration} et~al.}{2016b}]{planck16}
{Planck Collaboration} et~al., 2016b, \mndoi [\aap]
  {10.1051/0004-6361/201525823}, \href
  {https://ui.adsabs.harvard.edu/abs/2016A&A...594A..27P} {594, A27}

\bibitem[\protect\citeauthoryear{{Planck Collaboration} et~al.,}{{Planck
  Collaboration} et~al.}{2016c}]{Planck2016}
{Planck Collaboration} et~al., 2016c, \mndoi [\aap]
  {10.1051/0004-6361/201525823}, \href
  {https://ui.adsabs.harvard.edu/abs/2016A&A...594A..27P} {594, A27}

\bibitem[\protect\citeauthoryear{{Poole}, {Fardal}, {Babul}, {McCarthy},
  {Quinn}  \& {Wadsley}}{{Poole} et~al.}{2006}]{Poole2006}
{Poole} G.~B.,  {Fardal} M.~A.,  {Babul} A.,  {McCarthy} I.~G.,  {Quinn} T.,
  {Wadsley} J.,  2006, \mndoi [\mnras] {10.1111/j.1365-2966.2006.10916.x},
  \href {https://ui.adsabs.harvard.edu/abs/2006MNRAS.373..881P} {373, 881}

\bibitem[\protect\citeauthoryear{{Pratt}, {Croston}, {Arnaud}  \&
  {B{\"o}hringer}}{{Pratt} et~al.}{2009}]{pcab09}
{Pratt} G.~W.,  {Croston} J.~H.,  {Arnaud} M.,   {B{\"o}hringer} H.,  2009,
  \mndoi [\aap] {10.1051/0004-6361/200810994}, \href
  {http://adsabs.harvard.edu/abs/2009A%26A...498..361P} {498, 361}

\bibitem[\protect\citeauthoryear{{Price-Whelan} et~al.,}{{Price-Whelan}
  et~al.}{2018}]{astropy:2018}
{Price-Whelan} A.~M.,  et~al., 2018, \mndoi [\aj] {10.3847/1538-3881/aabc4f},
  \href {https://ui.adsabs.harvard.edu/#abs/2018AJ....156..123T} {156, 123}

\bibitem[\protect\citeauthoryear{{Raja} et~al.,}{{Raja}
  et~al.}{2020}]{Raja2020}
{Raja} R.,  et~al., 2020, \mndoi [\mnras] {10.1093/mnrasl/slaa002}, \href
  {https://ui.adsabs.harvard.edu/abs/2020MNRAS.493L..28R} {493, L28}

\bibitem[\protect\citeauthoryear{{Raja}, {Rahaman}, {Datta}, {van Weeren},
  {Intema}  \& {Paul}}{{Raja} et~al.}{2021}]{Raja2021}
{Raja} R.,  {Rahaman} M.,  {Datta} A.,  {van Weeren} R.~J.,  {Intema} H.~T.,
  {Paul} S.,  2021, \mndoi [\mnras] {10.1093/mnras/staa3432}, \href
  {https://ui.adsabs.harvard.edu/abs/2021MNRAS.500.2236R} {500, 2236}

\bibitem[\protect\citeauthoryear{{Rajpurohit} et~al.,}{{Rajpurohit}
  et~al.}{2018}]{Rajpurohit2018}
{Rajpurohit} K.,  et~al., 2018, \mndoi [\apj] {10.3847/1538-4357/aa9f13}, \href
  {https://ui.adsabs.harvard.edu/abs/2018ApJ...852...65R} {852, 65}

\bibitem[\protect\citeauthoryear{{Rajpurohit} et~al.,}{{Rajpurohit}
  et~al.}{2020}]{Rajpurohit2020a}
{Rajpurohit} K.,  et~al., 2020, \mndoi [\aap] {10.1051/0004-6361/201937139},
  \href {https://ui.adsabs.harvard.edu/abs/2020A&A...636A..30R} {636, A30}

\bibitem[\protect\citeauthoryear{{Rajpurohit} et~al.,}{{Rajpurohit}
  et~al.}{2021}]{Rajpurohit2021a}
{Rajpurohit} K.,  et~al., 2021, \mndoi [\aap] {10.1051/0004-6361/202039591},
  \href {https://ui.adsabs.harvard.edu/abs/2021A&A...646A.135R} {646, A135}

\bibitem[\protect\citeauthoryear{{Reichardt} et~al.,}{{Reichardt}
  et~al.}{2013}]{spt2}
{Reichardt} C.~L.,  et~al., 2013, \mndoi [\apj] {10.1088/0004-637X/763/2/127},
  \href {http://cdsads.u-strasbg.fr/abs/2013ApJ...763..127R} {763, 127}

\bibitem[\protect\citeauthoryear{{Robitaille} \& {Bressert}}{{Robitaille} \&
  {Bressert}}{2012}]{Robitaille2012}
{Robitaille} T.,  {Bressert} E.,  2012, {APLpy: Astronomical Plotting Library
  in Python}, Astrophysics Source Code Library (\mn@eprint {ascl} {1208.017})

\bibitem[\protect\citeauthoryear{{Santos}, {Rosati}, {Tozzi}, {B{\"o}hringer},
  {Ettori}  \& {Bignamini}}{{Santos} et~al.}{2008}]{Santos2008}
{Santos} J.~S.,  {Rosati} P.,  {Tozzi} P.,  {B{\"o}hringer} H.,  {Ettori} S.,
  {Bignamini} A.,  2008, \mndoi [\aap] {10.1051/0004-6361:20078815}, \href
  {https://ui.adsabs.harvard.edu/abs/2008A%26A...483...35S} {483, 35}

\bibitem[\protect\citeauthoryear{{Sarazin}}{{Sarazin}}{2002}]{Sarazin2002}
{Sarazin} C.~L.,  2002, {The Physics of Cluster Mergers}.
pp 1--38, \mndoi{10.1007/0-306-48096-4_1}

\bibitem[\protect\citeauthoryear{{Sault}, {Teuben}  \& {Wright}}{{Sault}
  et~al.}{1995}]{stw95}
{Sault} R.~J.,  {Teuben} P.~J.,   {Wright} M.~C.~H.,  1995, in {Shaw} R.~A.,
  {Payne} H.~E.,   {Hayes} J.~J.~E.,  eds,  Astronomical Society of the Pacific
  Conference Series Vol. 77, Astronomical Data Analysis Software and Systems
  IV. p.~433 (\mn@eprint {} {astro-ph/0612759})

\bibitem[\protect\citeauthoryear{{Shakouri}, {Johnston-Hollitt}  \&
  {Pratt}}{{Shakouri} et~al.}{2016}]{sjp16}
{Shakouri} S.,  {Johnston-Hollitt} M.,   {Pratt} G.~W.,  2016, \mndoi [\mnras]
  {10.1093/mnras/stw812}, \href
  {http://adsabs.harvard.edu/abs/2016MNRAS.459.2525S} {459, 2525}

\bibitem[\protect\citeauthoryear{{Shimwel}, {Stroe}  \& {Hoang}}{{Shimwel}
  et~al.}{2015}]{atca:a141}
{Shimwel} T.,  {Stroe} A.,   {Hoang} D.,  2015, Ultra-deep observations to map
  the diffuse radio emission from a sample of merging galaxy clusters (C2915),
  \url {https://atoa.atnf.csiro.au/}

\bibitem[\protect\citeauthoryear{{Shimwell}, {Brown}, {Feain}, {Feretti},
  {Gaensler}  \& {Lage}}{{Shimwell} et~al.}{2014}]{Shimwell2014}
{Shimwell} T.~W.,  {Brown} S.,  {Feain} I.~J.,  {Feretti} L.,  {Gaensler}
  B.~M.,   {Lage} C.,  2014, \mndoi [\mnras] {10.1093/mnras/stu467}, \href
  {https://ui.adsabs.harvard.edu/abs/2014MNRAS.440.2901S} {440, 2901}

\bibitem[\protect\citeauthoryear{{Slee}, {Roy}, {Murgia}, {Andernach}  \&
  {Ehle}}{{Slee} et~al.}{2001}]{srm+01}
{Slee} O.~B.,  {Roy} A.~L.,  {Murgia} M.,  {Andernach} H.,   {Ehle} M.,  2001,
  \mndoi [\aj] {10.1086/322105}, \href
  {http://adsabs.harvard.edu/abs/2001AJ....122.1172S} {122, 1172}

\bibitem[\protect\citeauthoryear{{Sokolowski} et~al.,}{{Sokolowski}
  et~al.}{2017}]{Sokolowski2017}
{Sokolowski} M.,  et~al., 2017, \mndoi [\pasa] {10.1017/pasa.2017.54}, \href
  {https://ui.adsabs.harvard.edu/abs/2017PASA...34...62S} {34, e062}

\bibitem[\protect\citeauthoryear{{Song} et~al.,}{{Song} et~al.}{2012}]{spt1}
{Song} J.,  et~al., 2012, \mndoi [\apj] {10.1088/0004-637X/761/1/22}, \href
  {http://cdsads.u-strasbg.fr/abs/2012ApJ...761...22S} {761, 22}

\bibitem[\protect\citeauthoryear{{Stroe} et~al.,}{{Stroe}
  et~al.}{2016}]{Stroe2016}
{Stroe} A.,  et~al., 2016, \mndoi [\mnras] {10.1093/mnras/stv2472}, \href
  {https://ui.adsabs.harvard.edu/abs/2016MNRAS.455.2402S} {455, 2402}

\bibitem[\protect\citeauthoryear{{Struble} \& {Rood}}{{Struble} \&
  {Rood}}{1999}]{sr99}
{Struble} M.~F.,  {Rood} H.~J.,  1999, \mndoi [\apjs] {10.1086/313274}, \href
  {http://adsabs.harvard.edu/abs/1999ApJS..125...35S} {125, 35}

\bibitem[\protect\citeauthoryear{{Tingay} et~al.,}{{Tingay}
  et~al.}{2013}]{tgb+13}
{Tingay} S.~J.,  et~al., 2013, \mndoi [\pasa] {10.1017/pasa.2012.007}, \href
  {http://adsabs.harvard.edu/abs/2013PASA...30....7T} {30, 7}

\bibitem[\protect\citeauthoryear{{Tonry} et~al.,}{{Tonry}
  et~al.}{2012}]{tsl+12}
{Tonry} J.~L.,  et~al., 2012, \mndoi [\apj] {10.1088/0004-637X/750/2/99}, \href
  {http://adsabs.harvard.edu/abs/2012ApJ...750...99T} {750, 99}

\bibitem[\protect\citeauthoryear{{Vazza}, {Wittor}, {Brunetti}  \&
  {Br{\"u}ggen}}{{Vazza} et~al.}{2021}]{Vazza2021}
{Vazza} F.,  {Wittor} D.,  {Brunetti} G.,   {Br{\"u}ggen} M.,  2021, arXiv
  e-prints, \href {https://ui.adsabs.harvard.edu/abs/2021arXiv210204193V} {p.
  arXiv:2102.04193}

\bibitem[\protect\citeauthoryear{{Venturi}, {Giacintucci}, {Brunetti},
  {Cassano}, {Bardelli}, {Dallacasa}  \& {Setti}}{{Venturi}
  et~al.}{2007}]{Venturi2007}
{Venturi} T.,  {Giacintucci} S.,  {Brunetti} G.,  {Cassano} R.,  {Bardelli} S.,
   {Dallacasa} D.,   {Setti} G.,  2007, \mndoi [\aap]
  {10.1051/0004-6361:20065961}, \href
  {https://ui.adsabs.harvard.edu/abs/2007A&A...463..937V} {463, 937}

\bibitem[\protect\citeauthoryear{{Venturi}, {Giacintucci}, {Dallacasa},
  {Cassano}, {Brunetti}, {Bardelli}  \& {Setti}}{{Venturi}
  et~al.}{2008}]{Venturi2008}
{Venturi} T.,  {Giacintucci} S.,  {Dallacasa} D.,  {Cassano} R.,  {Brunetti}
  G.,  {Bardelli} S.,   {Setti} G.,  2008, \mndoi [\aap]
  {10.1051/0004-6361:200809622}, \href
  {https://ui.adsabs.harvard.edu/abs/2008A&A...484..327V} {484, 327}

\bibitem[\protect\citeauthoryear{{Wayth} et~al.,}{{Wayth}
  et~al.}{2015}]{wlb+15}
{Wayth} R.~B.,  et~al., 2015, \mndoi [\pasa] {10.1017/pasa.2015.26}, \href
  {http://adsabs.harvard.edu/abs/2015PASA...32...25W} {32, 25}

\bibitem[\protect\citeauthoryear{{Wayth} et~al.,}{{Wayth}
  et~al.}{2018}]{wtt+18}
{Wayth} R.~B.,  et~al., 2018, \mndoi [\pasa] {10.1017/pasa.2018.37}, \href
  {http://adsabs.harvard.edu/abs/2018PASA...35...33W} {35}

\bibitem[\protect\citeauthoryear{{Wilber} et~al.,}{{Wilber}
  et~al.}{2018}]{Wilber2018}
{Wilber} A.,  et~al., 2018, \mndoi [\mnras] {10.1093/mnras/stx2568}, \href
  {https://ui.adsabs.harvard.edu/abs/2018MNRAS.473.3536W} {473, 3536}

\bibitem[\protect\citeauthoryear{{Wilber}, {Johnston-Hollitt}, {Duchesne},
  {Tasse}, {Akamatsu}, {Intema}  \& {Hodgson}}{{Wilber}
  et~al.}{2020}]{Wilber2020}
{Wilber} A.~G.,  {Johnston-Hollitt} M.,  {Duchesne} S.~W.,  {Tasse} C.,
  {Akamatsu} H.,  {Intema} H.,   {Hodgson} T.,  2020, \mndoi [\pasa]
  {10.1017/pasa.2020.34}, \href
  {https://ui.adsabs.harvard.edu/abs/2020PASA...37...40W} {37, e040}

\bibitem[\protect\citeauthoryear{{Wilson} et~al.,}{{Wilson}
  et~al.}{2011}]{cabb}
{Wilson} W.~E.,  et~al., 2011, \mndoi [\mnras]
  {10.1111/j.1365-2966.2011.19054.x}, \href
  {http://adsabs.harvard.edu/abs/2011MNRAS.416..832W} {416, 832}

\bibitem[\protect\citeauthoryear{{Xie} et~al.,}{{Xie} et~al.}{2020}]{Xie2020}
{Xie} C.,  et~al., 2020, \mndoi [\aap] {10.1051/0004-6361/201936953}, \href
  {https://ui.adsabs.harvard.edu/abs/2020A&A...636A...3X} {636, A3}

\bibitem[\protect\citeauthoryear{{de Gasperin}, {van Weeren}, {Br{\"u}ggen},
  {Vazza}, {Bonafede}  \& {Intema}}{{de Gasperin} et~al.}{2014}]{dvb+14}
{de Gasperin} F.,  {van Weeren} R.~J.,  {Br{\"u}ggen} M.,  {Vazza} F.,
  {Bonafede} A.,   {Intema} H.~T.,  2014, \mndoi [\mnras]
  {10.1093/mnras/stu1658}, \href
  {http://adsabs.harvard.edu/abs/2014MNRAS.444.3130D} {444, 3130}

\bibitem[\protect\citeauthoryear{{de Gasperin}, {Ogrean}, {van Weeren},
  {Dawson}, {Br{\"u}ggen}, {Bonafede}  \& {Simionescu}}{{de Gasperin}
  et~al.}{2015}]{deGasperin2015b}
{de Gasperin} F.,  {Ogrean} G.~A.,  {van Weeren} R.~J.,  {Dawson} W.~A.,
  {Br{\"u}ggen} M.,  {Bonafede} A.,   {Simionescu} A.,  2015, \mndoi [\mnras]
  {10.1093/mnras/stv129}, \href
  {https://ui.adsabs.harvard.edu/abs/2015MNRAS.448.2197D} {448, 2197}

\bibitem[\protect\citeauthoryear{{de Gasperin} et~al.,}{{de Gasperin}
  et~al.}{2017}]{deGasperin2017}
{de Gasperin} F.,  et~al., 2017, \mndoi [Science Advances]
  {10.1126/sciadv.1701634}, \href
  {https://ui.adsabs.harvard.edu/abs/2017SciA....3E1634D} {3, e1701634}

\bibitem[\protect\citeauthoryear{{van Weeren}, {R{\"o}ttgering}, {Br{\"u}ggen}
  \& {Hoeft}}{{van Weeren} et~al.}{2010}]{vanWeeren2010}
{van Weeren} R.~J.,  {R{\"o}ttgering} H. J.~A.,  {Br{\"u}ggen} M.,   {Hoeft}
  M.,  2010, \mndoi [Science] {10.1126/science.1194293}, \href
  {https://ui.adsabs.harvard.edu/abs/2010Sci...330..347V} {330, 347}

\bibitem[\protect\citeauthoryear{{van Weeren} et~al.,}{{van Weeren}
  et~al.}{2016}]{vanWeeren2016}
{van Weeren} R.~J.,  et~al., 2016, \mndoi [\apj] {10.3847/0004-637X/818/2/204},
  \href {https://ui.adsabs.harvard.edu/abs/2016ApJ...818..204V} {818, 204}

\bibitem[\protect\citeauthoryear{{van Weeren} et~al.,}{{van Weeren}
  et~al.}{2017}]{vanWeeren2017}
{van Weeren} R.~J.,  et~al., 2017, \mndoi [Nature Astronomy]
  {10.1038/s41550-016-0005}, \href
  {https://ui.adsabs.harvard.edu/abs/2017NatAs...1E...5V} {1, 0005}

\bibitem[\protect\citeauthoryear{{van Weeren}, {de Gasperin}, {Akamatsu},
  {Br{\"u}ggen}, {Feretti}, {Kang}, {Stroe}  \& {Zandanel}}{{van Weeren}
  et~al.}{2019}]{vda+19}
{van Weeren} R.~J.,  {de Gasperin} F.,  {Akamatsu} H.,  {Br{\"u}ggen} M.,
  {Feretti} L.,  {Kang} H.,  {Stroe} A.,   {Zandanel} F.,  2019, \mndoi [\ssr]
  {10.1007/s11214-019-0584-z}, \href
  {http://adsabs.harvard.edu/abs/2019SSRv..215...16V} {215, 16}

\bibitem[\protect\citeauthoryear{{van Weeren} et~al.,}{{van Weeren}
  et~al.}{2020}]{vanWeeren2020}
{van Weeren} R.~J.,  et~al., 2020, arXiv e-prints, \href
  {https://ui.adsabs.harvard.edu/abs/2020arXiv201102387V} {p. arXiv:2011.02387}

\bibitem[\protect\citeauthoryear{van~der Walt, Colbert  \& Varoquaux}{van~der
  Walt et~al.}{2011}]{Numpy2011}
van~der Walt S.,  Colbert S.~C.,   Varoquaux G.,  2011, \mndoi [Computing in
  Science Engineering] {10.1109/MCSE.2011.37}, 13, 22

\makeatother
\end{thebibliography}
}

\begin{appendix}

\section{\emph{u}--\emph{v} coverage plots}\label{appendix:uv}
\begin{figure*}
    \centering
    \begin{subfigure}{0.33\linewidth}
    \includegraphics[draft=false,width=1\linewidth]{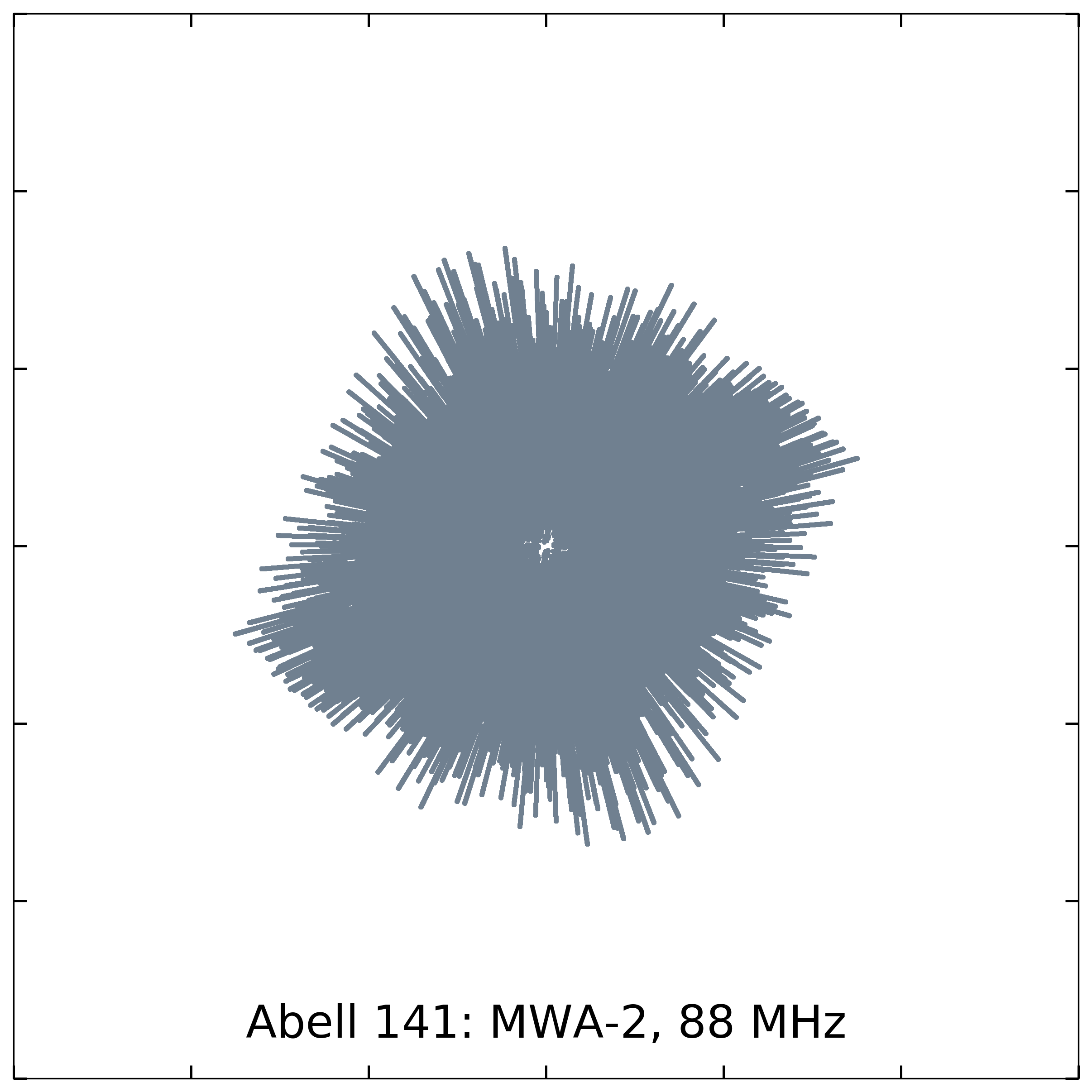}
    \subcaption{\label{fig:uv:a141:mwa:c69}}
    \end{subfigure}%
    \begin{subfigure}{0.33\linewidth}
    \includegraphics[draft=false,width=1\linewidth]{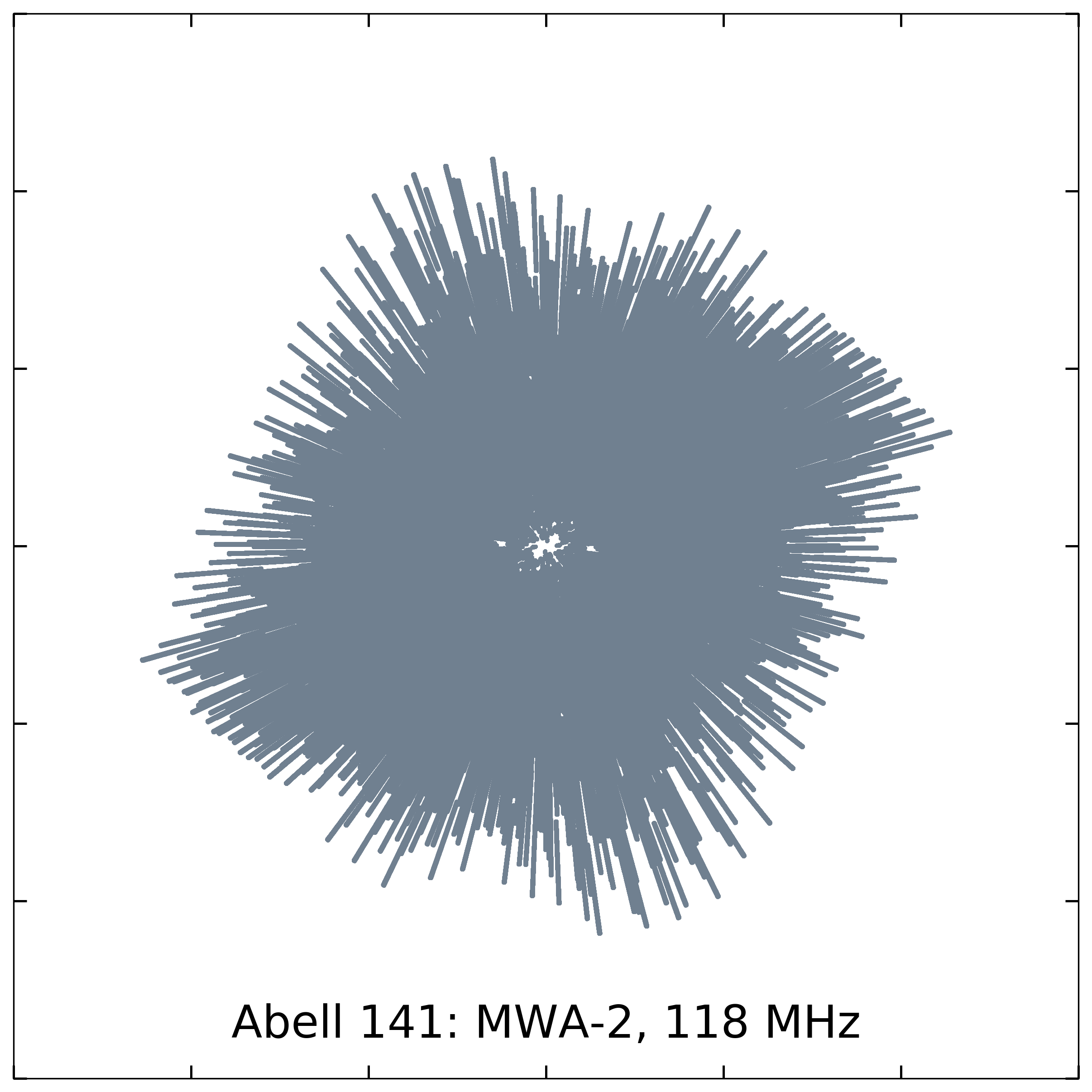}
    \subcaption{\label{fig:uv:a141:mwa:c93}}
    \end{subfigure}%
    \begin{subfigure}{0.33\linewidth}
    \includegraphics[draft=false,width=1\linewidth]{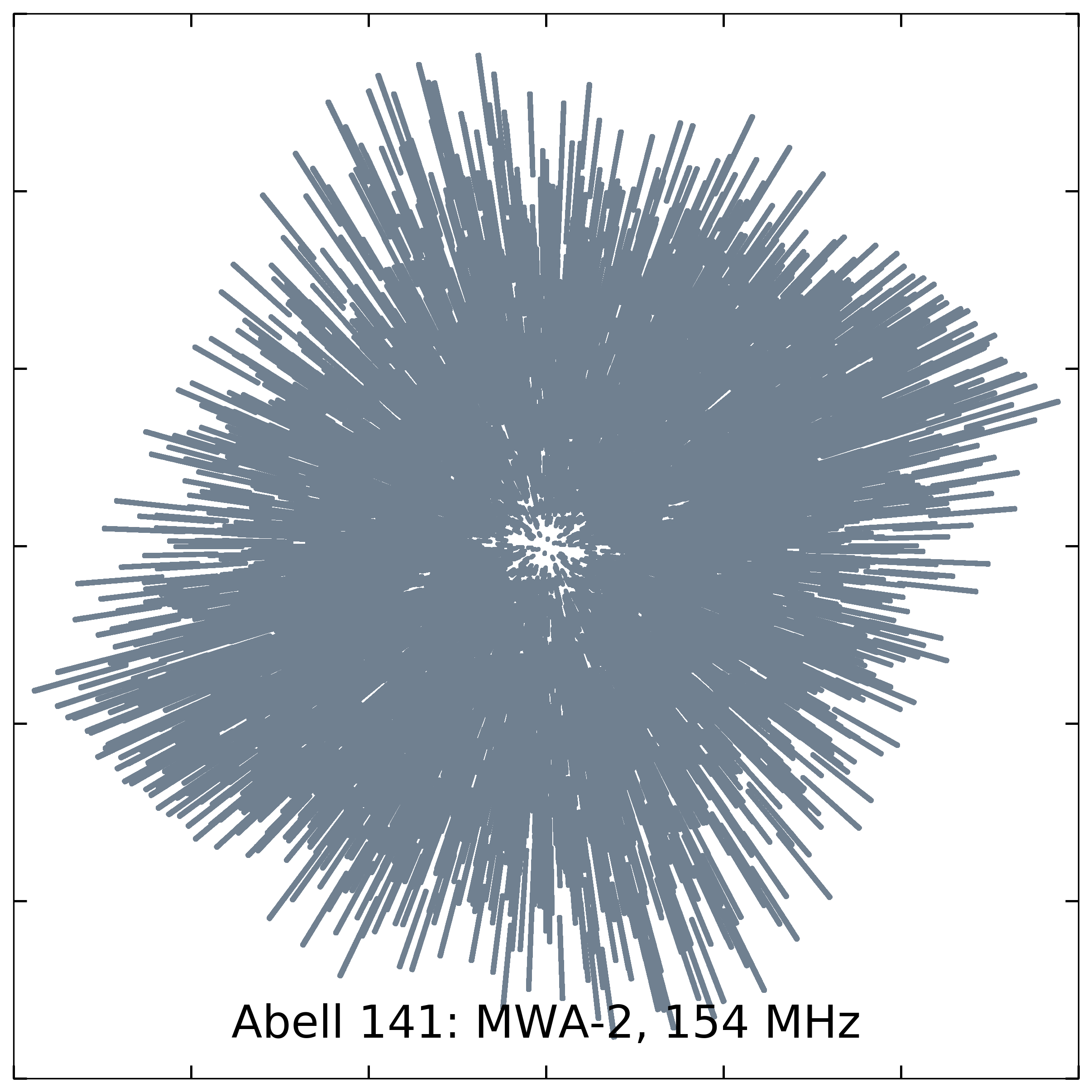}
    \subcaption{\label{fig:uv:a141:mwa:c121}}
    \end{subfigure}\\[0.5em]%
    \begin{subfigure}{0.33\linewidth}
    \includegraphics[draft=false,width=1\linewidth]{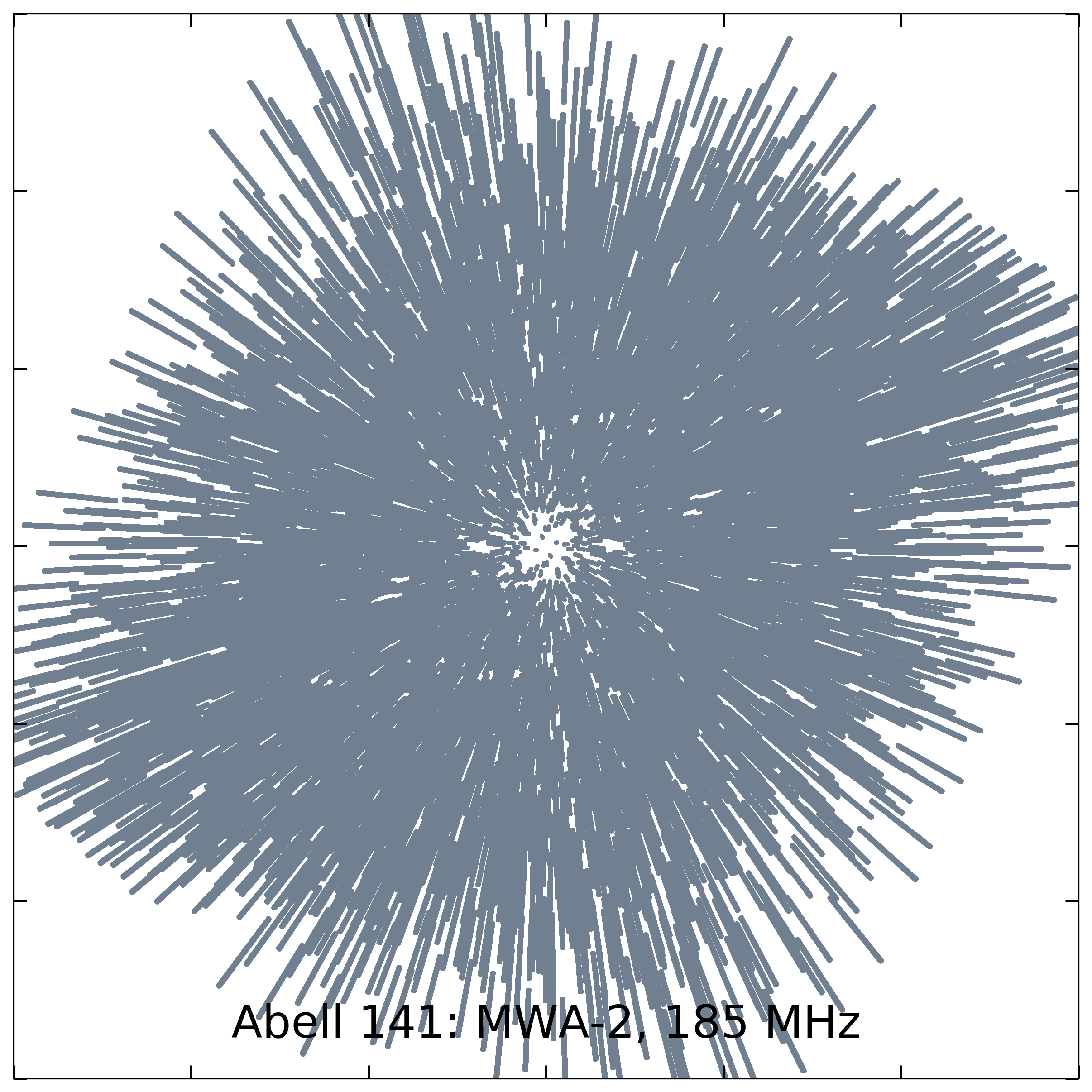}
    \subcaption{\label{fig:uv:a141:mwa:c145}}
    \end{subfigure}%
    \begin{subfigure}{0.33\linewidth}
    \includegraphics[draft=false,width=1\linewidth]{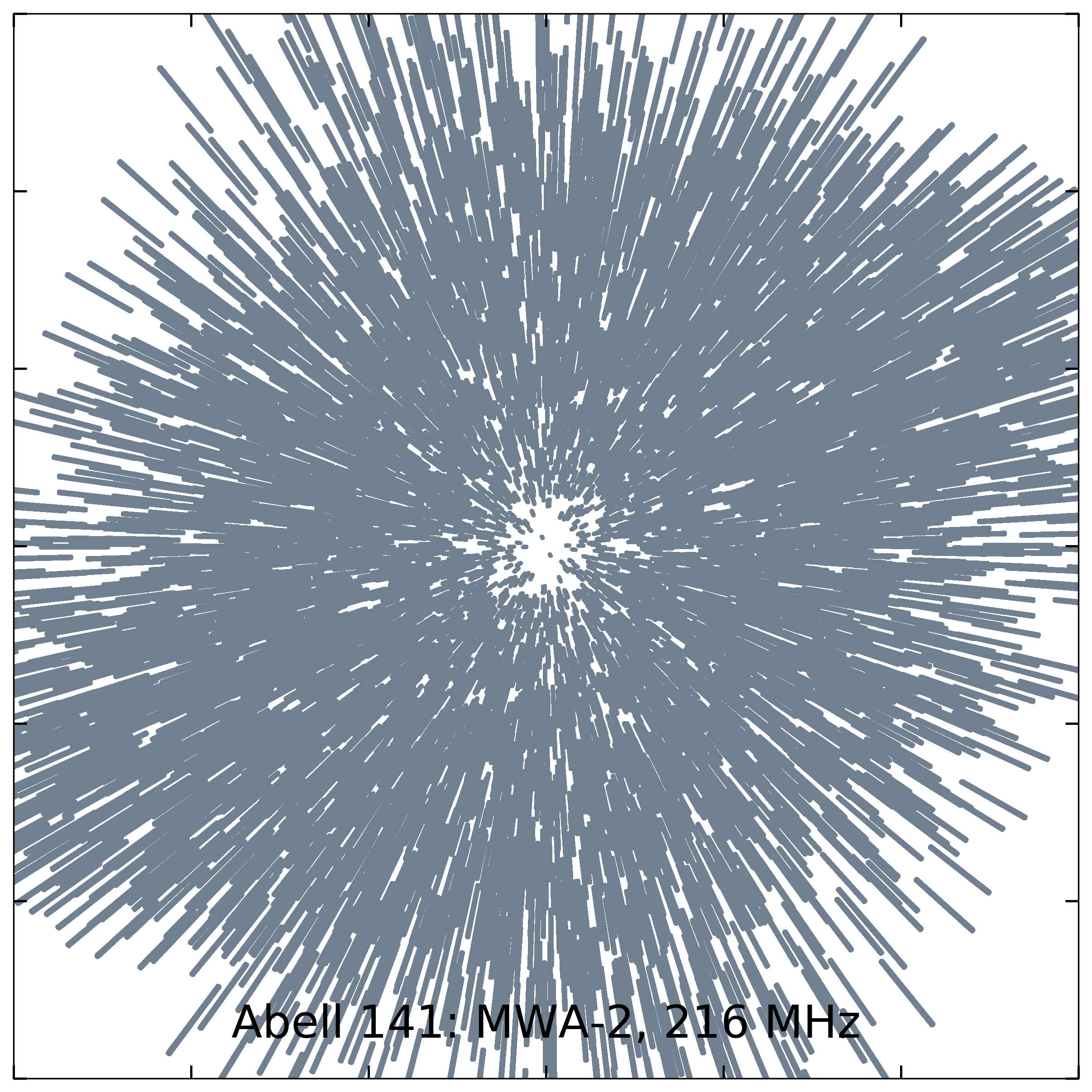}
    \subcaption{\label{fig:uv:a141:mwa:c169}}
    \end{subfigure}%
    \begin{subfigure}{0.33\linewidth}
    \includegraphics[draft=false,width=1\linewidth]{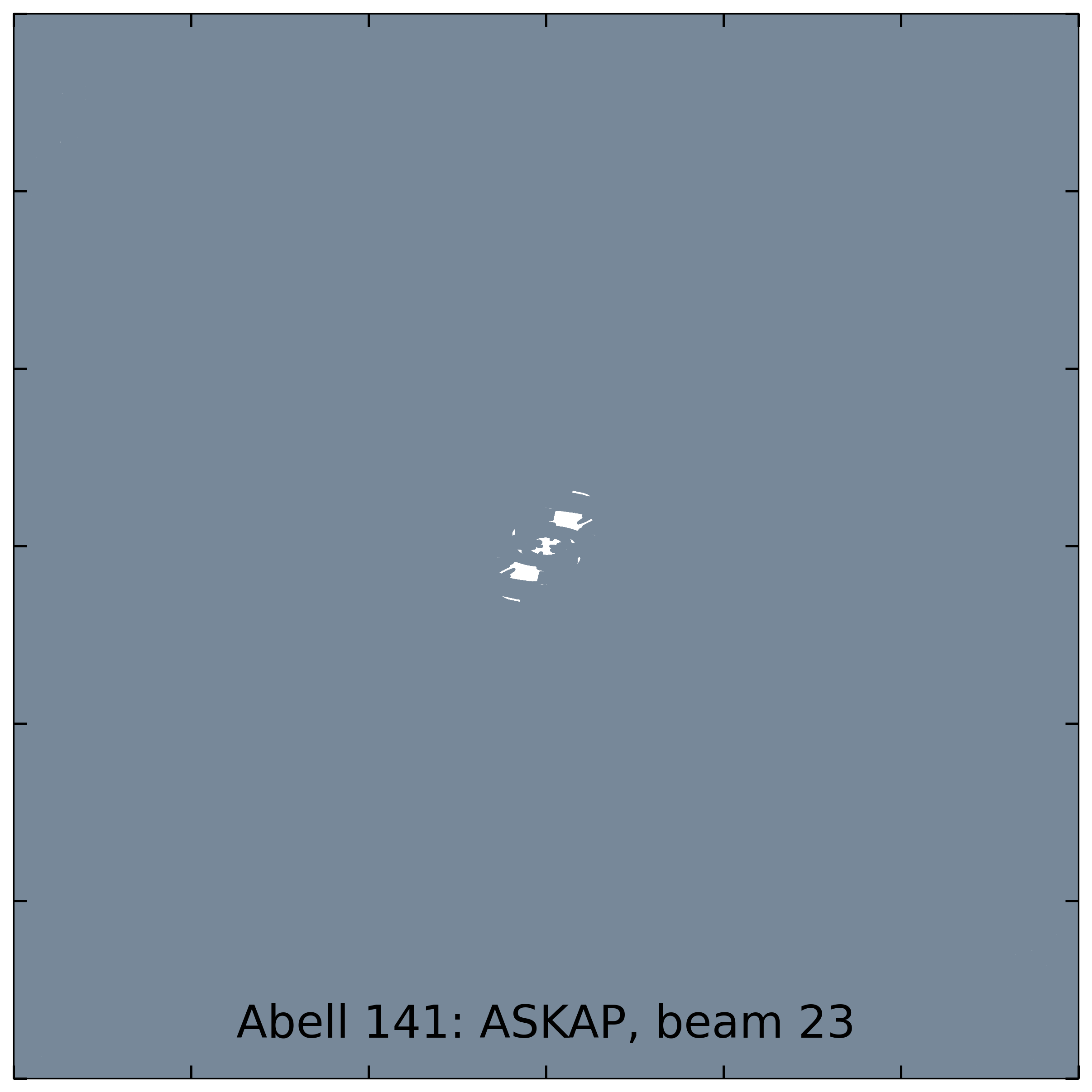}
    \subcaption{\label{fig:uv:a141:askap}}
    \end{subfigure}\\[0.5em]%
    \centering
    \begin{subfigure}{0.33\linewidth}
    \includegraphics[draft=false,width=1\linewidth]{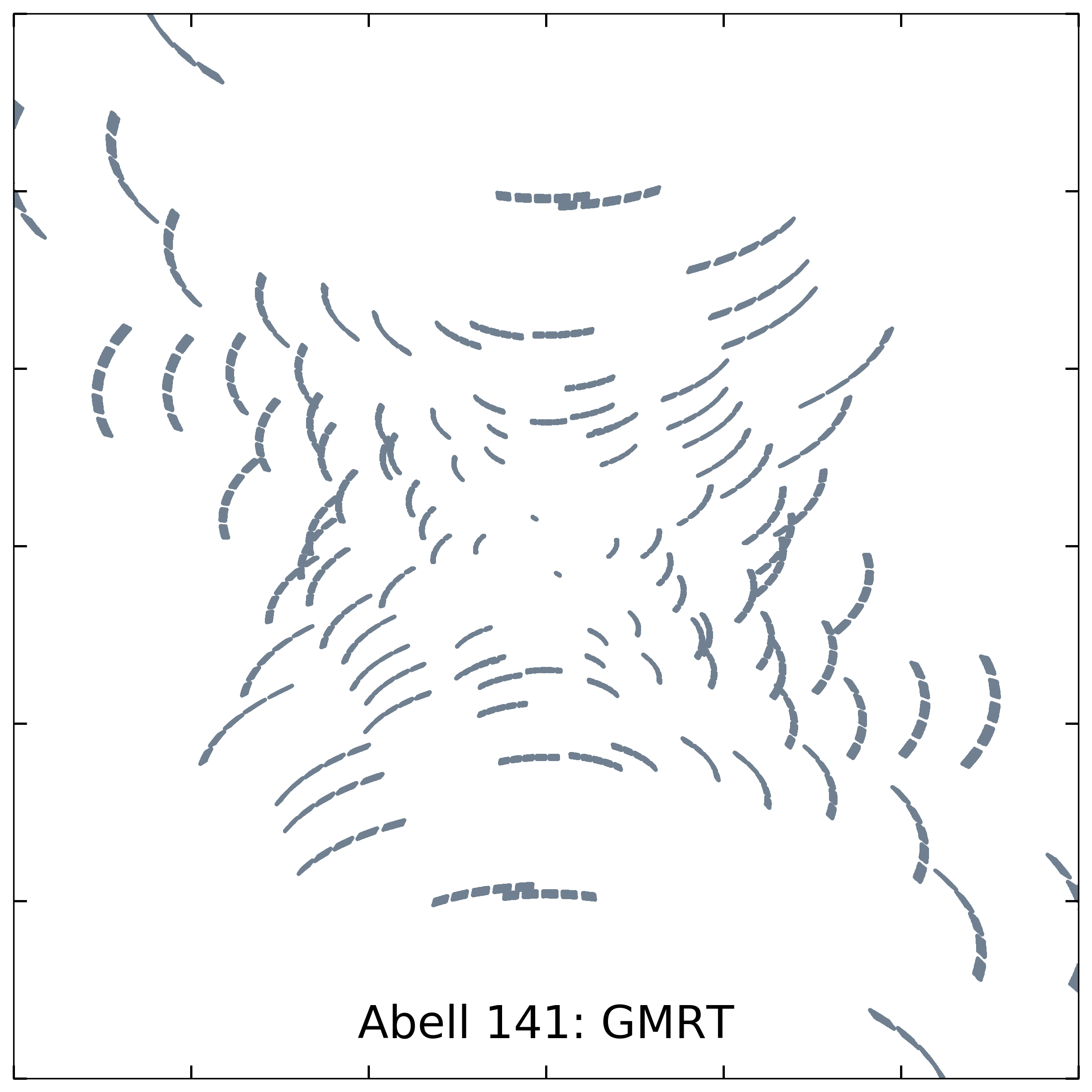}
    \subcaption{\label{fig:uv:a141:gmrt}}
    \end{subfigure}%
    \begin{subfigure}{0.33\linewidth}
    \includegraphics[draft=false,width=1\linewidth]{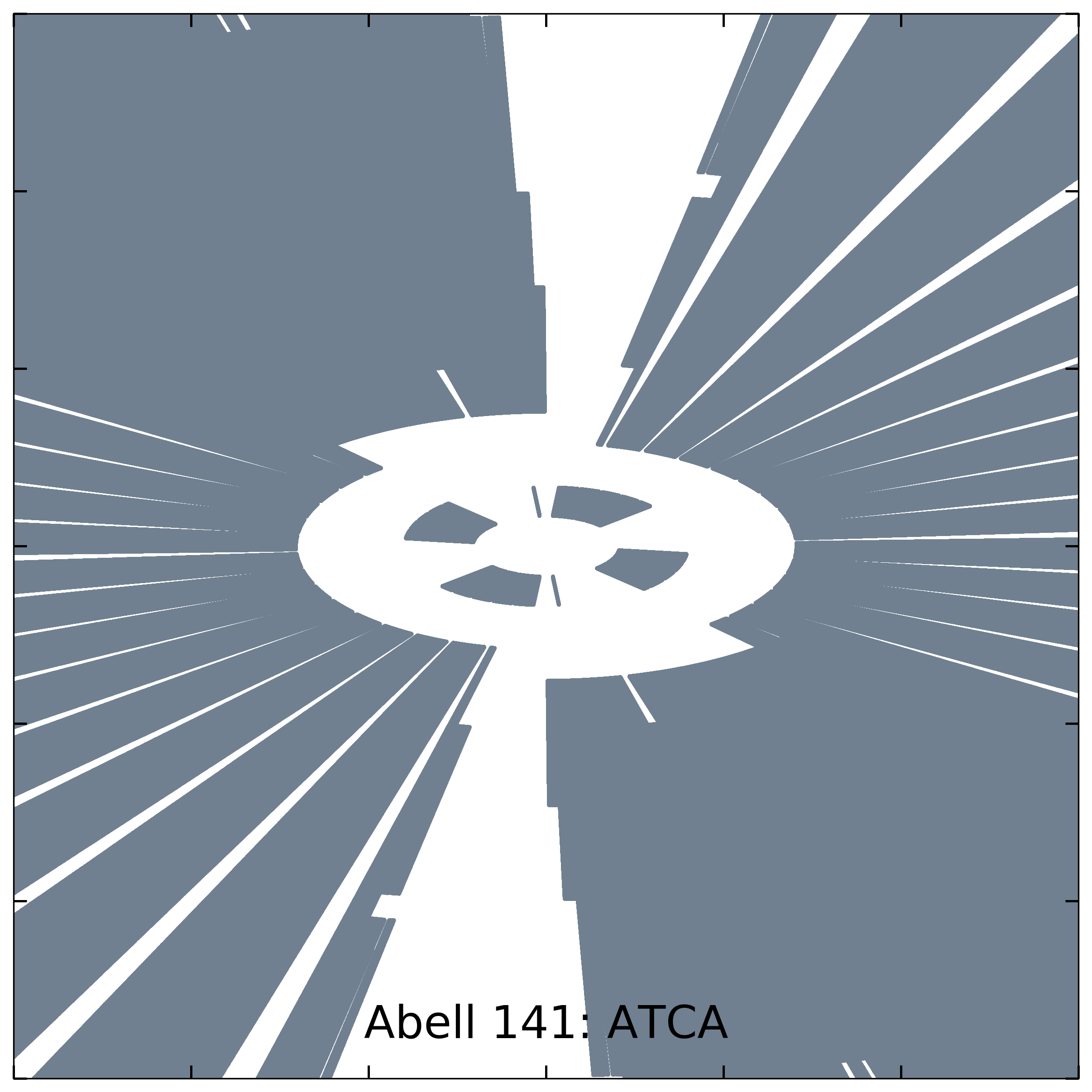}
    \subcaption{\label{fig:uv:a141:cabb}}
    \end{subfigure}%
    \caption{\label{fig:uv:a141} $u$--$v$ coverage plots for Abell~141 data. Axes are centered on zero and range from $-3000\lambda$ to $3000\lambda$. Note that the MWA-2 data are of single 2-min snapshots, representative of the snapshot observations. The true $u$--$v$ coverage is slightly more filled in. \CORRS{The observation used for the ASKAP example is SB15191.}}
\end{figure*}

\begin{figure*}

    \centering
    \begin{subfigure}{0.33\linewidth}
    \includegraphics[draft=false,width=1\linewidth]{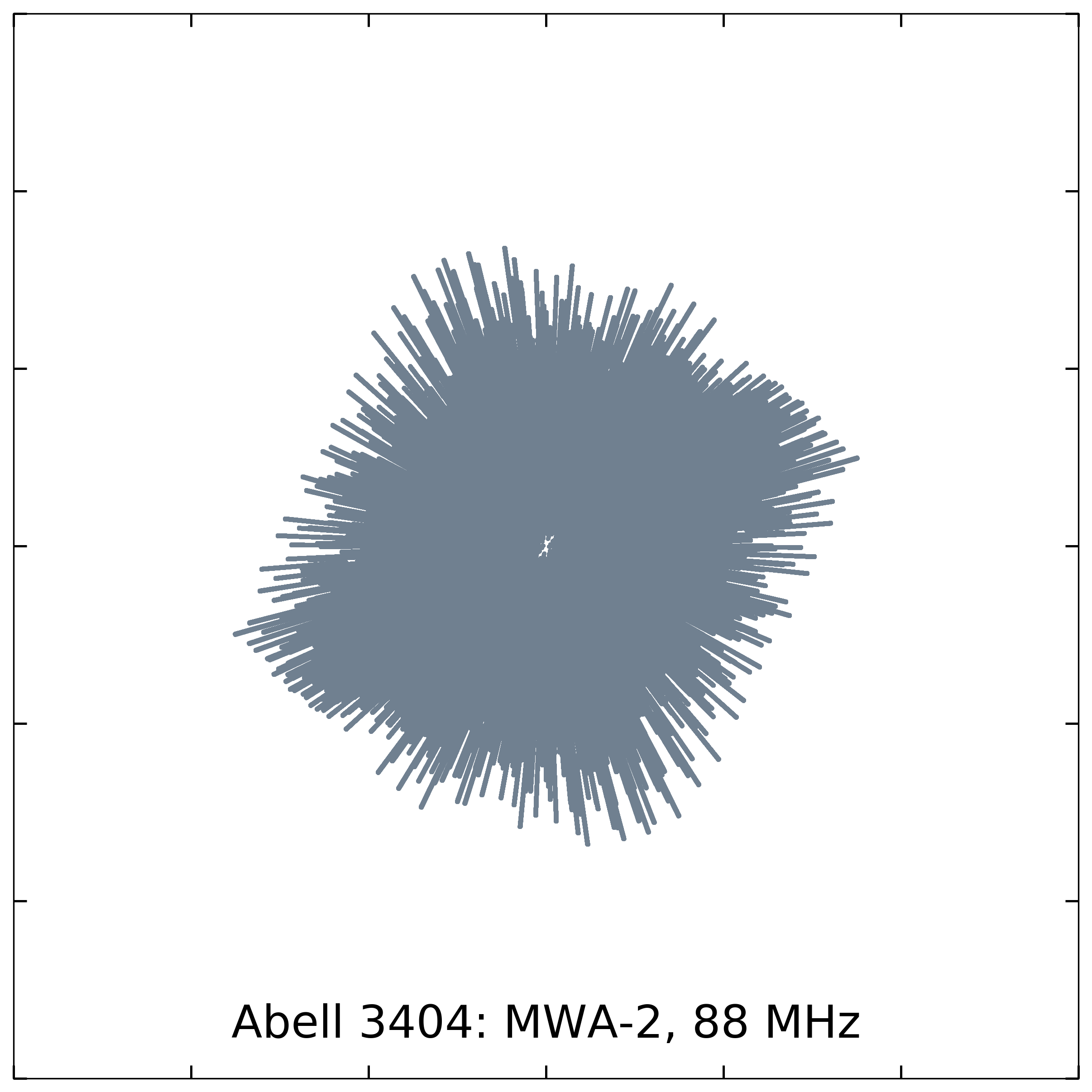}
    \subcaption{\label{fig:uv:a3404:mwa:c69}}
    \end{subfigure}%
    \begin{subfigure}{0.33\linewidth}
    \includegraphics[draft=false,width=1\linewidth]{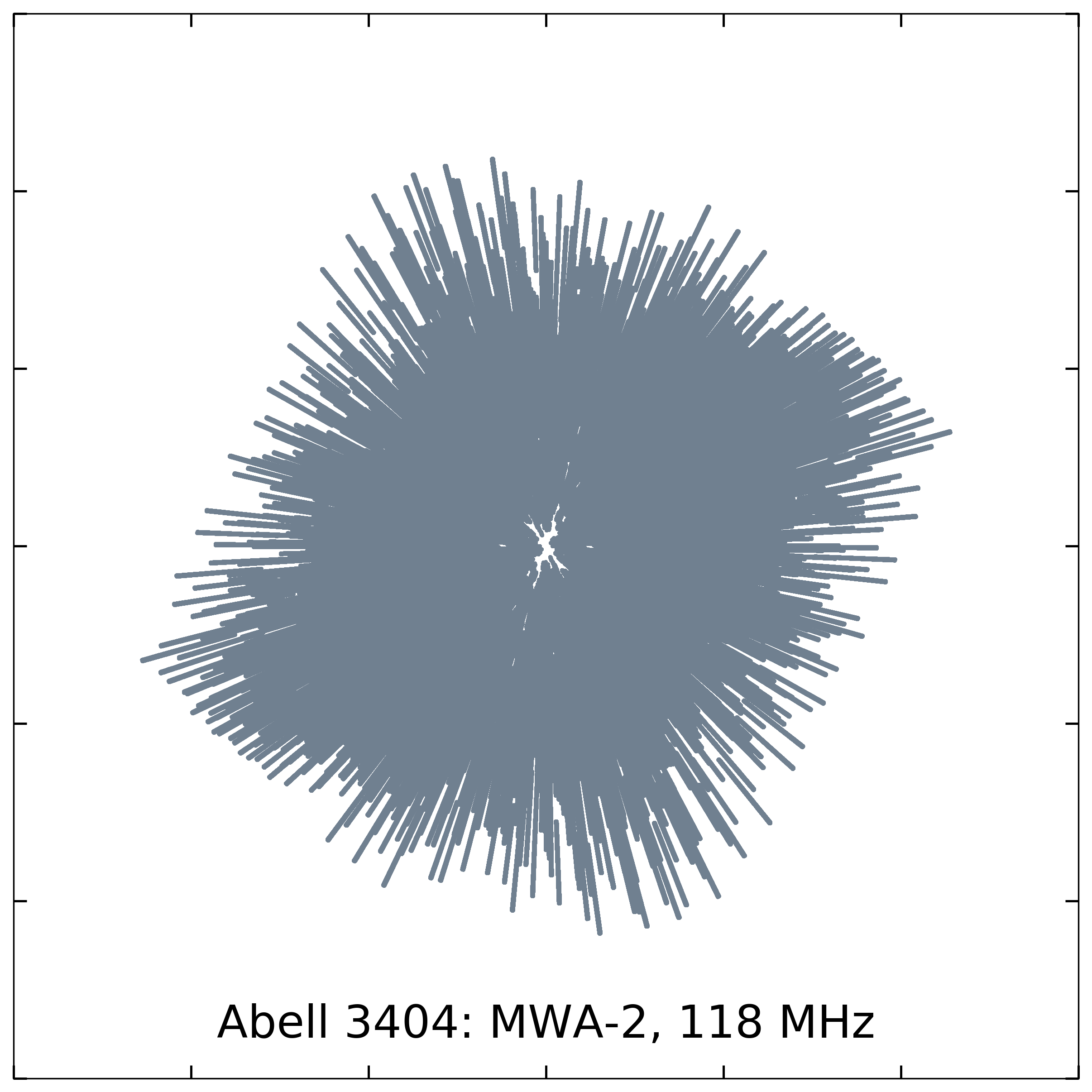}
    \subcaption{\label{fig:uv:a3404:mwa:c93}}
    \end{subfigure}%
    \begin{subfigure}{0.33\linewidth}
    \includegraphics[draft=false,width=1\linewidth]{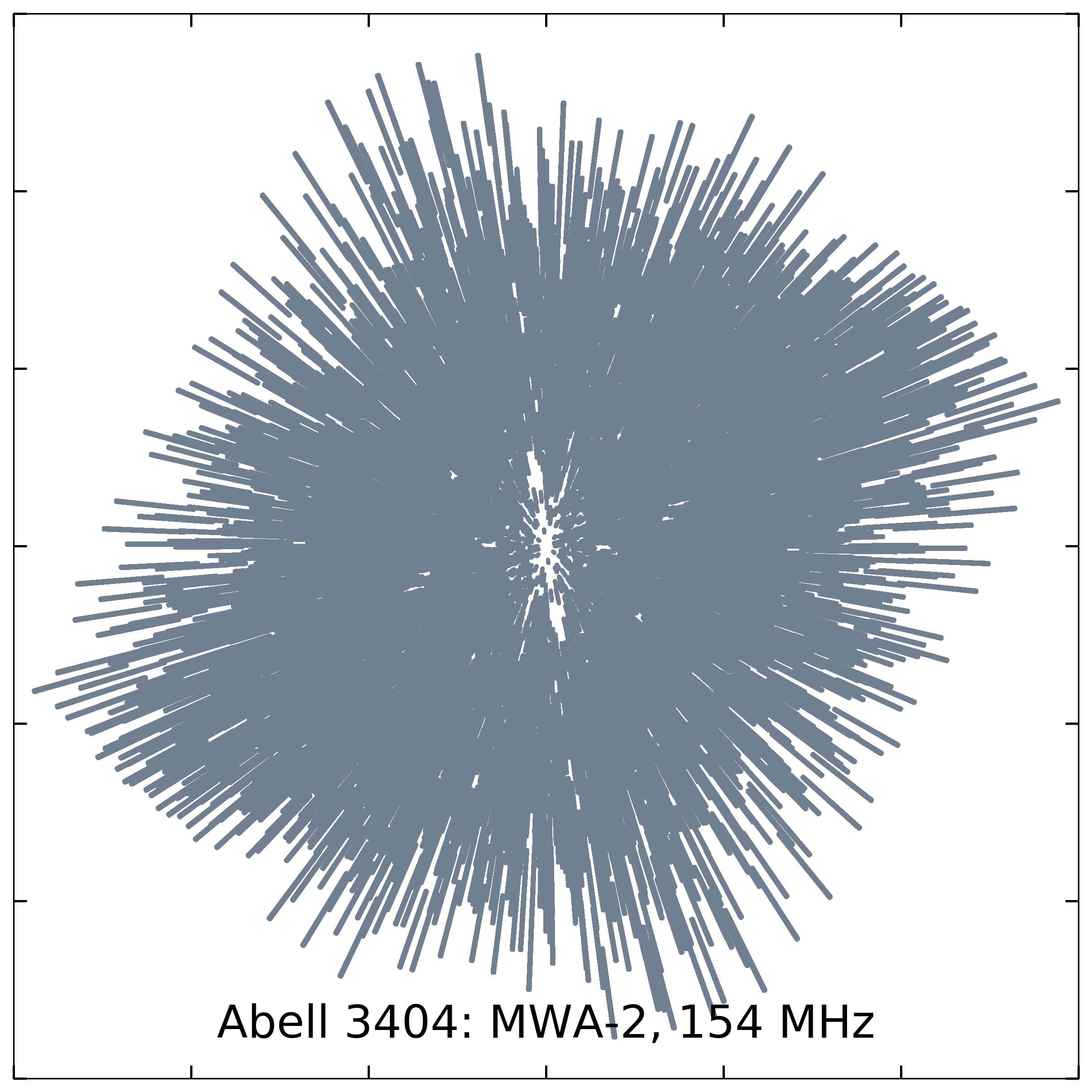}
    \subcaption{\label{fig:uv:a3404:mwa:c121}}
    \end{subfigure}\\[0.5em]%
    \begin{subfigure}{0.33\linewidth}
    \includegraphics[draft=false,width=1\linewidth]{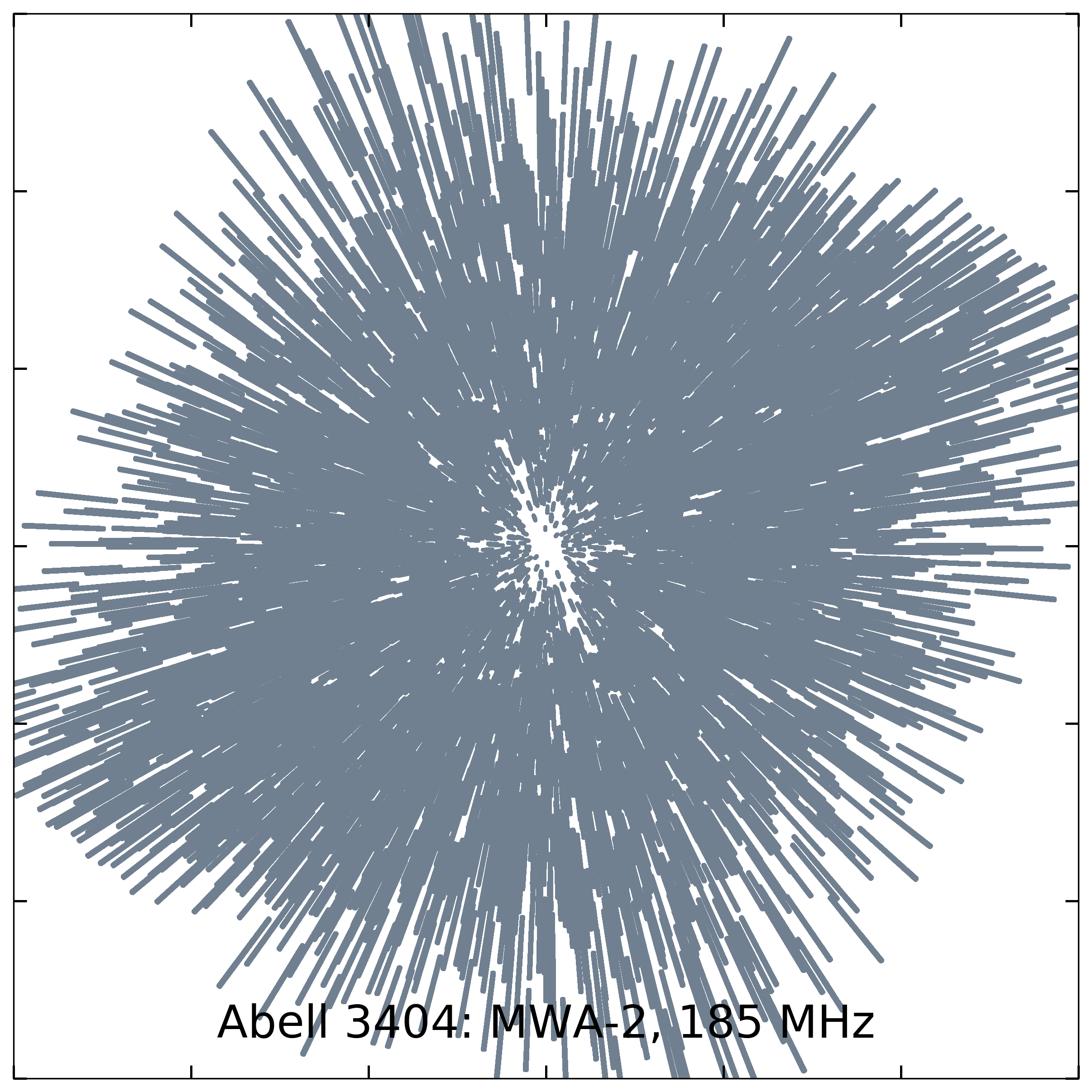}
    \subcaption{\label{fig:uv:a3404:mwa:c145}}
    \end{subfigure}%
    \begin{subfigure}{0.33\linewidth}
    \includegraphics[draft=false,width=1\linewidth]{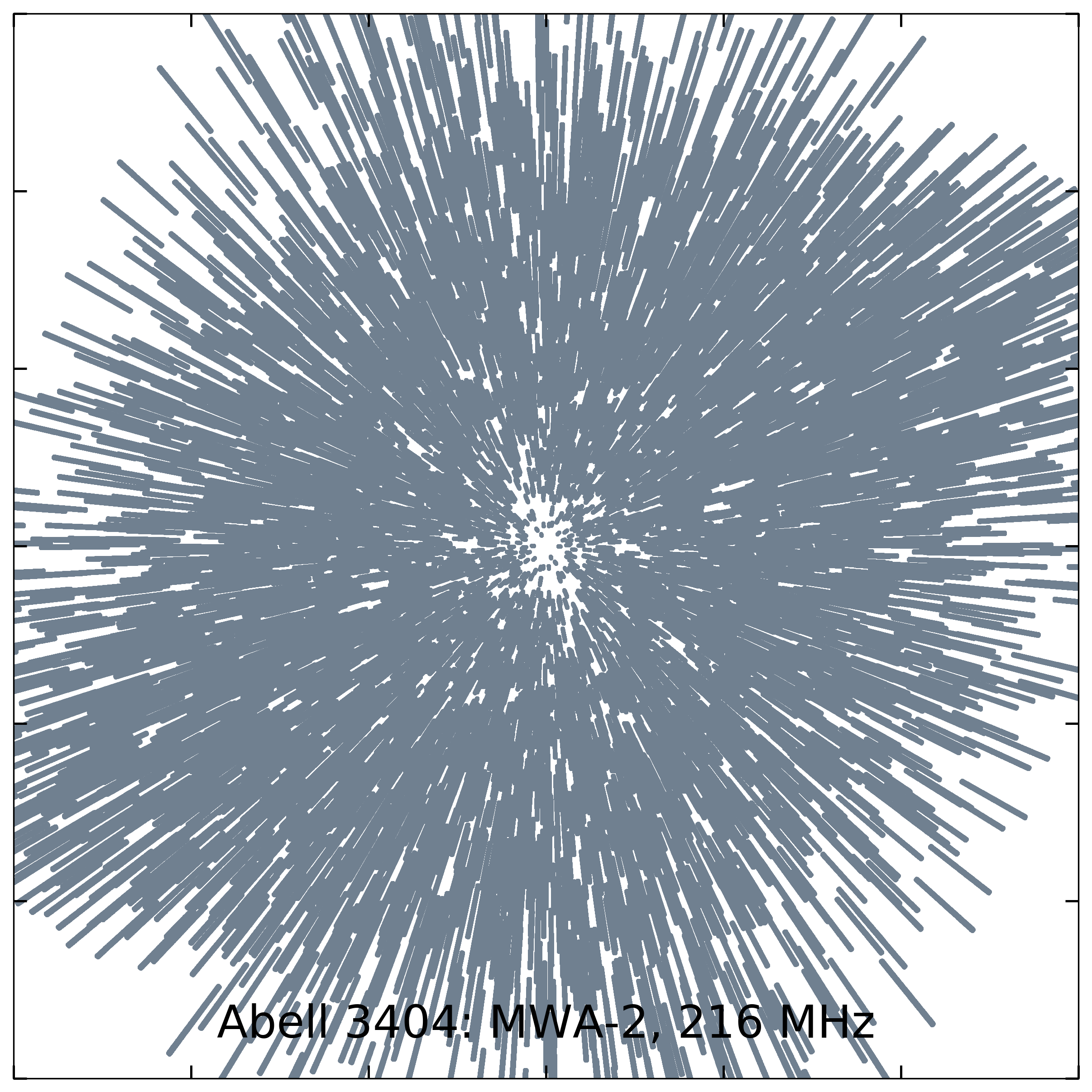}
    \subcaption{\label{fig:uv:a3404:mwa:c169}}
    \end{subfigure}%
    \begin{subfigure}{0.33\linewidth}
    \includegraphics[draft=false,width=1\linewidth]{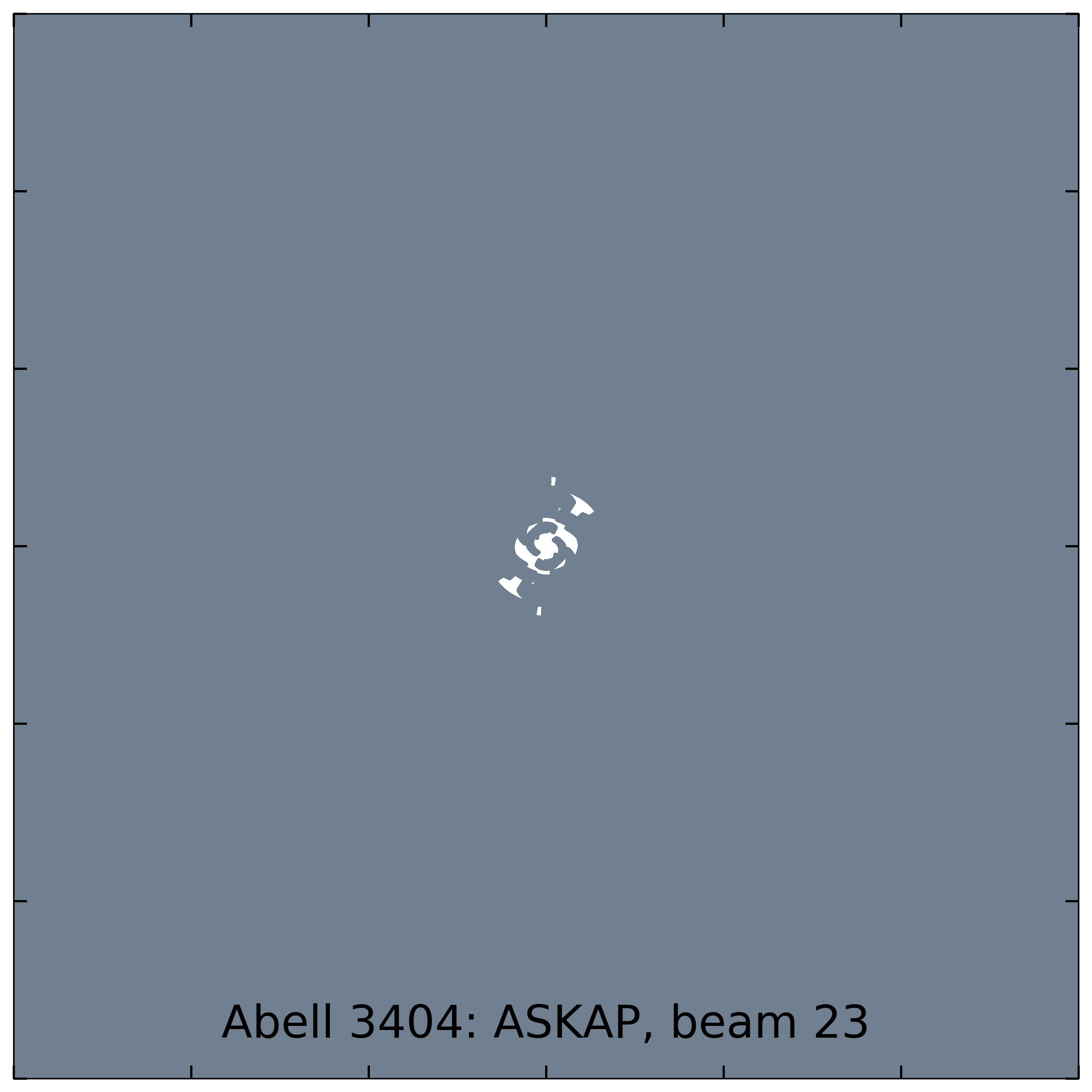}
    \subcaption{\label{fig:uv:a3404:askap}}
    \end{subfigure}\\[0.5em]%
    \begin{subfigure}{0.33\linewidth}
    \includegraphics[draft=false,width=1\linewidth]{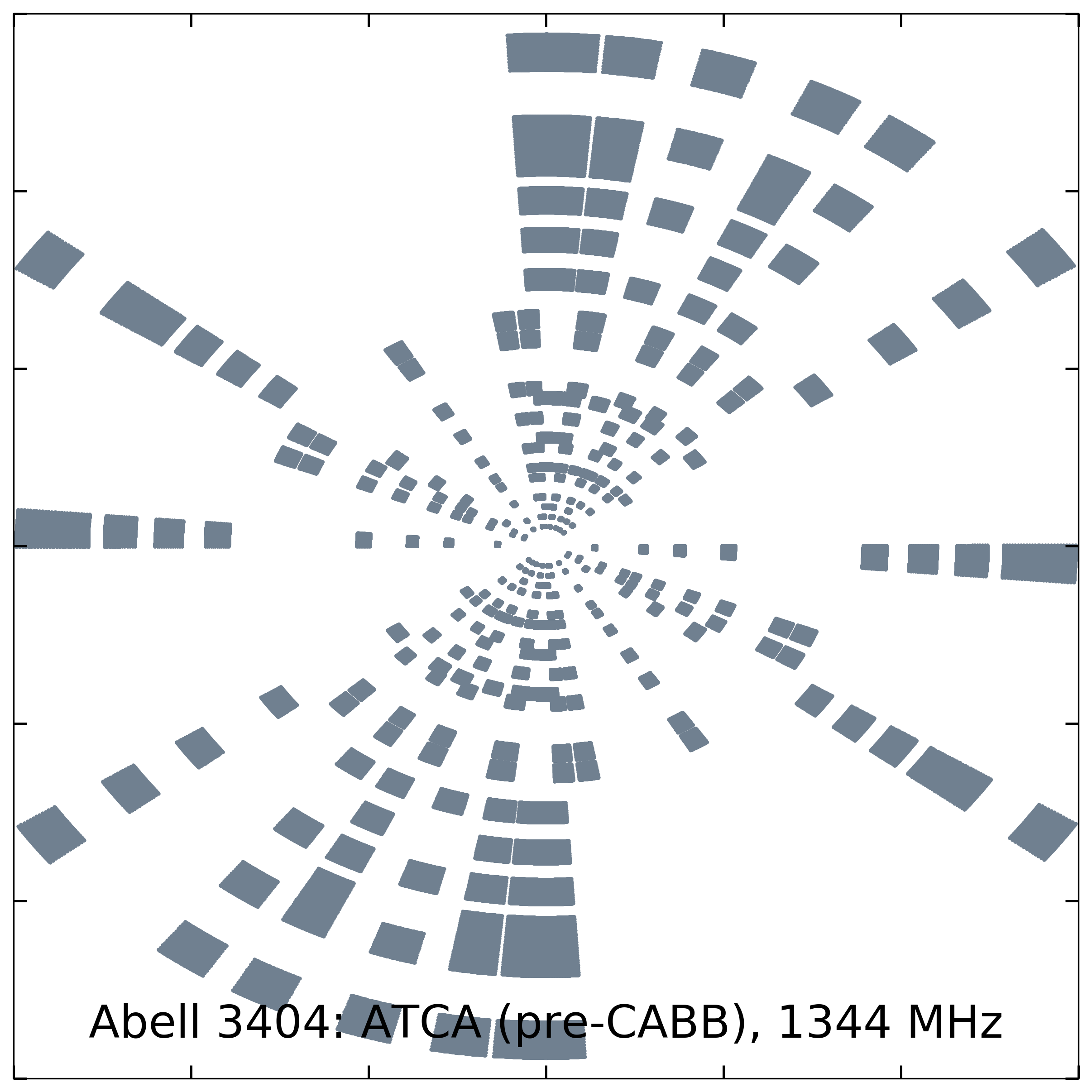}
    \subcaption{\label{fig:uv:a3404:precabb:1344}}
    \end{subfigure}%
    \begin{subfigure}{0.33\linewidth}
    \includegraphics[draft=false,width=1\linewidth]{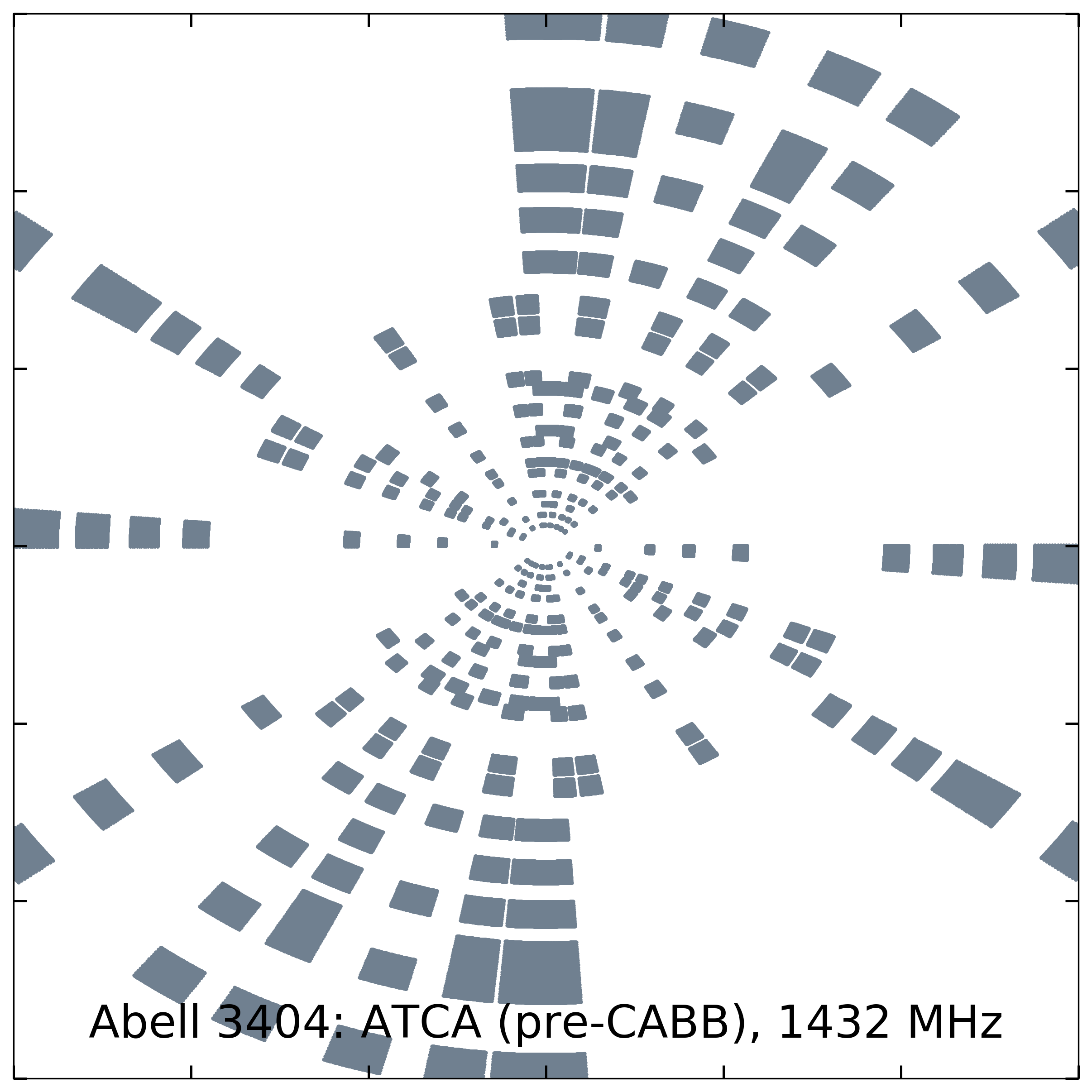}
    \subcaption{\label{fig:uv:a3404:precabb:1432}}
    \end{subfigure}%
    \begin{subfigure}{0.33\linewidth}
    \includegraphics[draft=false,width=1\linewidth]{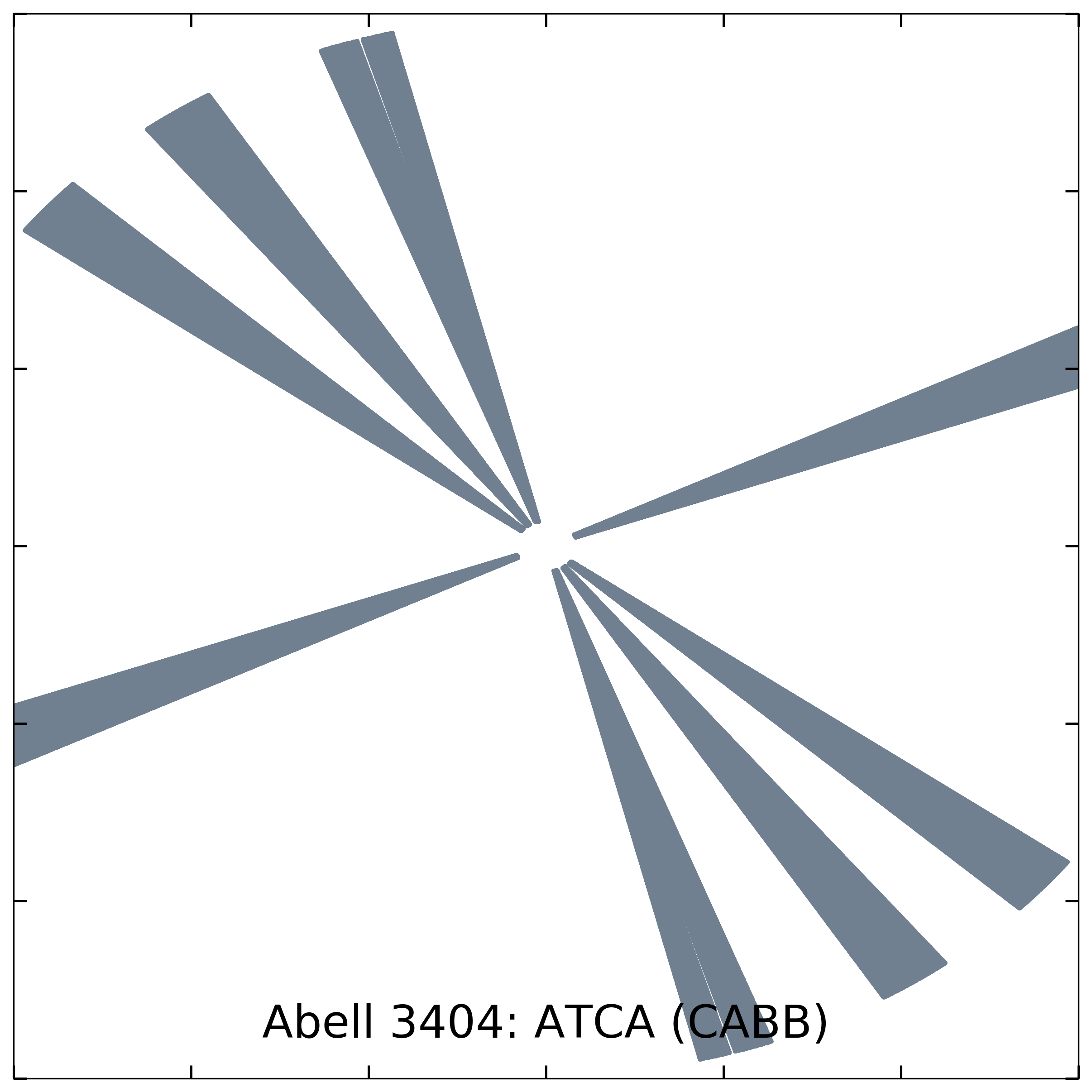}
    \subcaption{\label{fig:uv:a3404:cabb}}
    \end{subfigure}%
    \caption{\label{fig:uv:a3404} $u$--$v$ coverage plots for Abell~3404 data. Axes are centered on zero and range from $-3000\lambda$ to $3000\lambda$.}
\end{figure*}

Fig. \ref{fig:uv:a141:mwa:c69}--\ref{fig:uv:a141:cabb} shows $u$--$v$ coverage (in $\lambda$, from $-3000\lambda$--$3000\lambda$) for the Abell~141 data used in this work, excluding MWA-1 data. Similarly Fig. \ref{fig:uv:a3404:mwa:c69}--\ref{fig:uv:a3404:cabb} show the $u$--$v$ coverage for observations of Abell~3404. We include these plots to highlight the coverage offered by both the MWA and ASKAP in comparison to ATCA and GMRT.

\end{appendix}

% A year is not so long, just once around the Sun.
\end{document}